\documentclass[12pt]{article}

\usepackage{amssymb,esvect,amsmath,graphicx,latexsym,amsthm,slashed,eso-pic,hyperref}
\usepackage[final]{pdfpages}
\usepackage[titletoc]{appendix}
\usepackage{epsfig,amsmath,amssymb,verbatim,mathrsfs,hyperref}
\usepackage{xspace}
\usepackage{xcolor}
\usepackage{slashed}
\usepackage{dcolumn}
\usepackage{multirow}
\usepackage{setspace}
\graphicspath{{./}{Figs/}}

\def\beq{\begin{eqnarray}}
\def\eeq{\end{eqnarray}}
\def\bea{\begin{eqnarray}}
\def\eea{\end{eqnarray}}

\newcommand{\gsim}{\lower.7ex\hbox{$\;\stackrel{\textstyle>}{\sim}\;$}}
\newcommand{\lsim}{\lower.7ex\hbox{$\;\stackrel{\textstyle<}{\sim}\;$}}

\newcommand{\newc}{\newcommand}
\newc{\hfpi}{f_{\tilde\pi}}
\newc{\fpi}{f_{\pi}}
\newc{\hpi}{\widetilde\pi}
\newc{\hl}{\widetilde\Lambda}
\newc{\lqcd}{\Lambda_{\rm QCD}}
\newc{\tqcd}{\theta_{\rm QCD}}
\newc{\nhc}{N_{\widetilde c}}
\newc{\hg}{\widetilde g}
\newc{\heta}{\widetilde\eta}
\newc{\etap}{\eta^{\, \prime}}
\newc{\hetap}{\widetilde\eta^{\, \prime}}
\newc{\hrho}{\widetilde\rho}
\newc{\octet}{{\widetilde{\pi}}^8}
\newc{\thetatwo}{{\overline{\theta}_2}}
\newc{\pizero}{{\widetilde{\pi}^0}}
\newc{\hth}{\widetilde \theta}
\newc{\bto}{\overline{\theta}_1}
\newc{\llll}{\langle\lambda\lambda\rangle}
\newc{\FFd}{F^a\tilde F^a}
\newc{\qbar}{{\overline q}}
\newc{\TR}{{\rm Tr}}
\newc{\Kahler}{K\"ahler }
\newc{\Zbb}{{\mathbb Z}}
\newc{\Rt}{{\mathbb R}^3}
\newc{\Rf}{{\mathbb R}^4}
\newc{\So}{{\mathbb S}^1}
\newc{\zt}{{\mathbb Z}_2}
\newc{\RtSo}{{\mathbb R}^3\times{\mathbb S}^1}
\newc{\scriminus}{{\cal I}^-}
\newc{\scriplus}{{\cal I}^+}
\newc{\Ricci}{\mathcal{R}}
\newc{\bv}{\phi}
\newc{\calU}{{\cal U}}
\newc{\calK}{K}
\newc{\calUi}{{\cal U}^{-1}}
\newc{\calG}{{\cal G}}
\newc{\calO}{{\cal O}}
\newc{\calOb}{{\cal O}^\dagger}
\newc{\hphi}{{\hat\phi}}
\newc{\qcdp}{{\rm QCD}^\prime}

\begin{document}
\begin{titlepage}
\begin{flushright}
{\large 
ACFI-T18-01\\
}
\end{flushright}

\noindent
\begin{center}
  \begin{LARGE}
    \begin{bf}
Theta in new QCD-like sectors
     \end{bf}
  \end{LARGE}
\end{center}
\vspace{0.3cm}
\begin{center}
\begin{large}
Patrick Draper, Jonathan Kozaczuk, and Jiang-Hao Yu
\end{large}
\vspace{1cm}\\
\begin{it}
Amherst Center for Fundamental Interactions, Department of Physics,\\ 
University of Massachusetts, Amherst, MA 01003, USA

\vspace{0.2cm}
\end{it}
\end{center}
\center{\today}

\begin{abstract}
New QCD-like ``hypercolor" sectors can generate a broad class of new signatures at hadron colliders and furnish a variety of dark matter candidates. Paired diboson resonances are a particularly important collider signature, arising both from $CP$-conserving vector hypermeson decays of the form $\hrho\rightarrow\widetilde\pi\widetilde\pi\rightarrow 4V$ and from $CP$-violating pseudoscalar hypermeson decays of the form $\widetilde\eta\rightarrow \widetilde\pi\widetilde\pi\rightarrow 4V$. The latter are sensitive to the vacuum angle $\hth$ in the hypercolor sector. We study single- and paired-diboson resonance signatures in final states involving gluons and photons at the LHC and a future 100 TeV $pp$ collider, illustrating the discovery potential at both colliders in simple benchmark models. We also describe some of the theoretical and cosmological consequences of $\hth$. If $CP$-violating hypermeson decays are observable at hadron colliders, ordinary QCD must have an axion. Such scenarios also provide a natural setting for a dark pion component of dark matter, with its relic abundance set by $CP$-violating annihilations. If the new vacuum angle is relaxed to zero by a dark axion, the relic density can instead be a mixture of axions and dark axions. Overproduction of dark axions is most easily avoided if the universe underwent a period of early matter domination.

\end{abstract}

\vskip 1.0 cm

\end{titlepage}

\setcounter{footnote}{0} \setcounter{page}{2}
\setcounter{section}{0} \setcounter{subsection}{0}
\setcounter{subsubsection}{0}
\setcounter{figure}{0}

\tableofcontents


\section{Introduction\label{sec:intro}}

New strongly coupled gauge sectors are an important family of new physics models encompassing a broad range of collider and cosmological signatures. Apart from phenomenological interest, they may also be associated with mechanisms of electroweak naturalness, dark matter, and other fundamental problems. Moreover, the discovery of such a sector would present a valuable new handle on strongly coupled 4D gauge theories, which remain theoretically challenging and of which we have only one other experimental example.

 Vectorlike Confinement (VC) models, or ``hypercolor" sectors, are a  rich class of models of new strong dynamics with basic properties similar to those of QCD~\cite{kilic1,kilic2}. In VC, new matter fields charged under hypercolor obey approximate chiral symmetries that are spontaneously broken near the confinement scale, leading to a tower of hadron-like states. The  hyperquarks are also given vector-like charges under the Standard Model (SM) gauge interactions, providing couplings between the new bound states and SM fields. The collider phenomenology of VC models has received much attention in the past and motivates a variety of searches at the LHC (see, for example,~\cite{kilic2, Kilic:2008pm, Kilic:2008ub, Bai:2011mr, Schumann:2011ji, baibargerberger, craigdraperkilicthomas}.)

The purpose of this work is to explore some of the phenomenological and cosmological roles of the vacuum angle in hypercolor sectors. In QCD, the vacuum angle $\theta$ is known to be very small~\cite{ramsey,baker}, and explaining this small number dynamically is a deep and open problem. In a new QCD-like sector, the new vacuum angle $\hth$ might again be small, or it might be ${\cal O}(1)$. The latter can lead to interesting new collider signatures, and both cases have implications for cosmology.

In QCD, if $\theta$ were large, chiral perturbation theory (ChPT) predicts that the $\eta$ meson would exhibit $CP$-violating decays into pairs of charged and neutral pions. In the neutral case, we would then observe $4\gamma$ final states reconstructing one $4\gamma$ parent resonance and two $2\gamma$ daughter resonances,
\begin{align}
\eta\rightarrow\pi^0\pi^0\rightarrow (2\gamma)(2\gamma)\;.
\end{align}
Analogously, in a hypercolor sector, heavier spin-0 hyperpions can decay to pairs of lighter hyperpions through $\hth$-dependent interactions. The decays of the lighter states are typically much richer than the QCD example, involving other SM diboson pairs through anomaly-induced couplings. Here we will focus on the paired diboson/$4V$ resonance topology with $V\,=$ gluons (jets) and photons.

The resonant 4-boson topology is also of broader relevance to hypercolor sector phenomenology. The new sector should possess (a family of) spin-1 mesons analogous to the $\rho$ of QCD, which decay primarily into pairs of hyperpions. We will therefore include $\rho\rightarrow\pi\pi\rightarrow4V$-type topologies among our analyses. These are $CP$-conserving decays and insensitive to $\hth$, but are important in the spirit of motivating new search channels at colliders. 

In the first part of this paper, we study $4g$, $2g2\gamma$, and $3g1\gamma$ resonance signatures of new QCD-like sectors at the high-luminosity LHC and a future 100 TeV $pp$ collider. We work in simple VC models with states analogous to the QCD $\pi^0,\eta,\rho,$ and $\eta^\prime$, as well as color octet hyperpions. Single-hyperpion couplings to the SM are induced by chiral anomalies and allow for resonant production in gluon fusion as well as diboson decays. We find that $CP$-violating decays of the $\eta$-like state can provide a powerful probe of these models at the LHC, with the sensitivity in the $2g2\gamma$ final state surpassing that of more conventional diboson ($gg$, $g\gamma$, $\gamma \gamma$) searches for $\mathcal{O}(1)$ vacuum angles. $CP$-violating $\eta^{\prime}$-like decays will be more difficult to observe at the LHC given current limits from diboson searches on these models; similar conclusions hold for $CP$-conserving decays of the $\rho$-like state. However, a future 100 TeV collider will have an opportunity to probe new QCD-like sectors with confinement scales up to $\sim 40$ TeV via $\eta$, $\eta^{\prime}$, and $\rho$-like decays to the hyperpions.  The interplay between the various tetraboson and diboson searches is summarized in Sec.~\ref{sec:implications}. 

In the latter part of this work, we examine the interplay between a new $\hth$ angle and three dark matter candidates: the QCD axion, a stable hyperpion, and a ``dark axion" coupled to the hypercolor sector. First, we show that when hypermesons can be resonantly produced in gluon fusion, an observably large $\hth$ is incompatible with UV solutions to the strong $CP$ problem, providing indirect evidence for the QCD axion. Secondly, hypercolor models tend to exhibit accidental symmetries which can be promoted to stabilize a variety of dark matter candidates~\cite{baihill,strumia1,strumia2,craigdraperkilicthomas}, and it has previously been noted that $\hth$-dependent couplings can provide the dominant annihilation channels setting the relic abundance of stable hyperpions~\cite{strumia1}. Such models thus naturally accommodate mixed WIMP-axion dark matter, and we note that  thermal broadening of the hyperpion annihilation rates favors the regime where $\heta\rightarrow\hpi\hpi$-type decays are most readily observable. Finally, we consider the possibility that both QCD and hypercolor couple to axions. In this case, $\hth$ is relaxed and does not give rise to collider-observable processes. In a conventional radiation-dominated history, this scenario is tightly constrained by dark axion cosmology, but we show that an acceptable relic abundance from misalignment can be achieved if the universe experienced a period of early matter domination.

In Sec.~\ref{sec:concl} we summarize and conclude. Some details of the  models discussed in Sec.~\ref{sec:models} and utilized in Secs.~\ref{sec:paireddiboson}-\ref{sec:cosmo} are collected in the Appendix.

\section{Models} \label{sec:models}
For illustration, we employ two simple models of new confining $SU(\nhc)$ gauge sectors in this work. In numerical results we take the number of hypercolors $\nhc=3$, but we keep it general in analytic expressions. The models differ in their matter content and both are minimal in different senses. We refer to the new sectors generically as hypercolor or $\qcdp$. Here we summarize their salient properties; expressions for the relevant masses and couplings are given in the Appendix.

Each model includes light hyperquarks $\psi_i,\bar\psi_i$ in the fundamental representation of hypercolor, and the hyperquarks carry vectorlike charges under SM gauge groups. Since we will focus on gluon and photon final states at colliders, for simplicity, we only give $SU(3)_c$ and $U(1)_Y$ quantum numbers to the hyperquarks, but this can be generalized to produce couplings to $W$s (see e.g.~Ref.~\cite{baidobrescu} for a recent study of related multi-boson topologies involving $W$s and $Z$s.)  The models are assumed to confine and spontaneously break chiral symmetries, leading to a low-energy effective description in terms of a chiral Lagrangian with cutoff $\hl$.  

\subsection{$U(5)$}
The first model possesses one color triplet hyperquark, which we take to be in the $(3,1)_{4/3}$ representation of the SM gauge groups, and two SM-neutral hyperquarks $(2\times(1,1)_0)$. The model has an approximate classical $U(5)\times U(5)$ global symmetry spontaneously broken to $U(5)_V$, of which an $SU(3)$ subgroup is identified with color.\footnote{We will include the anomalous axial $U(1)$ and associated $\hetap$ in some of our chiral perturbation theory analysis, with the understanding that it is only a heuristic model subject to analytically incalculable ${\cal O}(1)$ corrections unless $\nhc$ is large.}

The pseudo-Goldstone spectrum of the $U(5)$ model includes light neutral hyperpion states $\hpi^0$ and $\heta$, analogous to the $\pi^0$ and $\eta$ of QCD, as well as an intermediate-mass color octet hyperpion $\hpi^8$. It also includes neutral and QCD triplet states charged under accidental global ``species" symmetries. Higher-dimension operators can be added to allow the triplets to decay, and we will not consider them further in this work.\footnote{Dimension-6 operators are possible with $Y=2/3$ and $Y=4/3$; we choose the latter as a benchmark to maximize the reach of photon channels.} The neutral stable states can be dark matter candidates and we label them  $\hpi_{\rm DM}$. We will not discuss the $\hpi_{\rm DM}$ states at colliders, but some of their cosmological aspects are discussed in Sec.~\ref{sec:cosmo} below.  Near the chiral Lagrangian cutoff, the model also includes an assortment of $\hrho$ mesons, an $\hetap$, hyperbaryons, and other heavy resonances. 

The $U(5)$ model has the minimal field content exhibiting:
\begin{itemize}
\item couplings to QCD (allowing for significant production of some states at colliders), and 
\item $\hth$-dependent parity-violating decays in the calculable framework of chiral perturbation theory. 
\end{itemize}
It is also the minimal model with both couplings to QCD and a dark pion dark matter candidate (see e.g.~Ref.~\cite{Berlin:2018tvf} for a discussion of dark pion dark matter models without direct couplings to QCD).

\begin{figure}[t!]
\begin{center}
\includegraphics[width=0.5\linewidth]{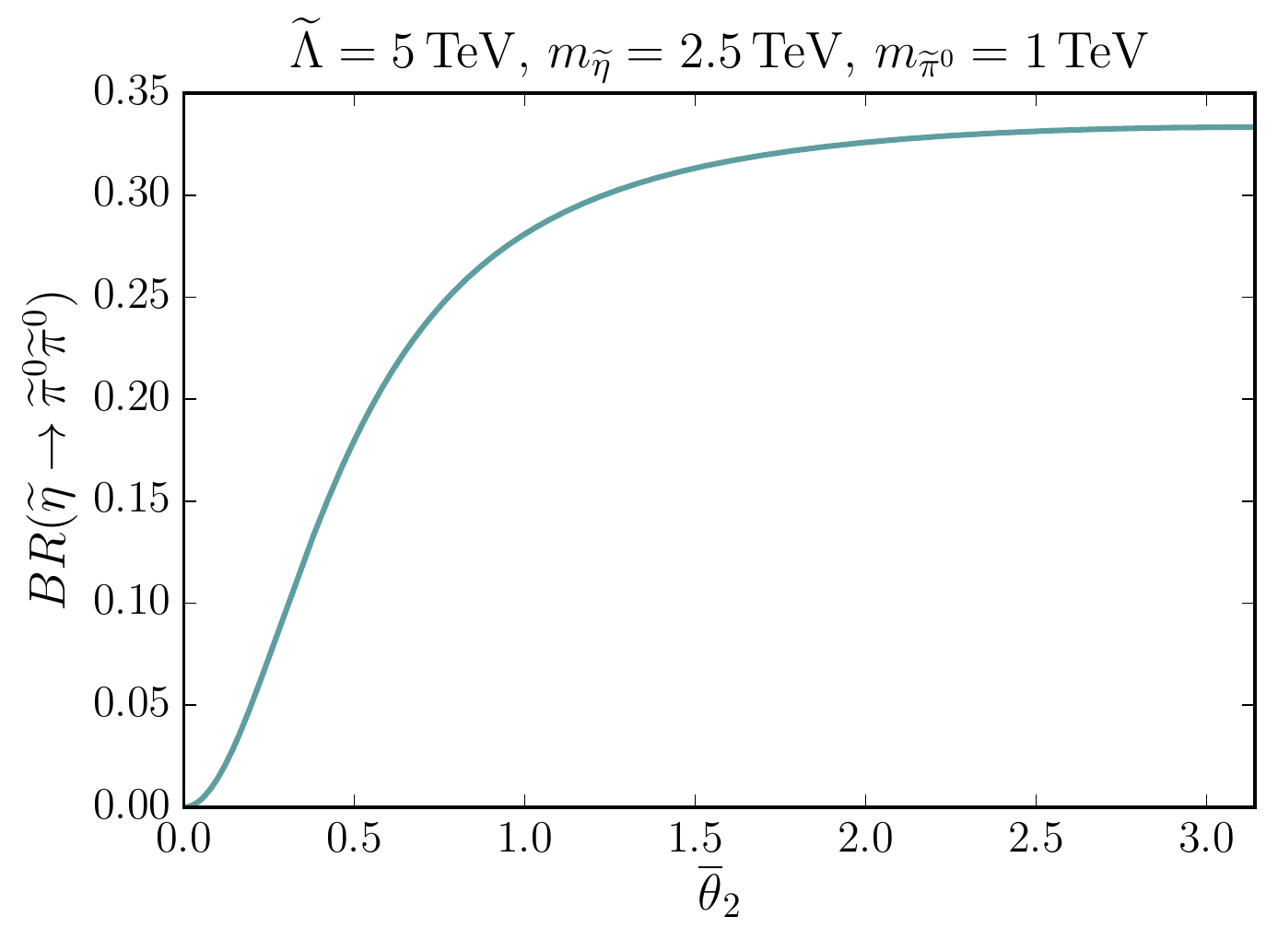}\,\includegraphics[width=0.5\linewidth]{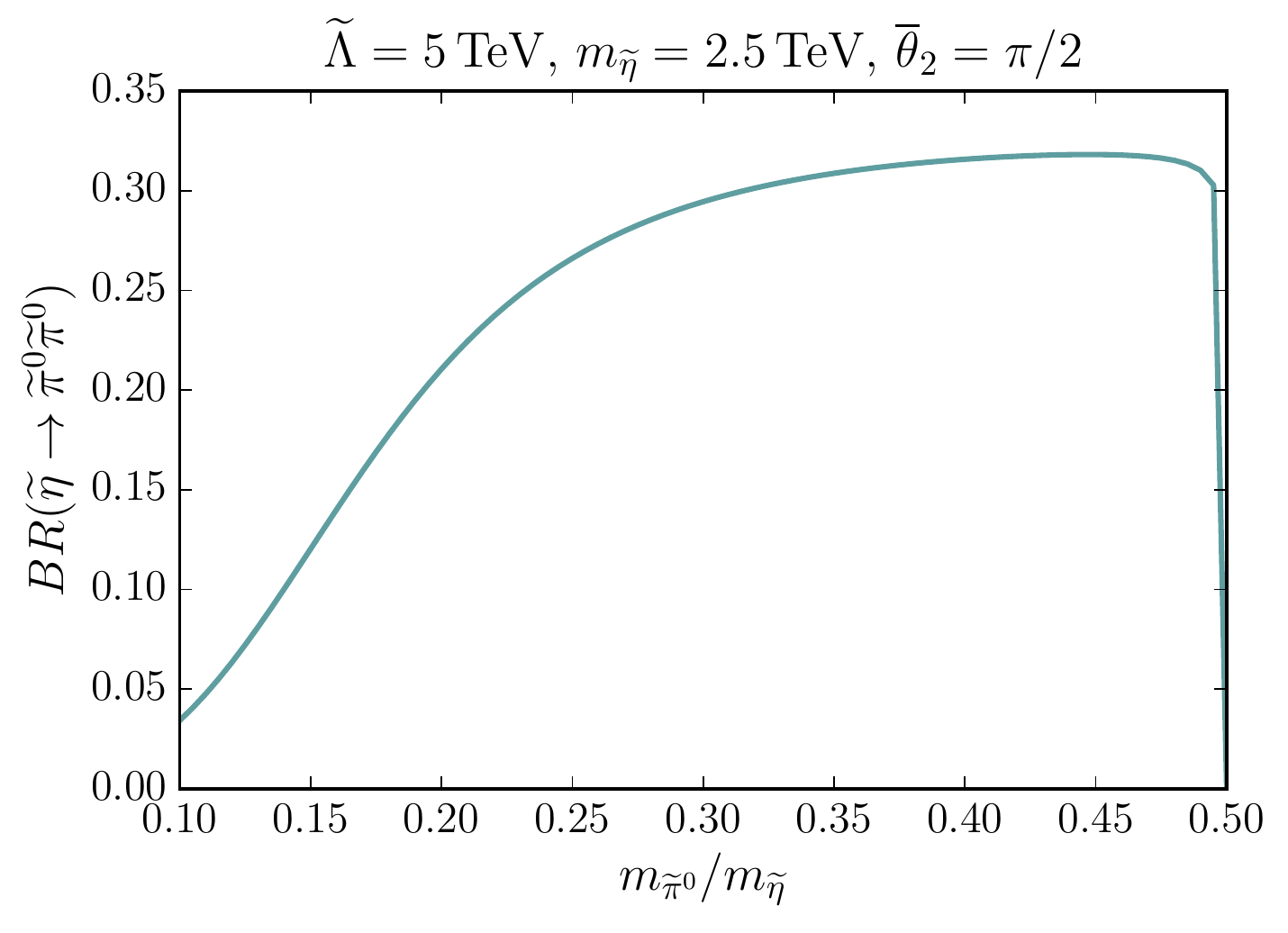}
\caption{BR($\heta\rightarrow\hpi^0\hpi^0$) at benchmark points in the $U(5)$ model. Left: as a function of $\hth$, fixing $m_{\heta}=2.5$ TeV, $m_{\hpi^0}=m_{\rm DM}=1$ TeV. Right: as a function of $m_{\hpi^0}/m_{\heta}$, fixing $m_{\heta}=2.5$ TeV, $m_{\hpi^0}=m_{\rm DM}$, and $\hth=\pi/2$. In both cases the branching ratios saturate at 1/3, reflecting the isospin symmetry and decays into $\hpi_{\rm DM}\hpi_{\rm DM}$.} 
\label{fig:u5BR}
\end{center}
\end{figure} 

We will focus primarily on the $\pizero$, $\heta$, and $\hpi_{\rm DM}$ states in the $U(5)$ model, choosing parameters so that the $\heta$ is heavier than the $\pizero$ and $\hpi_{\rm DM}$. (This is analogous to the $m_s\gg m_{u,d}$ property of ordinary QCD.) The $\heta$ state couples to QCD and QED through  chiral anomalies,
\begin{align}
{\cal L}\sim \heta G\widetilde{G}+ \heta F\widetilde{F}\;,
\end{align}
with precise couplings given in the Appendix. These interactions allow resonant single production as well as dijet and diphoton decay modes,
\begin{align}
gg\rightarrow \heta\rightarrow gg,\gamma\gamma\;.
\end{align}
The $\hpi^0$ state inherits the same couplings to the SM through small ``isospin"-suppressed mixing with the $\heta$,
\begin{align} \label{eq:mixing_angle}
\theta_{\hpi-\heta}\sim \frac{m_1-m_2}{m_3},
\end{align}
where $m_{1,2}$ are the neutral hyperquark masses primarily controlling the $\hpi^0$ mass, and $m_3>m_{1,2}$ is the QCD triplet hyperquark mass primarily responsible for the $\heta$ mass. Therefore, while we will assume the $\hpi^0$ is more kinematically accessible than the $\heta$, due to mixing suppression the production rate for the $\heta$ is larger. 

The neutral hyperpions also interact through  triple-pion couplings of the form $\heta \hpi^0\hpi^0+ \heta \hpi_{\rm DM}\hpi_{\rm DM}$, which are proportional to the lightest hyperquark mass and $\sin(\hth)$. More properly, the vacuum angle that appears is $\thetatwo$, an anomalous field-redefinition-invariant angle with contributions from $\hth$ and phases in hyperquark masses. $\thetatwo$ is defined for each model in the Appendix, but the distinction between it and $\hth$ is not important for our work and we will refer to them interchangeably. These $CP$-violating couplings induce $\heta\rightarrow\hpi\hpi$ decays when $m_{\heta}>2m_{\hpi}$. In Fig.~\ref{fig:u5BR} we show  $\heta\rightarrow\hpi^0\hpi^0$ branching ratios at representative benchmark points.

The anomalies, mixing, and $CP$-violating couplings give rise to the processes
\begin{align}
gg\rightarrow\heta\rightarrow \hpi^0\hpi^0\rightarrow (2g)(2g),(2g)(2\gamma),(2\gamma)(2\gamma)
\end{align}
that will be the focus of our collider studies below.

\subsection{$U(3)$}
The second  model has only one  color-triplet hyperquark which we again take to be in the $(3,1)_{4/3}$ representation of the SM. The model has an approximate classical $U(3)\times U(3)$ global symmetry spontaneously broken to $U(3)_V$, of which the $SU(3)_V$ is weakly gauged as ordinary QCD. The only pseudo-Goldstone hypermeson in this model is a QCD octet $\hpi^8$, which acquires mass both from the hyperquark mass and from QCD loops. In addition to this state, we will be interested in the vector octet $\hrho$ and the singlet $\hetap$ near the ChPT cutoff. 

\begin{figure}[t!]
\begin{center}
\includegraphics[width=0.5\linewidth]{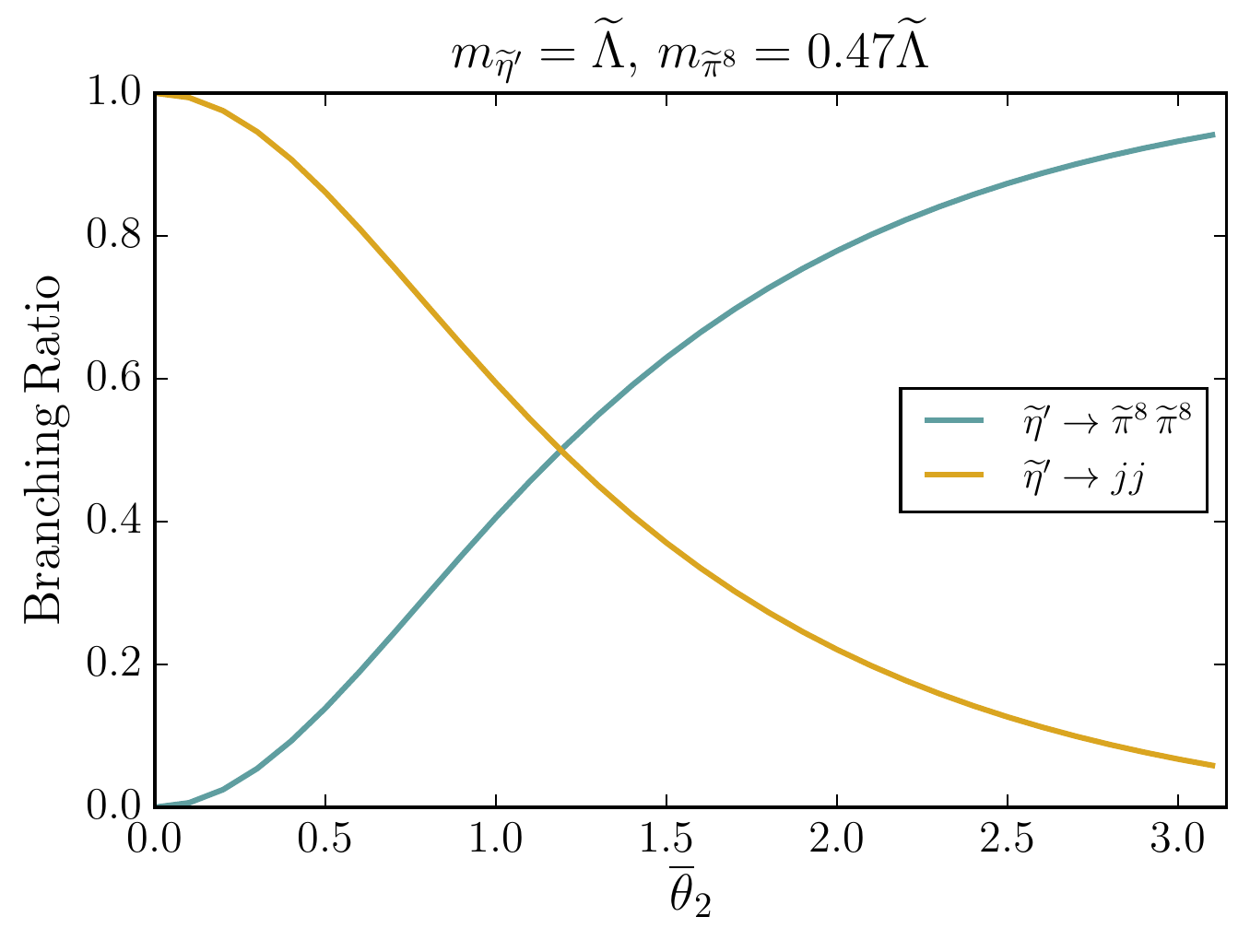}
\caption{BR($\hetap\rightarrow\hpi^8\hpi^8$) at a benchmark point in the $U(3)$ model with octet mass just below threshold. The rates are computed at leading order in ChPT and should only be considered qualitative.} 
\label{fig:u3BR}
\end{center}
\end{figure} 

The chiral Lagrangian description of the $U(3)$ model is rather trivial and limited to the $\hpi^8$ state. In addition to ordinary kinetic couplings of $\hpi^8$ pairs to QCD, there are anomaly-induced single-octet couplings to QCD and QED of the form
\begin{align} 
{\cal L}\sim \TR(\hpi^8 G\widetilde G)+\TR(\hpi^8 G)\widetilde F\;.
\end{align}
These interactions allow resonant single production as well as dijet and jet-photon decay modes,
\begin{align}
gg\rightarrow \hpi^8\rightarrow gg,g\gamma\;.
\end{align}

There is no hyperpion dark matter candidate in the $U(3)$ model. However, it is the minimal model exhibiting couplings to QCD, and $CP$-violating meson decays are still present, induced by $\hth$-dependent $\hetap\hpi^8\hpi^8$ couplings. (Again a field-redefinition-invariant vacuum angle $\thetatwo$ is defined for this model in the appendix and we use it interchangeably with $\hth$.) These $\hetap$ decay widths are not analytically calculable, but they can be modeled and estimated up to ${\cal O}(1)$ corrections by including the $\hetap$ in chiral perturbation theory. Sample branching ratios into dijets and octets are shown in Fig.~\ref{fig:u3BR}. Unlike the $\etap$ of QCD, these estimates suggest that the $\hetap$ may decay predominantly to hyperpions in this model.  It is therefore interesting to examine the processes
\begin{align}
gg\rightarrow\hetap\rightarrow \hpi^8\hpi^8\rightarrow (2g)(2g),(2g)(g\gamma),(g\gamma)(g\gamma)\;.
\label{eq:etapprocess}
\end{align}

These topologies are further motivated by the octet $\hrho$, which kinetically mixes with the gluon and decays primarily to $\hpi^8\hpi^8$ when this channel is kinematically open. We will thus consider also
\begin{align}
q\bar q\rightarrow\hrho\rightarrow \hpi^8\hpi^8\rightarrow (2g)(2g),(2g)(g\gamma),(g\gamma)(g\gamma)\;.
\label{eq:rhoprocess}
\end{align}
These processes are independent of $\hth$ and would in fact constitute a background in a measurement of  $\hth$. However, for the purpose of characterizing topologies motivated by new QCD-like sectors, the processes (\ref{eq:etapprocess}) and (\ref{eq:rhoprocess}) are complementary and we will include both below. Note that the $\hrho \rightarrow (2g)(2g)$ channel has been considered in previous work~\cite{Kilic:2008pm, Kilic:2008ub,Bai:2011mr,Schumann:2011ji}.

\section{Tetraboson Resonances}
\label{sec:paireddiboson}
The main signatures we will study are paired diboson resonances reconstructing a 4-boson parent resonance, encompassing an important and relatively unexplored class of processes predicted by new QCD-like sectors. To summarize the gluons+photons channels listed in the previous section, we have
\beq
\begin{aligned}
\heta \to \pizero \pizero &\to  (gg)(\gamma \gamma), \, (gg)(gg) \\
\hrho, \hetap \to \octet \octet &\to  (g\gamma)(g\gamma), \, (gg)(g\gamma),   (gg)(gg) \;.
\end{aligned}
\eeq
These processes above comprise the  primary manifestation of new vacuum angles at colliders through $CP$-violating $\hetap$ and $\heta$ decays, as well as the primary discovery modes for the $\hrho$ (provided decays to hyperpions are kinematically allowed). Below, we discuss the sensitivities to these channels at the LHC and future 100 TeV $pp$ collider.  Because of its small branching ratio, we do not consider the $\gamma \gamma + \gamma \gamma$ final state, although it may be worthwhile to revisit in the future.

Single diboson resonances are also  important signatures in this class of models. These have been more widely studied, and we defer their analysis to Sec.~\ref{sec:diboson} in order to compare with the paired signatures which are our primary focus. In some benchmark models, LHC diboson searches already provide strong constraints, suggesting the LHC is unlikely to observe the paired signatures in these models. In others there is still open parameter space for the LHC to explore in both classes of channels, and in all cases a 100 TeV collider would improve the reach substantially. However, a broader message of this work is that the paired channels are sufficiently novel and simple to motivate dedicated searches apart from our specific model frameworks.

\subsection{$jj+\gamma \gamma$} \label{sec:jjaa}

Let us begin with the $2j+2\gamma$ final state. We model these signatures with the process $gg\rightarrow \heta\rightarrow \hpi^0\hpi^0$ in the $U(5)$ model. Mixing between the $\heta$ and the $\pizero$  allows the $\pizero$ to decay to boson pairs. The relatively small backgrounds and good photon energy resolution make this a promising final state to observe the effects of $\hth$ in models like QCD with a spectrum of neutral pions. Since $m_{\widetilde{\rho}}\sim \hl$, and the non-resonant $\pizero$ pair production cross-section is suppressed by the $\heta-\pizero$ mixing angle, we can neglect these contributions to the $\pizero$ pair production cross-section in this analysis.

First, we estimate the expected reach in the $2j2\gamma$ channel at the high-luminosity LHC.  In all of our analyses, we simulate signal and background events in \texttt{Madgraph}~\cite{mg5}, with showering/hadronization in \texttt{Pythia}~\cite{pythia} and fast detector simulation in \texttt{DELPHES}~\cite{delphes}. For the LHC analysis, we use the default ATLAS detector card. Signal model files are prepared using the \texttt{Feynrules} package~\cite{feynrules}. All computations are performed at leading order throughout this study\footnote{Several of the processes we consider feature significant scale dependence in the production cross-section, in some cases inducing a factor of $\sim 2$ uncertainty in the rate. A next-to-leading order treatment would be useful to more precisely establish the discovery reach in the various channels. However, for the general arguments we are concerned with here, we expect a leading order determination of the rates to be sufficient, as our overall conclusions are not very sensitive to this variation.}. For this channel, we consider only the irreducible background from prompt $2j2\gamma$ events. Comparing parton-level cross-sections and assuming a jet-to-photon fake rate of $\sim \mathcal{O}(10^{-3})$, we expect fake photons from QCD jets to affect our backgrounds at the $\sim 10\%$ level, which will not significantly impact our conclusions. 

For a given $m_{\heta}$, we perform a cut-and-count analysis. We require at least two jets with $p_T>120$ GeV, $|\eta|<2.4$, and at least two photons with $p_T > 75$ GeV, $|\eta|<2.5$. We reconstruct the $2\gamma$ and $2j$ resonances from the leading photons and jets and compute the asymmetry parameter 
\beq
\mathcal{A}_{jj\gamma\gamma} \equiv \frac{|m_{jj} - m_{\gamma \gamma} |}{m_{jj} + m_{\gamma \gamma} },
\eeq
requiring $\mathcal{A}_{jj\gamma\gamma}<0.1$. For a given $\heta$ mass, we include an additional cut on the leading photon $p_T$, requiring
\beq
p_T({\gamma_1})  \geq \frac{1}{5}m_{\heta}.
\eeq
We then reconstruct the $\heta$ mass and cut in a window around the peak of the $m_{jj\gamma\gamma}$ distribution such that the signal falls to half its peak value at the edges of the window. We compute $S/\sqrt{B}$ in this window to estimate exclusion and discovery sensitivities in terms of the parton-level $\sigma \times BR$ into the $2g2\gamma$ final state. The signal-to-background ratio is $\sim 0.2$ or larger in the parameter space with $S/\sqrt{B}\geq 2$.

\begin{figure}[t!]
\begin{center}
\includegraphics[width=0.5\linewidth]{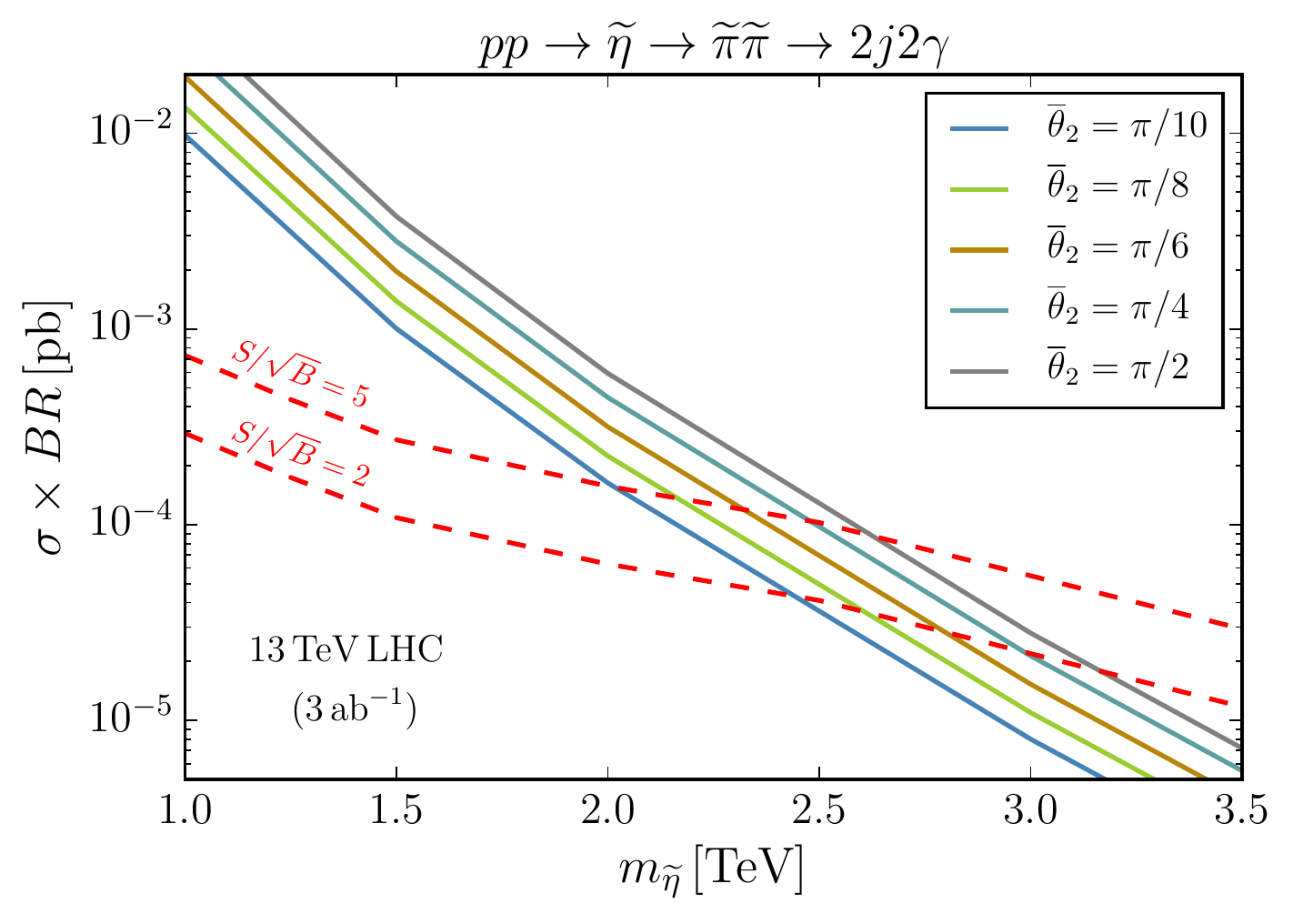}\,\includegraphics[width=0.5\linewidth]{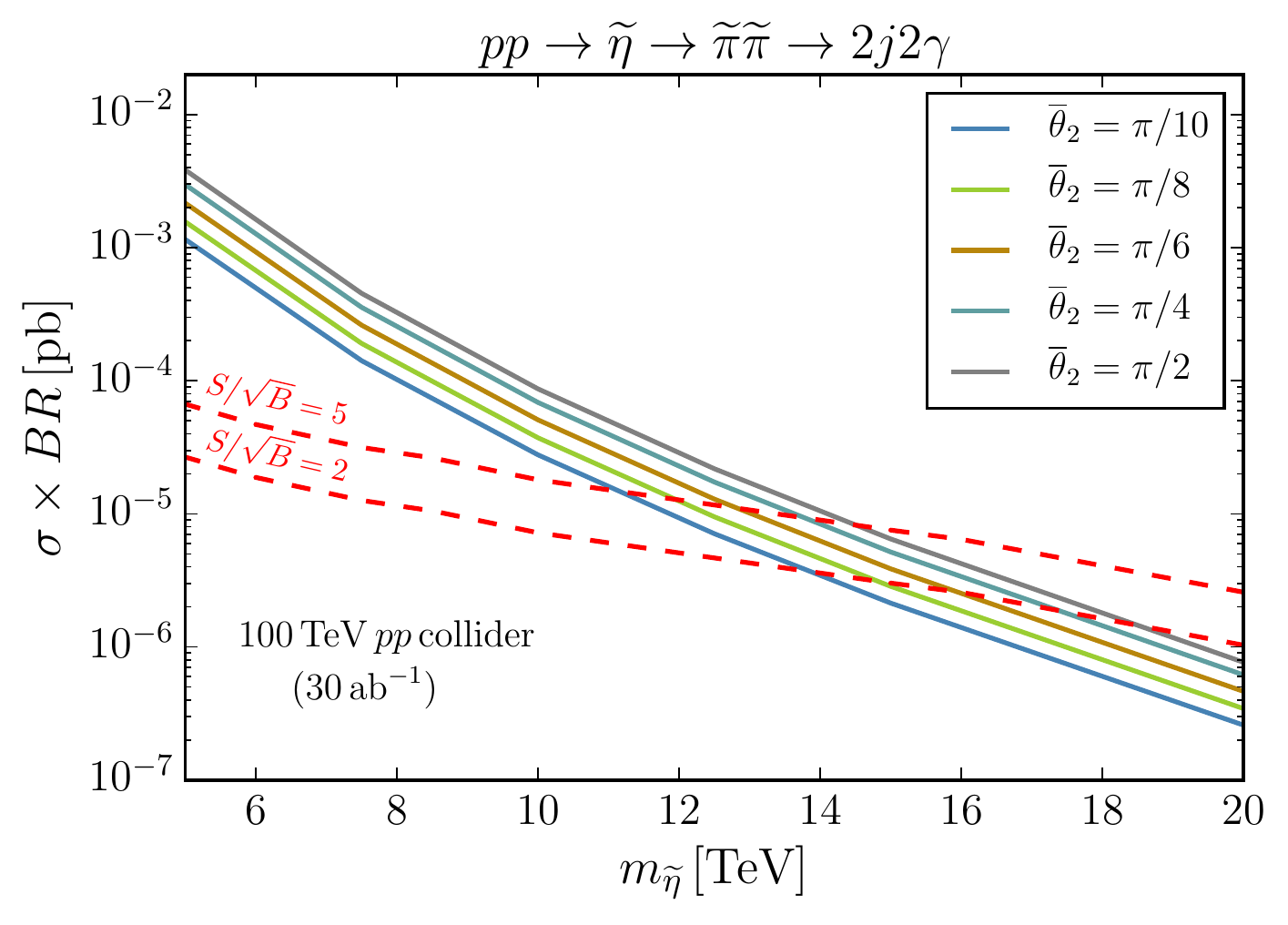}
\caption{$pp \to \heta \to 2j2\gamma$ cross-sections and projections in the $U(5)$ model for the 13 TeV LHC with 3 ab$^{-1}$ of integrated luminosity, and a future 100 TeV $pp$ collider with 30 ab$^{-1}$ integrated luminosity.  We have fixed $m_{\heta} = \hl/2$ and $m_{\pizero} = 0.4 m_{\heta}$. } 
\label{fig:2j2gam_model2}
\end{center}
\end{figure} 

Results for the 13 TeV LHC\footnote{Although the high-luminosity LHC is expected to run at a center-of-mass energy of 14 TeV, we show projections for 13 TeV since, when comparing against diboson sensitivities in Sec.~\ref{sec:diboson}, we utilize several existing LHC searches performed at 13 TeV. Increasing the center-of-mass energy will slightly improve the prospects for discovery relative to those shown, but not significantly affect our conclusions.} with $3\, {\rm ab}^{-1}$ integrated luminosity are shown in Fig.~\ref{fig:2j2gam_model2}. We have taken a benchmark point with $m_{\heta} = \hl/2$ and $m_{\pizero} = 0.4 m_{\heta}$, which are relative values similar to those in QCD. (The right-hand panel of Fig.~\ref{fig:u5BR} shows the sensitivity of $Br(\heta\to\pizero \pizero)$ to variations around this point; the branching ratio is independent of $\hl$ and varies modestly with $m_{\pizero}/m_{\heta}$.)  For these values, we see that $\thetatwo \gtrsim \pi/10$ and $\hl$ up to $\sim 4-6$ TeV can be observable at the high-luminosity LHC. When we compare to the reach in the various diboson channels in Sec.~\ref{sec:diboson}, we will see that this process may also be accompanied by a signal in the dijet, diphoton, and $j\gamma$ channels, but not necessarily.

For the 100 TeV analysis, we follow the same procedure with identical cuts, only now with the requirements $p_T>300$ GeV, $|\eta|<2.4$ for the jets, and $p_T > 200$ GeV, $|\eta|<2.5$ for the photons. We also utilize the FCC-hh \texttt{DELPHES} card without pileup for detector simulation. The results assuming 30 ab$^{-1}$ are shown on the right hand side in Fig.~\ref{fig:2j2gam_model2}. The projected reach in this channel is impressive: a 100 TeV $pp$ collider with 30 ab$^{-1}$ of integrated luminosity can probe new QCD-like sectors with $\hl$ up to $\sim 40$ TeV. We will compare this sensitivity to that from diboson resonance searches in Sec.~\ref{sec:diboson} below where we will show that this signature can be the most sensitive probe of new QCD-like sectors with $\mathcal{O}(1)$ vacuum angles, both at the LHC and 100 TeV.

\subsection{$jj + jj$}

As we have seen, searches for $2j2\gamma$ resonances can be a powerful probe of parity-violating interactions in new strongly-coupled sectors. A limitation of this channel is that it depends on the electric charges of the hyperquarks, which can vary from model to model (although we generally expect it to be non-zero to allow the triplet states to decay promptly on cosmological timescales). In contrast, if the unstable hyperpions can be produced in gluon fusion, they will always be able to decay to gluons with a fixed coupling, so $4j$ is an important and complementary final state. 

We focus on the process $pp\to \heta \to \pizero \pizero \to 4j$ in the $U(5)$ model. As in the previous subsection, for simplicity we assume that the cutoff is high enough that we can neglect contributions from heavy states like the $\widetilde{\rho}$, and that the non-resonant $\pizero \pizero$ contribution is negligible (it is suppressed by two dimension-5 operators and mixing with the $\heta$, so this is a very good approximation). The final state is a pair of dijet resonances of roughly the same invariant mass that themselves reconstruct a mass around $m_{\heta}$. 

Searches for paired dijets have been carried out at the LHC~\cite{Aaboud:2017nmi, Khachatryan:2014lpa}, without the parent mass requirement. However, the LHC is unlikely to observe the $4j$ parity-violating process in the $U(5)$ model. As shown in Sec.~\ref{sec:diboson} below, diboson searches already constrain the scalar octets to be heavier than about 3 TeV in this scenario (with the various parameters fixed as in the $jj\gamma \gamma$ analysis above). This in turn constrains $\hl$ and limits the cross section for $\heta$ production possible at the LHC. We find that the signal is swamped by the large QCD $4j$ background at the LHC for values of $\hl$ not already ruled out by diboson searches. (Note however that jet substructure techniques can improve the LHC reach for cases with a larger mass hierarchy between the $\heta$ and $\pizero$~\cite{Bai:2011mr}.)

\begin{figure}[t!]
\begin{center}
\includegraphics[width=0.65\linewidth]{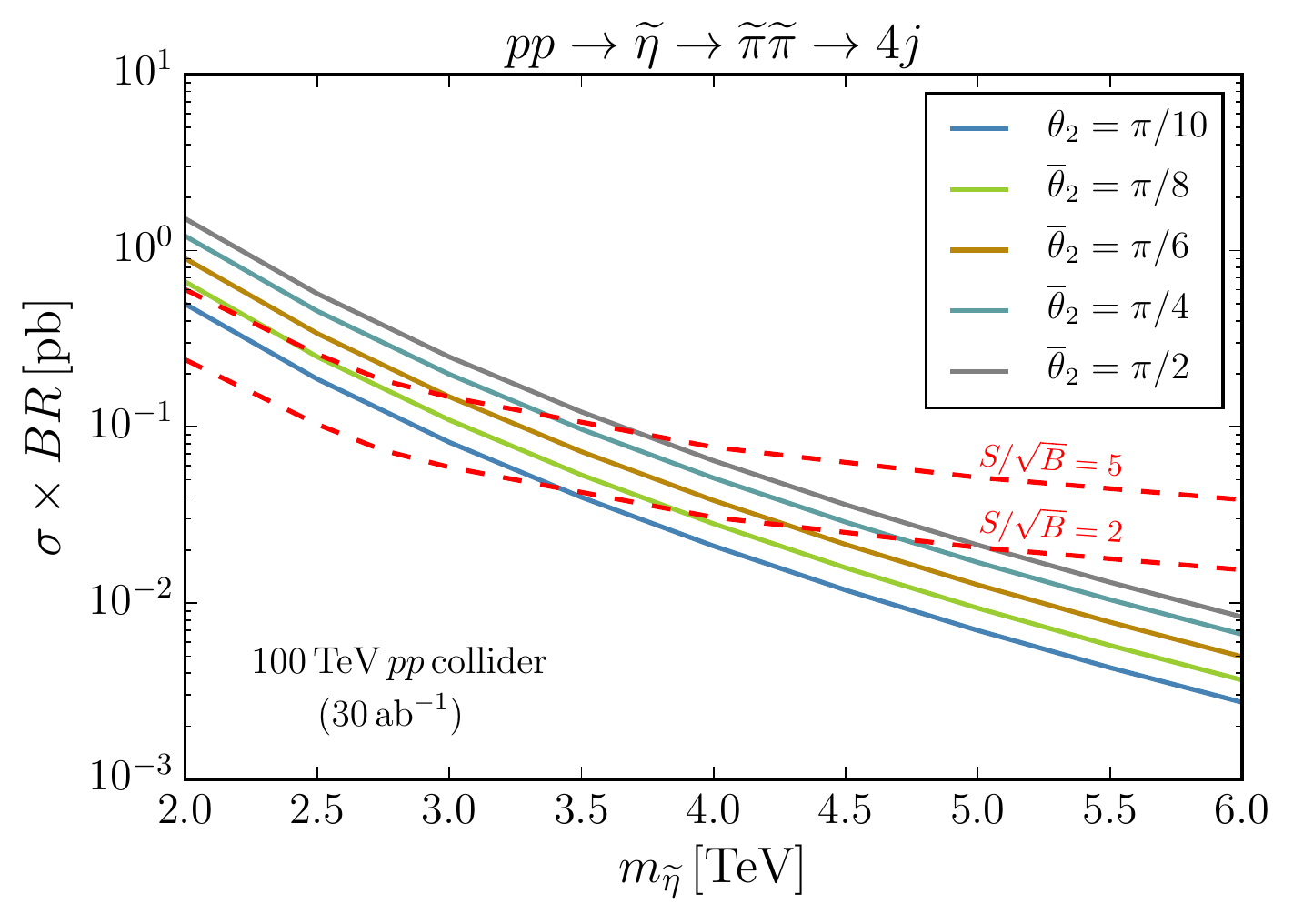}
\caption{Paired dijet cross-sections and projections at a future 100 TeV collider with 30 ab$^{-1}$ of integrated luminosity in the $U(5)$ model with $m_{\heta}=\hl/2$, $m_{\pizero}=0.4m_{\heta}$ and various values of $\thetatwo$.} 
\label{fig:4j}
\end{center}
\end{figure}

Therefore, for this channel we focus instead on the reach at a future 100 TeV collider. For concreteness we again consider the benchmark point $m_{\pizero} = 0.4 m_{\heta}$, $m_{\heta} = \hl/2$ and vary $\hl$ and $\thetatwo$. We simulate the signal process $pp\to \heta \to \pizero \pizero \to 4j$ and the QCD $4j$ background as described above. Our selection criteria are inspired by the ATLAS search in~\cite{Aaboud:2017nmi}. We require four jets with $p_T>300$ GeV and $|\eta|<2.5$. We construct two dijet pairs by finding the combination of the leading four jets that minimize the quantity
\beq
\Delta R_{\rm min} \equiv \operatorname{min}\left\{ \sum_i \left| \Delta R_i -1 \right| \right\}
\eeq
where $i=1,2$ are the two dijet pairs formed, and minimization is over the three possible pairings. Once the two dijet resonances are formed, we also impose cuts on the asymmetry parameter,
\beq
\mathcal{A}_{4j}\equiv \frac{\left| m_{jj, 1} - m_{jj,2} \right|}{m_{jj,1}+m_{jj,2}},
\eeq
where $m_{jj,1}$, $m_{jj,2}$ are the invariant masses of the two reconstructed dijets. We require $\mathcal{A}_{4j}\leq 0.4$ in our analysis. Following~\cite{Aaboud:2017nmi}, we define $m_{\rm avg}\equiv 1/2 \times (m_{jj,1}+m_{jj,2})$ and require
\beq
\Delta R_{\rm min} <0.0013 \times (m_{\rm avg}/{\rm GeV} - 225) + 0.72,
\eeq 
since signal events are expected to have larger angular separation at large masses.
We also compute the angle, $\theta^*$ between the two dijet resonances and the beam-line in the center-of-mass frame of the two resonances. The $t$-channel gluon QCD background tends to result in forward jets,  so we require $\left|\cos \theta^* \right| \leq 0.3$. Given these selection criteria, we  compute the four-jet invariant mass, $m_{4j}$, which should reconstruct to approximately $m_{\heta}$. A bump-hunt can then be performed on this distribution, similar to a conventional dijet search. To estimate the reach, we compare $S$ to $\sqrt{B}$ in a window with boundaries at the full-width half-maximum points of the $m_{4j}$ distribution for a given mass, assuming 30 ab$^{-1}$ of integrated luminosity. The results of this analysis are shown in Fig.~\ref{fig:4j} for various values of $\hl$ and $\thetatwo$.

Fig.~\ref{fig:4j} suggests that a 100 TeV collider would be able to discover parity-violating $\heta$ decays in the $4j$ final state for $m_{\heta}$ up to nearly 4 TeV for large $\thetatwo$, and exclude this process for $m_{\heta}$ up to $\sim 3.5-5$ TeV for $\thetatwo \gtrsim 0.3$. Improvements in detector technology and data analysis techniques could further extend this reach. Comparing to Fig.~\ref{fig:2j2gam_model2}, we see that the reach can be substantially better in models with couplings to QED. Nevertheless, the $4j$ signature is guaranteed to be present in these models, and can probe $CP$-violating hypermeson decays. 

\subsection{$j\gamma + j\gamma$} \label{sec:jaja}
We turn now to signatures associated with the scalar octet $\hpi^8$. Since the octet receives a large cutoff-sensitive contribution to its mass from radiative corrections,  relevant parent particles include the $\hetap$ and the vector octet $\hrho$, both of which have masses of order the cutoff. In this subsection we use the $U(3)$ model to study the processes
\begin{align}
gg&\rightarrow\hetap\rightarrow \octet\octet\rightarrow (g\gamma)(g\gamma)\nonumber\\
q\bar q&\rightarrow \hrho\rightarrow \octet\octet\rightarrow (g\gamma)(g\gamma)\;.
\end{align}
The former proceeds through $\thetatwo$-dependent interactions, while the latter is present even if $\thetatwo=0$. In general there is also contribution to this final state from non-resonant $gg\rightarrow \octet\octet$ production, which we must account for in addressing the discovery potential for $j\gamma+j\gamma$ resonances. 

For definiteness, we will take $m_{\hetap}= \hl$ and $m_{\hrho}=0.8 \hl$,  mimicking the corresponding meson mass ratio in QCD, and $\hl/3<m_{\octet}<m_{\hetap}/2$ so that $\hetap\rightarrow\octet\octet$ is kinematically open (the lower bound corresponds to the approximate contribution to $m_{\octet}$ from QCD loops.)  The signal then depends on whether the decay $\hrho\rightarrow\octet\octet$ proceeds on-shell or not. To cover both cases, we take two  benchmark points, 
\begin{align}
m_{\octet} &= 0.47 m_{\hetap}~~~~~~~{(\hrho\rightarrow\octet\octet {\rm~off-shell})}\nonumber\\
 m_{\octet}&=  0.47 m_{\hrho}~~~~~~~~{(\hrho\rightarrow\octet\octet {\rm~on-shell})}\;.
 \label{eq:benchmarks}
 \end{align}
 In terms of the cutoff, these  correspond to $m_{\octet} \simeq 0.47 \hl$ and $0.38 \hl$ respectively. In the former case, the $\hetap\rightarrow\octet\octet$ branching ratio is nearly maximal and close to unity for large $\thetatwo$ (see Fig.~\ref{fig:u3BR}), while $\hrho\rightarrow\octet\octet$ is off-shell. In the latter, both $\hetap\rightarrow\octet\octet$ and $\hrho\rightarrow\octet\octet$ proceed on-shell. In this case, because the hyperquark mass is smaller, the $\hetap \octet\octet$ coupling is suppressed, and the $\hetap\rightarrow\octet\octet$ branching ratio asymptotes to around $70\%$ for large $\thetatwo$. 

The simplest use of this channel is to search for octets that are pair-produced by any mechanism, in which case the $\hetap$, $\hrho$, and non-resonant contributions are all part of the signal. Here we are primarily interested in the first two processes, which probe new states near the confinement scale and produce a bump in the 4-object invariant mass. We will treat both of these processes as signal, and the non-resonant QCD-induced $\octet \octet$ contribution as background. As we will see below in Sec.~\ref{sec:diboson}, the $\octet$ will likely be observed in $j\gamma$ resonance searches before the $\hetap$ can be discovered in the $2j2\gamma$ channel, which is consistent with this treatment. We comment on possibilities for separating the individual $\hetap$ and $\hrho$ contributions to determine sensitivity to $\thetatwo$ below (note that this was not an issue in the $U(5)$ analyses above, since there was parametric separation between the $\heta$ and $\hrho$, $\octet$ masses).

To determine the sensitivity to $j\gamma + j\gamma$ resonances at the LHC and a 100 TeV collider, we simulate signal and background events using the \texttt{Madgraph} + \texttt{Pythia} + \texttt{DELPHES} chain as described above. The $\hrho$ is expected to have a large natural width ($m_{\rho}/\Gamma_{\rho}$ is $\sim 15\%$ in QCD), however, when considering the $2j2\gamma$ final state, we find that these widths are still below the $m_{j\gamma j\gamma}$ resolution implemented in \texttt{DELPHES}, and so we utilize the narrow width approximation in generating on-shell $\hrho$ events to obtain simple sensitivity estimates in this case. To isolate the signal from background, we require exactly two hard photons and place cuts on the $p_T$ of the four leading objects of interest, keeping events for which:
\beq
p_T(j_1) > m_{\hetap}/8, \quad p_T(\gamma_1) > m_{\hetap}/8, \quad p_T(j_2) > m_{\hetap}/12, \quad p_T(\gamma_2) > m_{\hetap} /12.
\eeq
We form two $j\gamma$ resonances from the leading two jets and photons by minimizing the asymmetry parameter 
\beq
\mathcal{A}_{j\gamma j \gamma}\equiv \frac{\left| m_{j\gamma, 1} - m_{j\gamma, 2} \right|}{m_{j\gamma,1}+m_{j\gamma,2}}.
\eeq
We  require $\mathcal{A}_{j\gamma j \gamma}<0.1$. After applying these cuts, for a given $\hl$, we define the signal ($S$) and background ($B$) by counting the remaining events in a window on the $m_{j\gamma j\gamma}$ distribution such that  $0.8 \hl < m_{j\gamma j \gamma} < 1.07 \hl$. 

\begin{figure}[t!]
\begin{center}
\includegraphics[width=0.49\linewidth]{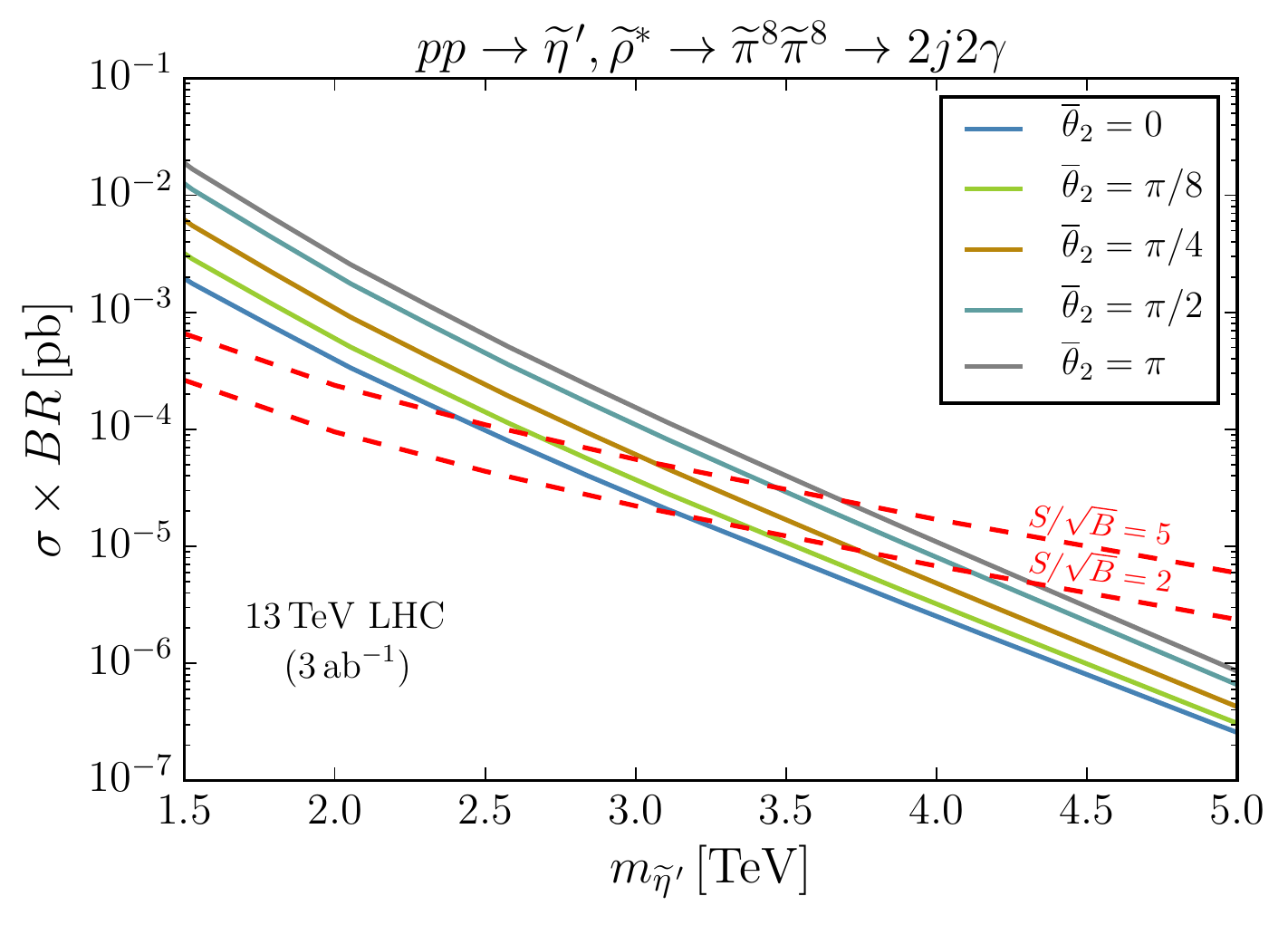}\,
\includegraphics[width=0.49\linewidth]{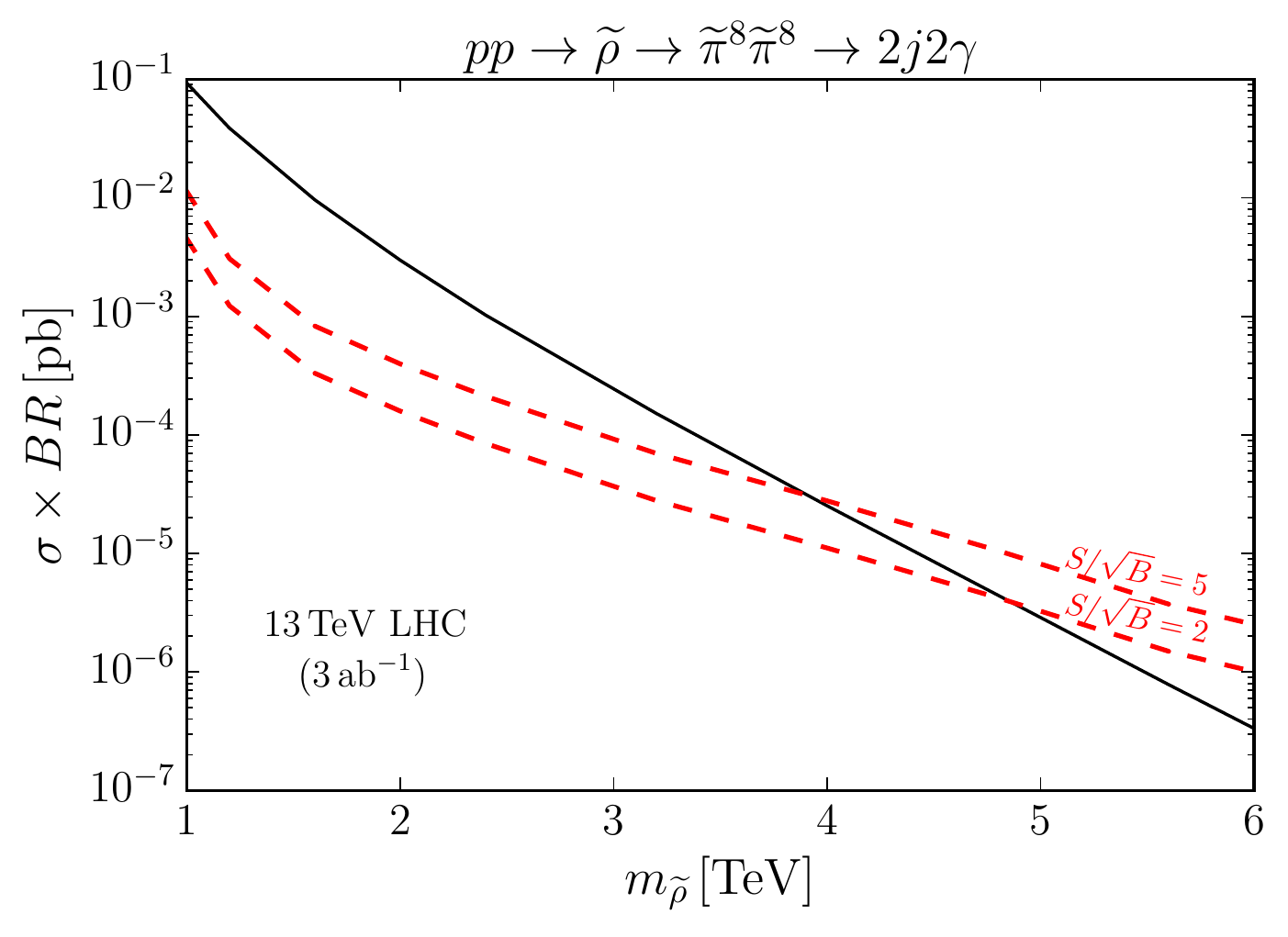}
\caption{Projections for discovery and exclusion of the $\hetap$ and $\hrho$ resonances at the LHC in the $j\gamma +j\gamma$ final state for the off-shell (left panel) and  on-shell (right panel) benchmark points in Eq.~(\ref{eq:benchmarks}) for the $U(3)$ model. On the left, the predicted signal is a combination of on-shell $\hetap \to \octet \octet \to 2j2\gamma$ decays and $\octet \octet$ production through an off-shell $\hrho$; predicted $\sigma \times BR$ values are shown for various values of $\thetatwo$ as a function of $m_{\hetap}$. On the right, the $CP$-conserving $\hrho$ decay proceeds on-shell and $\thetatwo$ is taken to be zero, so that the $\hetap$ does not contribute. The predicted $\sigma \times BR$ is shown as a function of $m_{\hrho}=0.8 \hl$. } 
\label{fig:jaja_LHC}
\end{center}
\end{figure} 

\begin{figure}[t!]
\begin{center}
\includegraphics[width=0.49\linewidth]{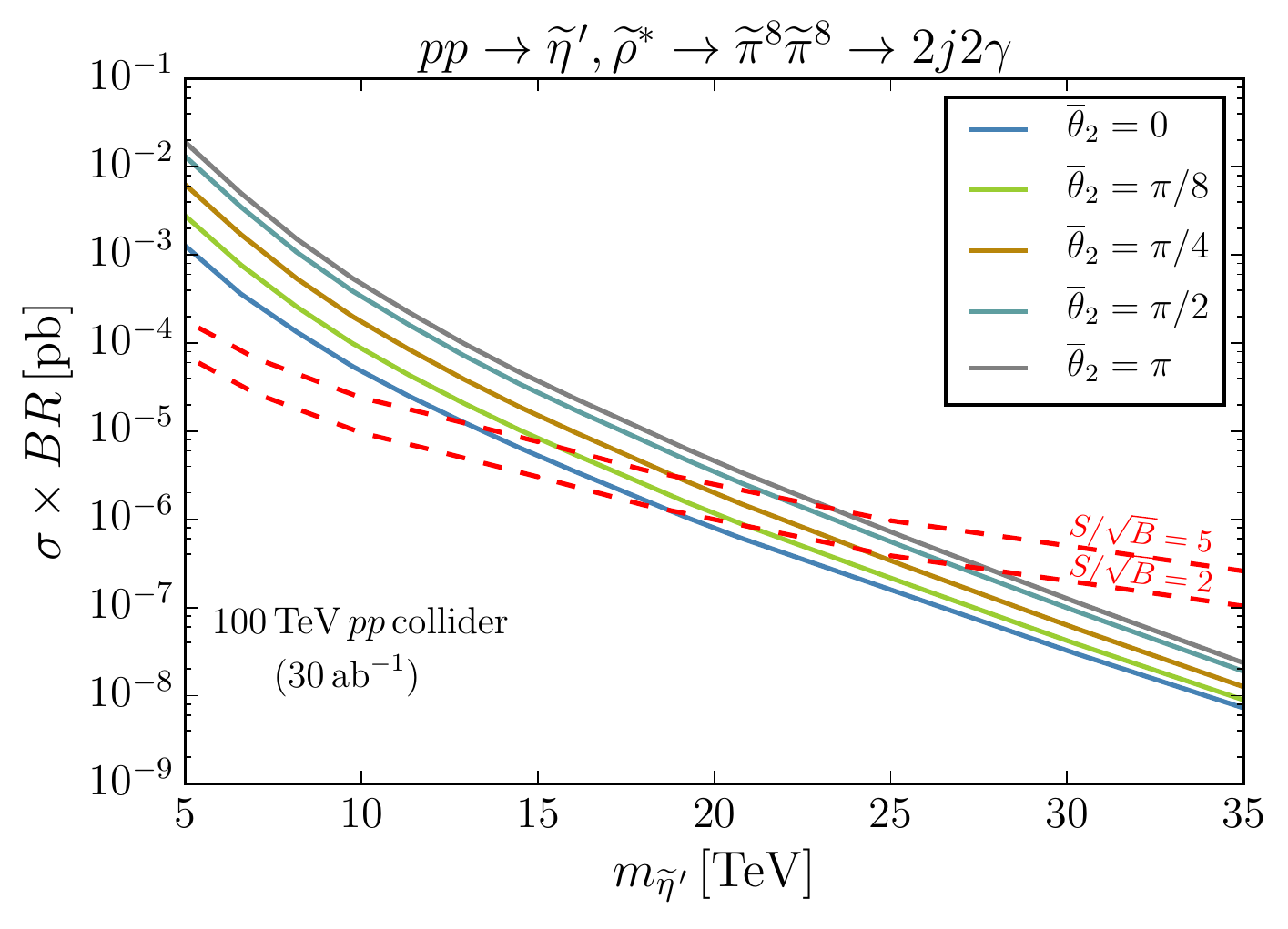}\,
\includegraphics[width=0.49\linewidth]{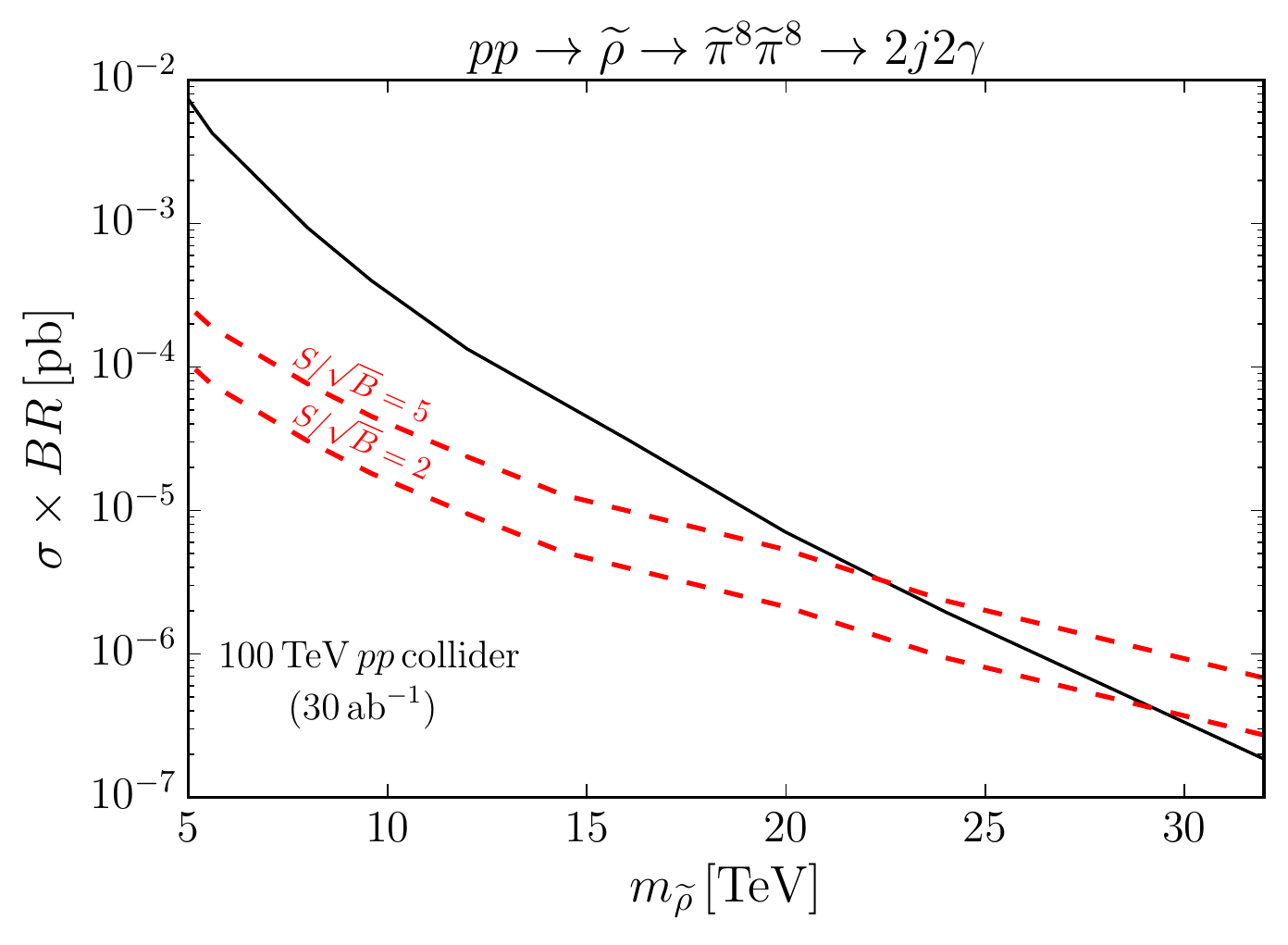}
\caption{As in Fig.~\ref{fig:jaja_LHC}, but at a future 100 TeV $pp$ collider, given 30 ab$^{-1}$ integrated luminosity. } 
\label{fig:jaja_FCC}
\end{center}
\end{figure}

The results of this procedure\footnote{The analysis described here differs slightly from that utilized in the $jj+\gamma\gamma$ case above, but both strategies yield similar sensitivity projections.} are shown in Fig.~\ref{fig:jaja_LHC} for the 13 TeV LHC with 3 ab$^{-1}$, and Fig.~\ref{fig:jaja_FCC} for a 100 TeV $pp$ collider with 30 ab$^{-1}$ of integrated luminosity. On the left-hand-side of these plots, we show the corresponding discovery and exclusion reach in this channel for the off-shell $\hrho \to \octet \octet$ benchmark in Eq.~(\ref{eq:benchmarks}), along with predicted $\sigma \times BR$ for various values of $\thetatwo$. As stated above, both the $\hetap$ and $\hrho^*$ contributions are treated as signal, so the predicted signal cross-section does not vanish at $\thetatwo = 0$. Note also that in deriving the sensitivity curves shown, we have assumed that the $\hetap$ and $\hrho$ efficiencies are the same (the sensitivity curve was obtained for $\thetatwo = 3\pi /4$); we have verified that the efficiencies agree within a factor of $\sim 1-2$ across the range of masses shown. On the right-hand-side of Figs.~\ref{fig:jaja_LHC} and~\ref{fig:jaja_FCC}, we show the corresponding reach for the on-shell $\hrho \to \octet \octet$ benchmark in Eq.~(\ref{eq:benchmarks}) at the LHC and a future 100 TeV $pp$ collider. In this case, we have assumed $\thetatwo=0$ so that the $\hetap$ does not contribute to the signal. 

Fig.~\ref{fig:jaja_LHC} shows that the LHC can be sensitive to $\hl \sim 3-5$ TeV via the $j\gamma + j \gamma$ channel in the off-shell $\hrho$ benchmark. The same channel can probe the on-shell $\hrho\to\octet\octet$ benchmark with $m_{\hrho}$ up to $\sim 5$ TeV.  At a future 100 TeV collider, Fig.~\ref{fig:jaja_FCC} indicates that the same process can probe $CP$-violating $\hetap$ decays for $\hl$ up to $\sim 25-30$ TeV, and on-shell $\hrho\to\octet\octet$  for $m_{\hrho}\lesssim 25-30$ TeV, providing a compelling target for future colliders.

Up to this point, we have considered both the $\hetap$ and $\hrho$ as contributing to the signal. In order to directly extract information about the new vacuum angle, however, it becomes necessary to distinguish between these two contributions. Since there is a large theoretical uncertainty on the normalization of the $\hrho$ production cross-section, differentiating the two channels is quite challenging. In the off-shell $\hrho$ case ($m_{\hrho}<2m_{\octet}$), one might hope to first observe the $\hrho$ in dijets, yielding information about its width and total production cross-section that could then be used to further disentangle the $\hetap$ and $\hrho^{(*)}$ contributions in $2j2\gamma$. Kinematic distributions beyond the reconstructed invariant masses could also be useful in separating these contributions in a boosted decision tree analysis. We expect that more sophisticated analysis techniques could further improve the sensitivity to $\hth$, but we leave further investigation of this possibility to future work.

\subsection{$jj + j\gamma$} \label{sec:jjja}

Decays of the $\hrho$ and the $\hetap$ into $\octet$ pairs can also result in a $jj + j \gamma$ resonance. The advantage of this channel over $j\gamma + j\gamma$ is that the $\octet \to jj$ branching ratio is larger. The disadvantage is that the jet energy resolution is not as good as the resolution for photons. Nevertheless, the $jj + j\gamma$ channel can be quite powerful in searching for hypermesons. We illustrate this by analyzing the discovery potential for the $\hrho$ in this channel.

\begin{figure}[t!]
\begin{center}
\includegraphics[width=0.5\linewidth]{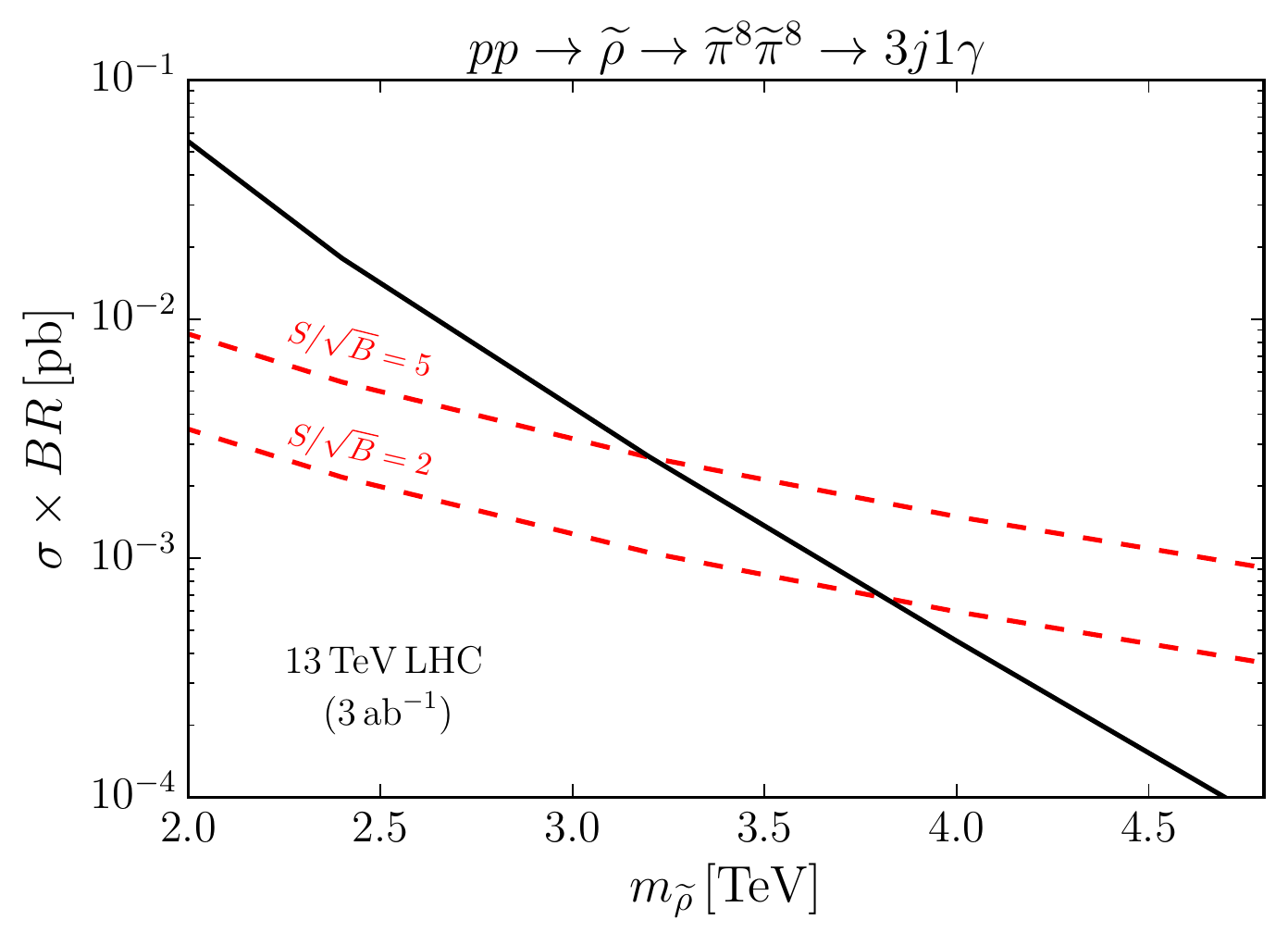}\,\includegraphics[width=0.5\linewidth]{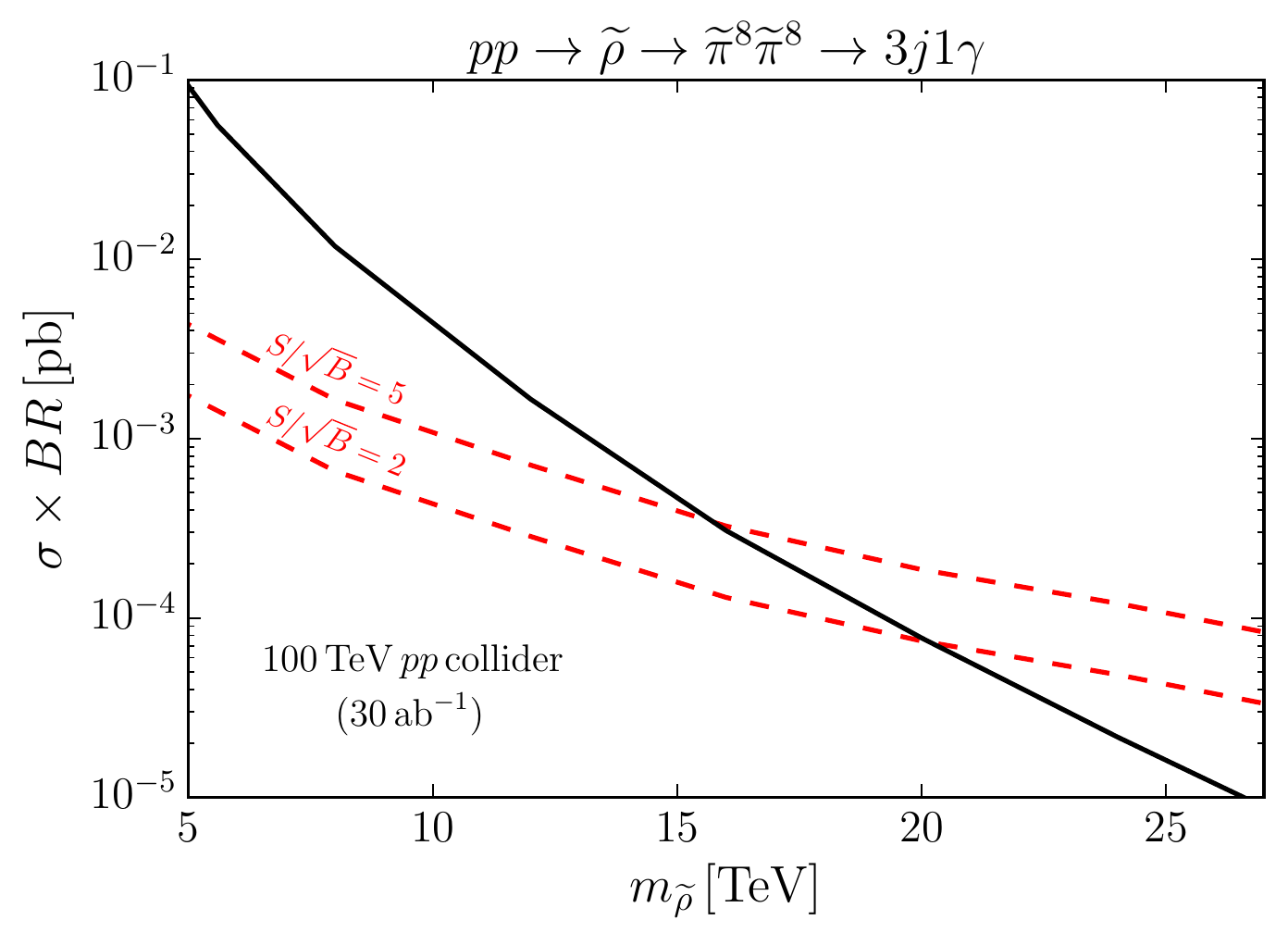}
\caption{Cross-sections and projected sensitivities in the $pp \to \hrho \to \octet \octet \to 3j 1\gamma$ channel for  at the 13 TeV LHC  (left) and a future 100 TeV collider (right). The cross-sections shown correspond to the on-shell $\hrho$ benchmark of Eq.~(\ref{eq:benchmarks}) in the $U(3)$ model. } 
\label{fig:3j1gam}
\end{center}
\end{figure} 

For concreteness, we again consider  the second benchmark of Eq.~(\ref{eq:benchmarks}), just below the $\hrho\rightarrow\octet\octet$ threshold, with $m_{\hrho}=0.8 \hl$. We neglect the $\hetap$ contribution for simplicity (i.e.~we assume that $\hth$ is negligible), and otherwise the analysis proceeds similarly to the previous channels, with the same basic $p_T$ selection requirements as in Sec.~\ref{sec:jjaa}. 

We consider the irreducible background from $3j1\gamma$ QCD events as well as non-resonant $\octet \octet$ production.  We construct the dijet and $j+\gamma$ resonances by finding the combination of the leading three jets and photon that minimize the quantity
\beq
\Delta R_{\rm min} \equiv \operatorname{min}\left\{\left| \Delta R_{jj} -1 \right|+\left| \Delta R_{j\gamma} -1 \right| \right\}
\eeq
where the minimization is over the three possible pairings. This choice is motivated by the selection criteria in the $4j$ search of Ref.~\cite{Aaboud:2017nmi}. Once the two resonances are formed, we impose cuts on the asymmetry parameter,
\beq
\mathcal{A}_{3j1\gamma}\equiv \frac{\left| m_{jj} - m_{j\gamma} \right|}{m_{jj}+m_{j\gamma}},
\eeq
requiring $\mathcal{A}_{3j1\gamma}\leq 0.1$. For a given $m_{\hrho}$, we also impose a cut on the $p_T$ of the leading photon, requiring $p_T(\gamma_1)  \geq m_{\hrho}/5$. We then cut in a window around the peak of the $m_{3j1\gamma}$ distribution such that the signal falls to half its peak value at the edges of the window. We compute signal and background in this window and find $S/B\sim 0.05$ or larger in the parameter space with $S/\sqrt{B}\geq 2$.

Results of this analysis for the LHC and a 100 TeV collider are shown in Fig.~\ref{fig:3j1gam}. The high-luminosity LHC will be able to probe hypercolor sectors with confinement scales up to $\sim 4-5$ TeV, while a 100 TeV collider can extend this reach up to $\sim 20-25$ TeV.  As we will see below, the octets will likely be discovered before this channel is observed, with our choices for the various parameters. However, given the large uncertainty on the $\hrho$ production cross-section, as well as the possibility of different charge assignments for the hyperquarks, this channel is still worthwhile to investigate at the LHC and 100 TeV, even without the presence of a corresponding $\octet$ signal in dijets or $j\gamma$. We are not aware of any existing searches dedicated to this topology.

\section{Diboson Resonances}
\label{sec:diboson}
Paired diboson resonances are of course not the only signature of new hypercolor sectors. The cross-section for single hypermeson production can be significant at hadron colliders,  so it is worthwhile to compare the corresponding collider sensitivity to our results for paired diboson resonances above. We focus on the processes
\beq
\begin{aligned}
&gg \to\octet \to gg, \, g\gamma  \\
&gg \to\hetap, \, \heta \to gg, \, \gamma \gamma
\end{aligned}
\eeq
at both the LHC and a 100 TeV $pp$ collider. We do not consider signatures involving the $\pizero$, since its production at colliders is suppressed by mixing with the $\heta$ (cf.~Eq.~(\ref{eq:mixing_angle})), which is small near the isospin limit. Note that $Z\gamma$ decays can also be of interest for the singlets, however the corresponding sensitivities in this channel are typically weaker than for $\gamma \gamma$, while the diphoton branching ratio is larger than that to $Z\gamma$ in the models of interest. We therefore do not consider such decays in what follows. Also, if the $\hrho \to \octet \octet$ decay is kinematically inaccessible, the $\hrho$ will decay to dijets and heavy quark pairs. In the cases of interest, however, the corresponding signals are subdominant to the diboson decays involving the $\heta$ and $\octet$, and so we do not consider these signatures further. 
 
\subsection{$jj$}
The CMS collaboration provides limits on parton-level cross-sections for narrow $gg$ resonances in Ref.~\cite{CMS-PAS-EXO-16-056}. These results are based on 13 TeV Run 2 data. The octet width is dominated by the decay to gluons, and is significantly smaller than the experimental mass resolution across the mass range considered. We are therefore justified in directly applying the limits derived in Ref.~\cite{CMS-PAS-EXO-16-056} to the octet in our models, as well as the narrow width approximation in estimating the corresponding dijet cross-sections, since $m_{\octet}/\Gamma_{\octet} \ll 1$. Similar conclusions hold for the $\hetap$ and $\heta$. The 13 TeV ATLAS dijet search~\cite{Aaboud:2017yvp} yields exclusions similar to Ref.~\cite{CMS-PAS-EXO-16-056}, so we only implement the latter.

Production of the octet  and singlets through gluon fusion arises from the dimension-5 anomaly couplings, which scale as $1/\hl$. The cross-sections also depend on the resonance mass. For the $\hetap$, we expect $m_{\hetap}\sim \hl$, but the $\octet$ and $\heta$ masses depend additionally on the hyperquark masses. For concreteness, we will show results for $m_{\octet} = 0.38$, 0.47, and 0.75$\hl$ and $m_{\heta}=0.5 \hl$. These choices correspond to those considered in our paired diboson analyses for the $U(3)$ and $U(5)$ models\footnote{For the $U(5)$ model, we took $m_{\heta} = \hl/2, m_{\hpi}=0.4 m_{\heta}$, which, via Eq.~(\ref{eq:U5masses}), yields $m_{\octet}\simeq 0.75 \hl$. } and allow for a straightforward comparison of results for the various channels. 

\begin{figure}[t!]
\begin{center}
\includegraphics[width=0.5\linewidth]{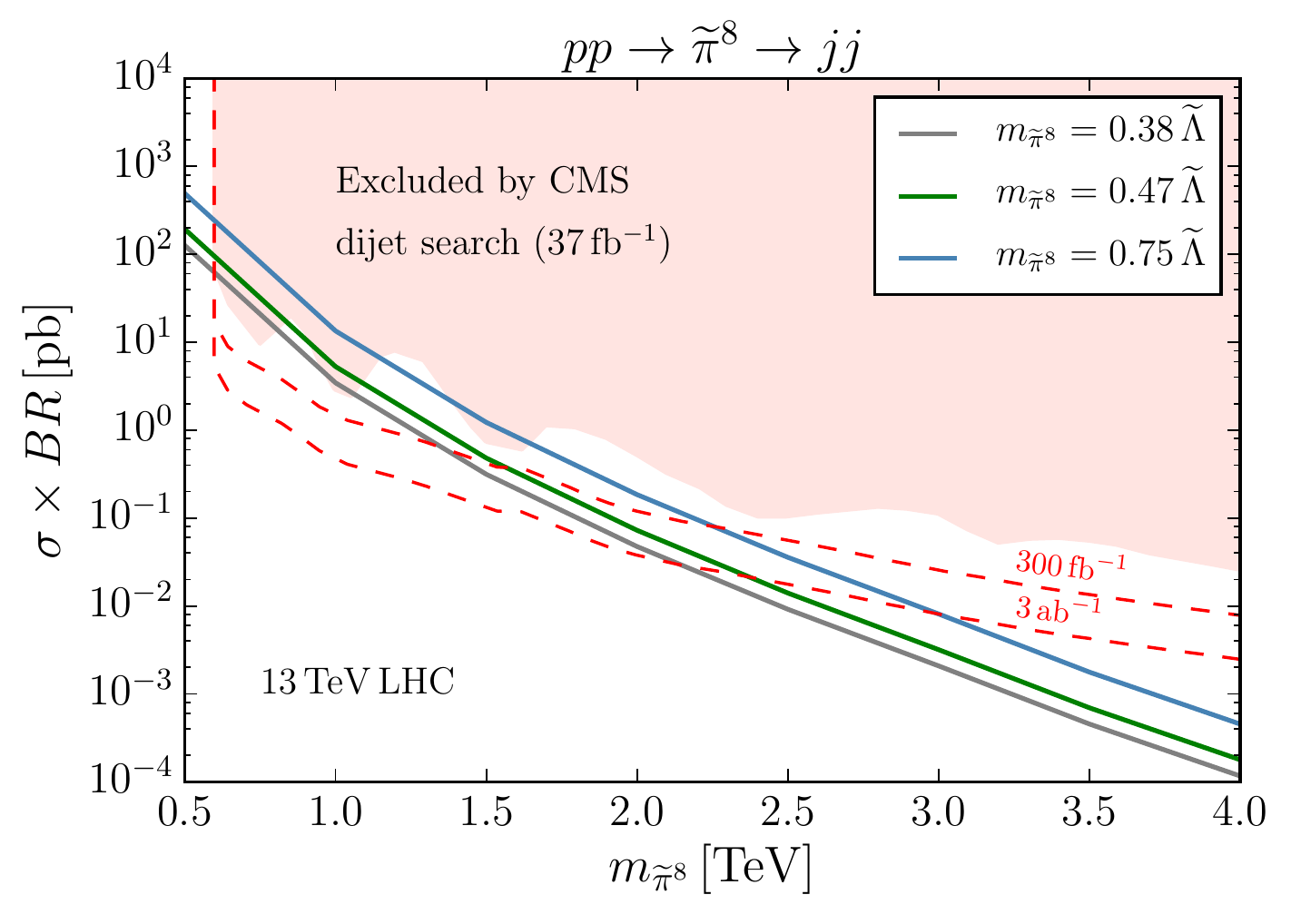}\,\includegraphics[width=0.5\linewidth]{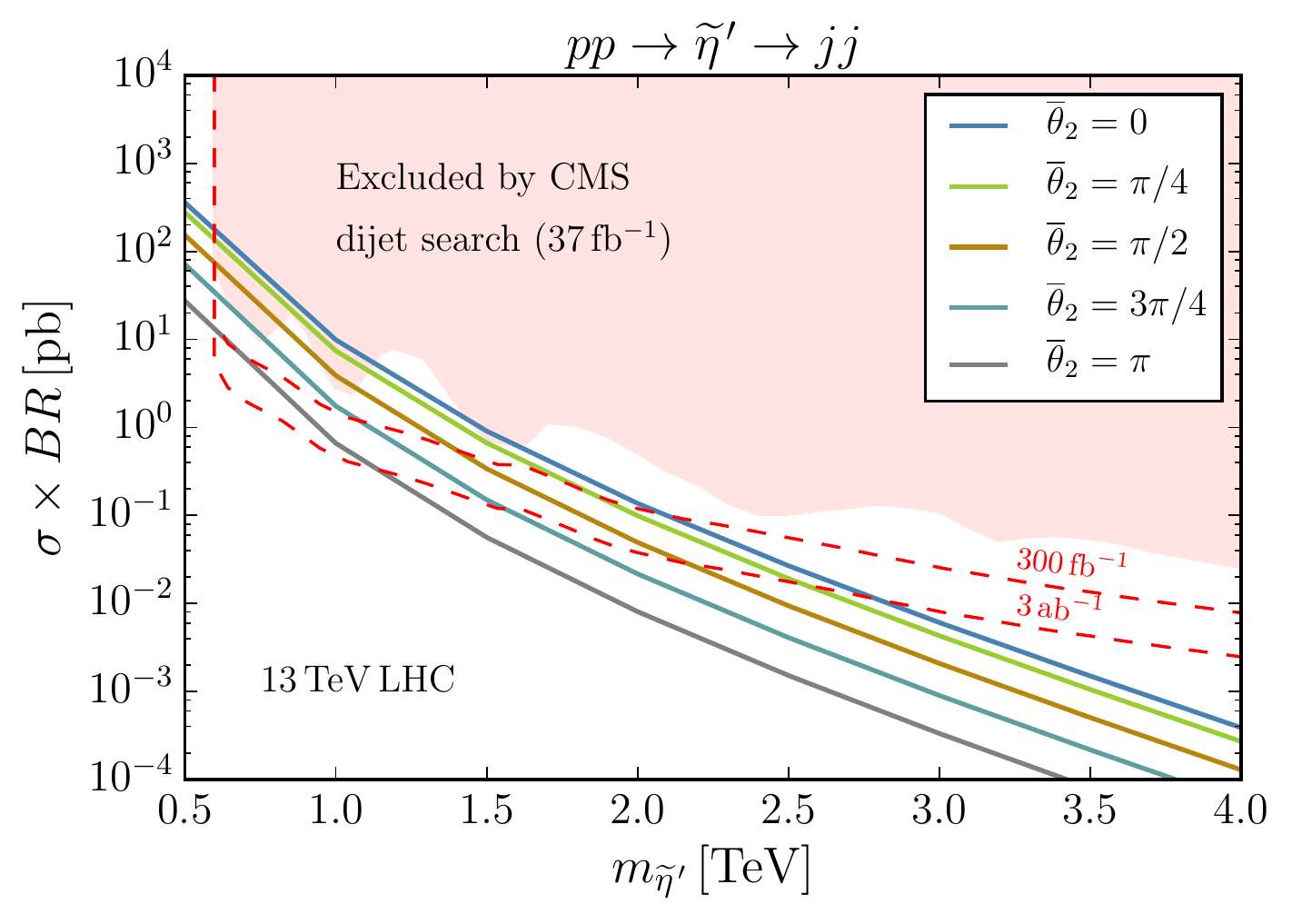}
\caption{Dijet cross-sections and limits at the 13 TeV LHC for octets (left) and singlets (right). For the singlet case, predictions are shown for $m_{\octet}=0.47 \hl$ (just below the $\hetap \to \octet \octet$ threshold) and various values of $\overline{\theta}_2$. Also plotted are estimates of the reach for 300 fb$^{-1}$ and 3 ab$^{-1}$ at 13 TeV, obtained by rescaling the expected bounds from Ref.~\cite{CMS-PAS-EXO-16-056}.} 
\label{fig:LHC}
\end{center}
\end{figure} 

The current LHC sensitivity to $\octet$ in the dijet channel is shown in Fig.~\ref{fig:LHC}. Applying the cuts specified in Ref.~\cite{CMS-PAS-EXO-16-056} corresponds to an acceptance $A \approx 0.6$ across the entire mass range. Limits on $\sigma \times BR$ can then be directly obtained from the 95\% C.L.~bounds on the parton-level $\sigma \times BR \times A$ for a dijet resonance decaying to gluons reported in Ref.~\cite{CMS-PAS-EXO-16-056}. Fig.~\ref{fig:LHC} shows that the LHC is currently sensitive to values of $m_{\octet}\lesssim 1-1.5$ TeV, corresponding to $\hl \lesssim 2-3$ TeV for the values of $m_{\octet}$ shown. Future LHC searches can extend this reach considerably higher, up to $m_{\octet}\sim 2-3$ TeV, $\hl \sim 4-5$ TeV in searches for the octet at the high-luminosity LHC. Estimates for 300 and 3000 fb$^{-1}$, corresponding to the dashed red curves in Fig.~\ref{fig:LHC}, were obtained by rescaling the expected sensitivities presented in Ref.~\cite{CMS-PAS-EXO-16-056} by the square root of the ratio of integrated luminosities, $\left(\int \mathcal{L} dt /36 \, {\rm fb}^{-1} \right)^{1/2}$. Increasing $\sqrt{s}$ to 14 TeV would of course slightly increase the reach.

On the right hand side of Fig.~\ref{fig:LHC}, we also show the LHC sensitivity to the singlet $\hetap$ in the $U(3)$ model, assuming $m_{\hetap}=\hl$, $m_{\octet}=0.47 m_{\hetap}$. The dijet branching ratio now depends on $\hth$, since the $CP$-violating $\hetap \to \octet \octet$ decay channel is open. For a fixed $\hl$, the resulting constraints are not as strong as those for the octet, due primarily to the higher mass of the $\hetap$ relative to $\hl$. Nevertheless, dijet production through the $\hetap$ excludes low values of $\hl$ for small $\thetatwo$, and the high-luminosity LHC will be able to significantly improve the sensitivity to the singlet.

\begin{figure}[t!]
\begin{center}
\includegraphics[width=0.5\linewidth]{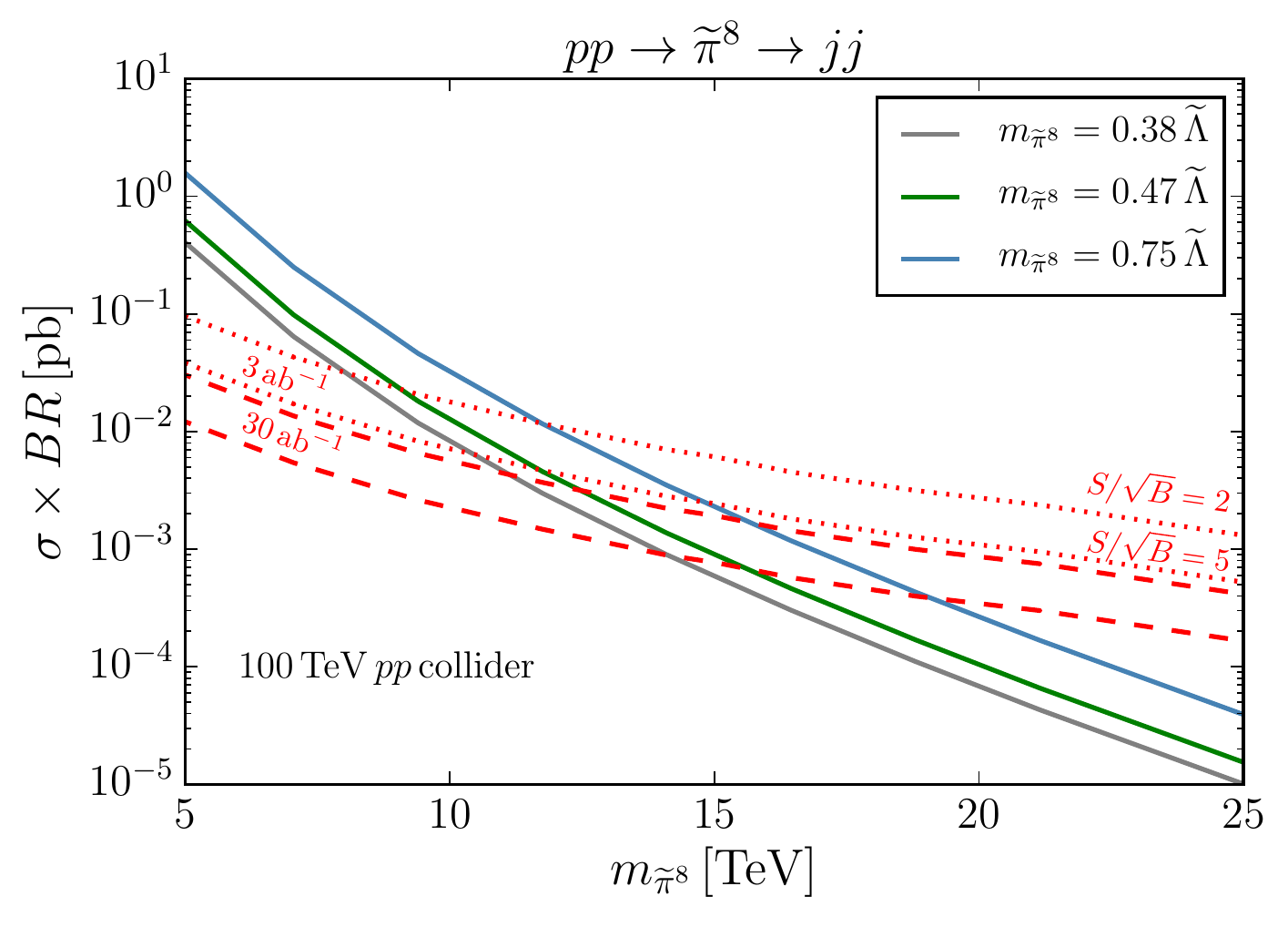}\,\includegraphics[width=0.5\linewidth]{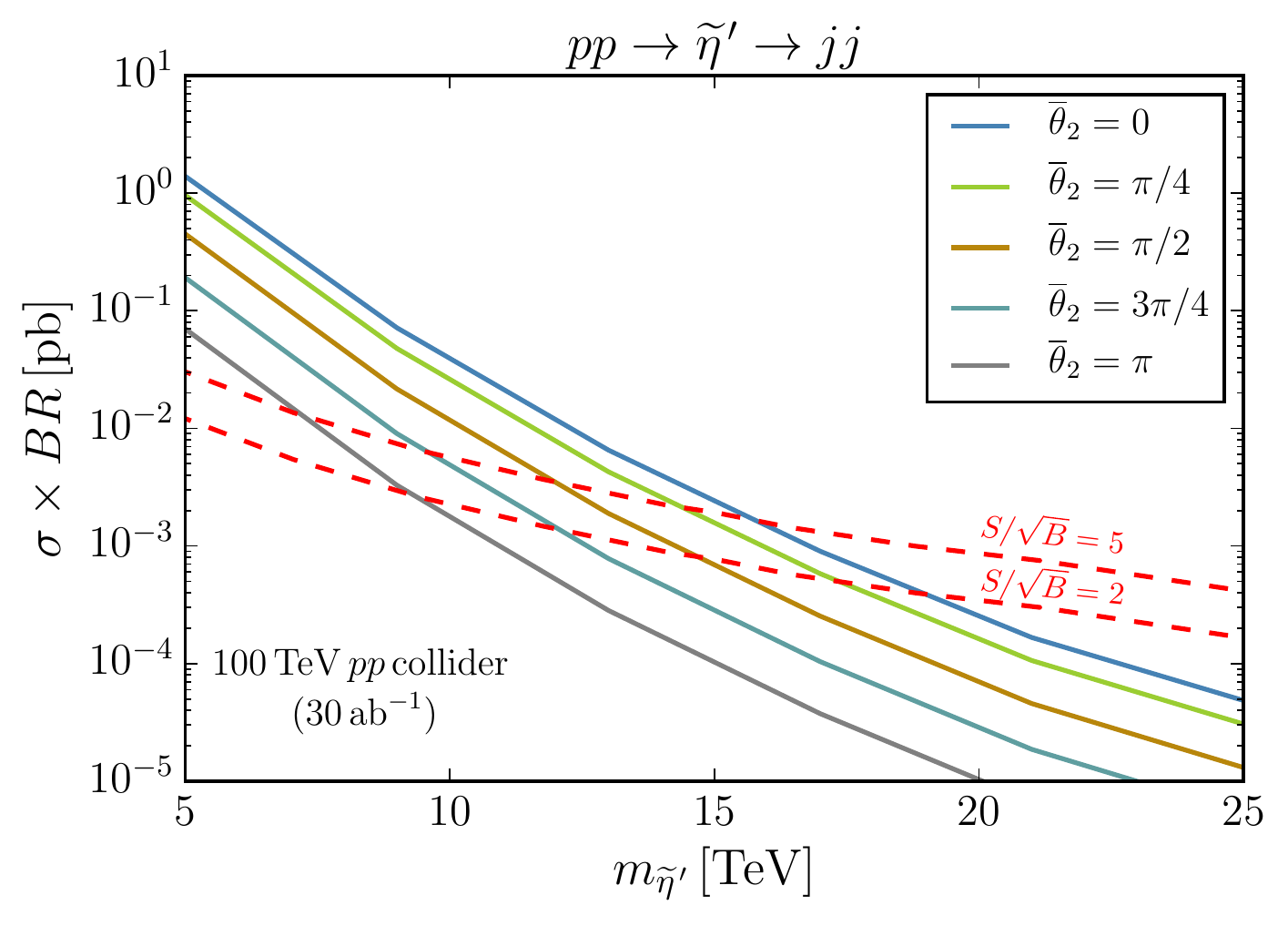}
\caption{Dijet cross-sections and projected sensitivities at the 100 TeV $pp$ collider for octets (left) and singlets (right). For the singlet case, predictions are shown for $m_{\octet}=0.47 \hl$ (just below the $\hetap \to \octet \octet$ threshold) and various values of $\overline{\theta}_2$. } 
\label{fig:FCC_dijet}
\end{center}
\end{figure} 

A 100 TeV $pp$ collider will have the opportunity to observe both the parametrically light and heavy states of a new hypercolor sector via dijets across a large range in $\hl$.  We show $\sigma \times BR$ for $p p \to \octet \to j j$ and $p p \to \hetap \to j j$ in Fig.~\ref{fig:FCC_dijet}, along with the projected sensitivities of a 100 TeV collider for different integrated luminosities. These curves were obtained by simulating dijet events for the signal and multi-jet QCD background as in the analyses of Sec.~\ref{sec:paireddiboson}. We impose the requirement $|\eta|<2.5$ on the pseudorapidity of the leading $p_T$ jets, as well as $|\Delta \eta(j_1 j_2)| \leq 1.3$ to reduce the background from $t$-channel QCD dijet events, as in Ref.~\cite{CMS-PAS-EXO-16-056}. We use the invariant mass of the leading $p_T$ jet pair to isolate the signal. For a given $\octet$ mass, we find the peak of the invariant mass distribution of the leading two jets, $m_{jj}^{\rm max}$, and apply a cut on $m_{jj}$ outside a double-sided window centered on $m_{jj}^{\rm max}$. The window extends out to values of $m_{jj}$ for which the distribution, $d\sigma/dm_{jj}$ falls to 80\% of its value at $m_{jj}^{\rm max}$. We then infer sensitivities from the number of signal and background events in this window. The sensitivity curves for the singlet are obtained analogously. We have verified that our results agree reasonably well with other 100 TeV dijet studies, such as Ref.~\cite{Yu:2013wta}. 

Fig.~\ref{fig:FCC_dijet} suggests that new hypercolor sectors up to $\hl \sim 20-30$ TeV could be discovered with 30 ab$^{-1}$ at a future 100 TeV collider through the $\octet$ dijet signal\footnote{Our analysis above neglected trigger effects, which could be an issue for reaching low masses (see e.g.~Ref.~\cite{Yu:2013wta} for a discussion). }. Interestingly, a future collider can also indirectly provide information about new vacuum angles in these models via dijets. If both the octet and singlet $\hetap$ are discovered in the dijet channel, and $m_{\hetap}>2m_{\octet}$, the observed value of $\sigma\times BR(pp \to \hetap\to jj)$ can be used to infer an upper bound on $\thetatwo$ in a given model. Obtaining a model-independent upper bound on new $CP$-violating effects in this way would require that the pseudoscalar nature of the $\hetap$ be inferred from e.g.~kinematic distributions of gluon fusion $\hetap$ production with additional ISR jets, as suggested in Ref.~\cite{Dolan:2014upa}, which may be difficult. 

\begin{figure}[t!]
\begin{center}
\includegraphics[width=0.5\linewidth]{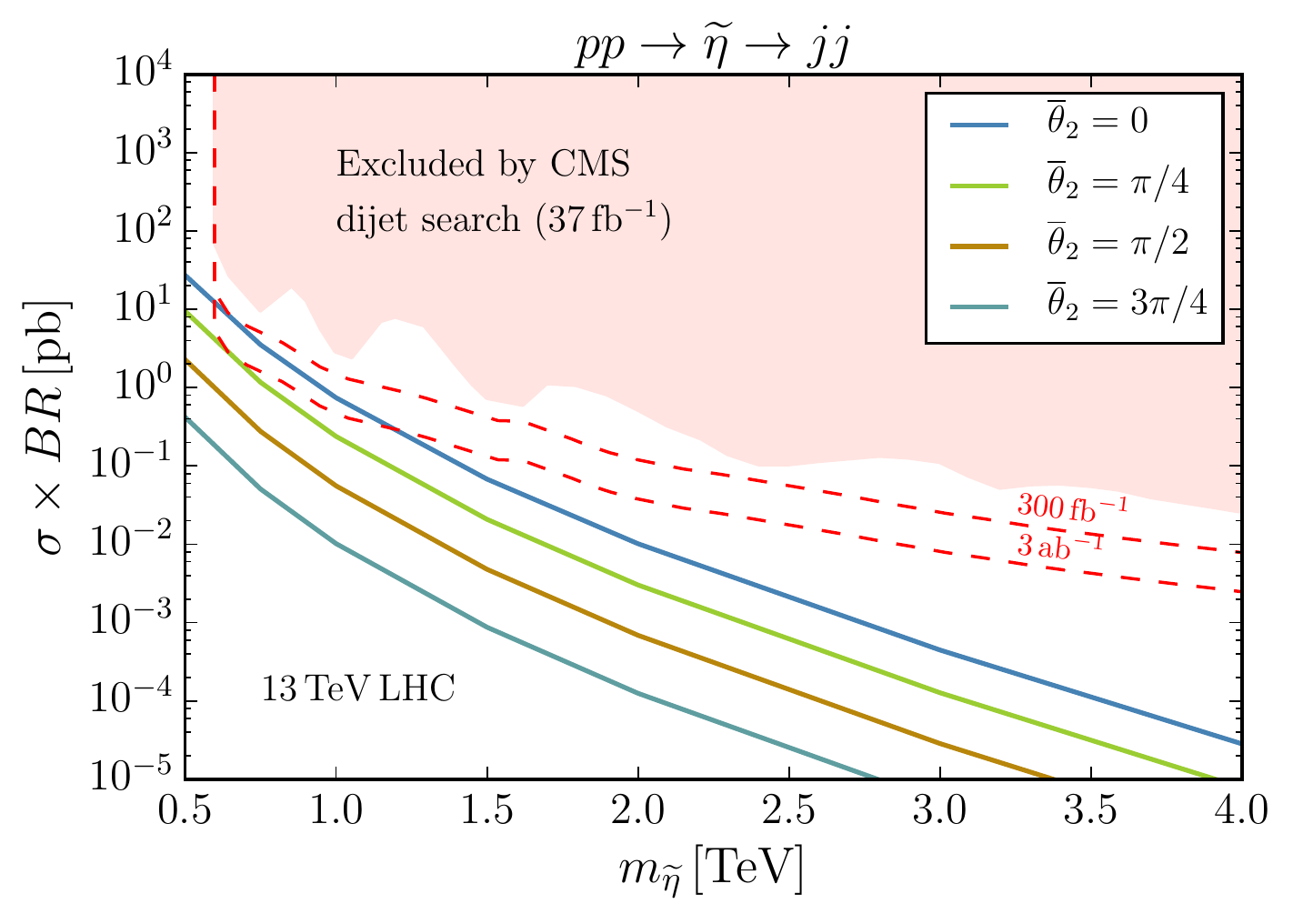}\,\includegraphics[width=0.5\linewidth]{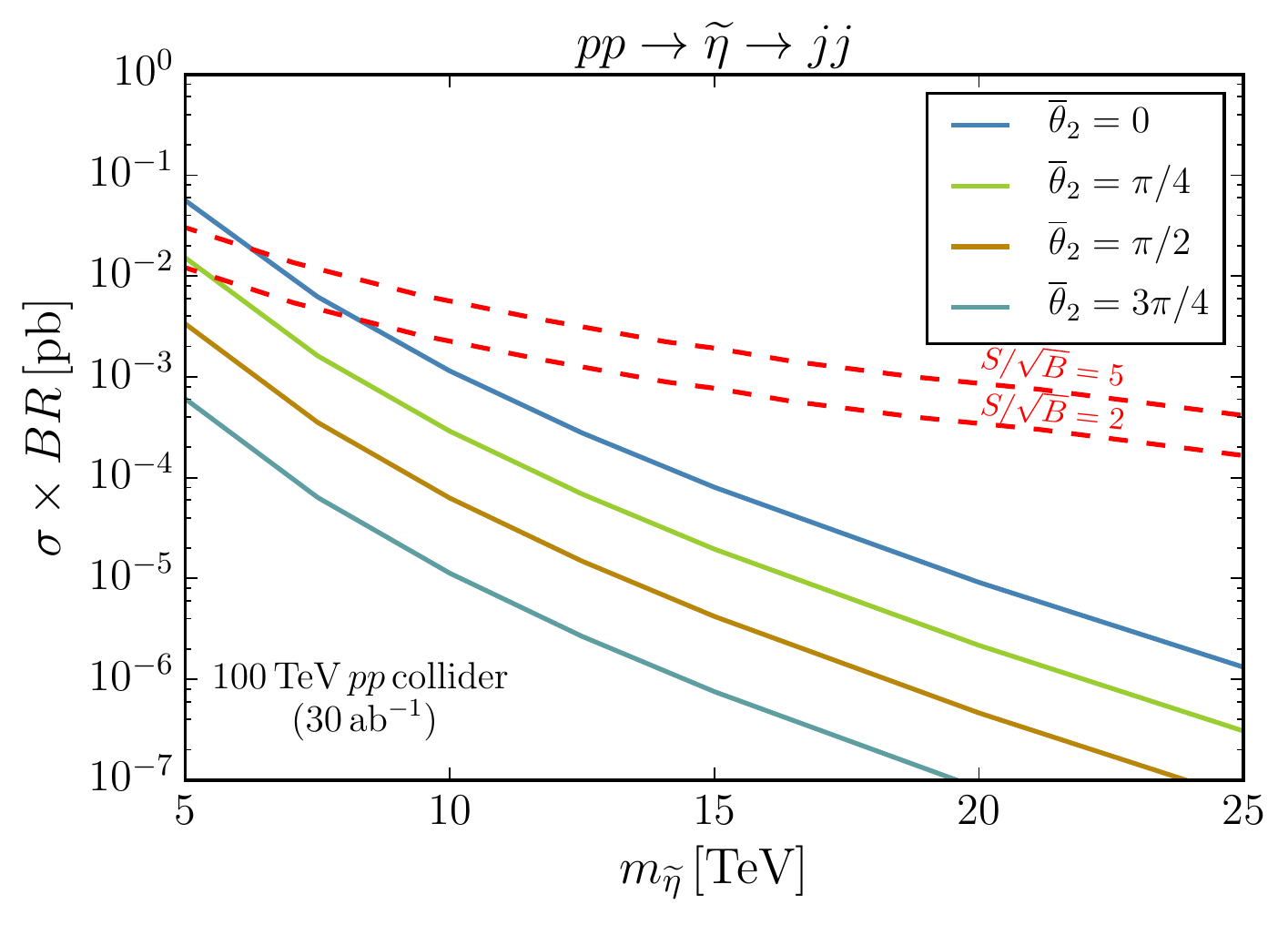}
\caption{Dijet cross-sections, limits, and projected sensitivities at the LHC and 100 TeV $pp$ collider for the $\heta$ of the $U(5)$ model. Predictions are shown for $m_{\heta}=\hl/2$, $m_{\pizero}=0.4 m_{\heta}$, and for various values of $\overline{\theta}_2$. } 
\label{fig:eta_dijets}
\end{center}
\end{figure} 

For completeness we also show results for the $\heta$ of the $U(5)$ model in dijets in Fig.~\ref{fig:eta_dijets}. There is less sensitivity in this channel compared to the corresponding results for the $\octet$ and $\hetap$ due to the relatively large $m_{\heta}/\hl$ ratio we have assumed, and because there is no color factor enhancement as for the $\octet$. We will see below that diphoton searches provide a significantly more powerful probe of the $\heta$ when the hyperquarks have non-zero hypercharge. 

\subsection{$j\gamma$}

We can proceed similarly for the $j\gamma$ final state. This channel has been recognized as a potentially powerful probe of new QCD-like sectors at the LHC~\cite{Bai:2010mn, Bai:2016czm}. Current searches already place limits on $\sigma \times BR (pp\to \octet \to j \gamma)$, such as the CMS study in Ref.~\cite{Sirunyan:2017fho}, and the ATLAS study in Ref.~\cite{Aaboud:2017nak}, which frame the results in terms of excited quark decays to $q+\gamma$. Simulating $pp \to \octet \to j \gamma$ events and applying the basic selection criteria outlined in Ref.~\cite{Sirunyan:2017fho}, we find a similar acceptance $\times$ efficiency as that reported by CMS. However, because the excited quarks decay to $q+\gamma$, and not $g+\gamma$, we also find that the reconstructed widths of the $\octet$ resonances in the $j\gamma$ invariant mass distribution are significantly larger than those for excited quarks. We therefore expect a direct application of the CMS and ATLAS bounds to our scenario to likely be overly pessimistic, but still illustrative.

\begin{figure}[t!]
\begin{center}
\includegraphics[width=0.5\linewidth]{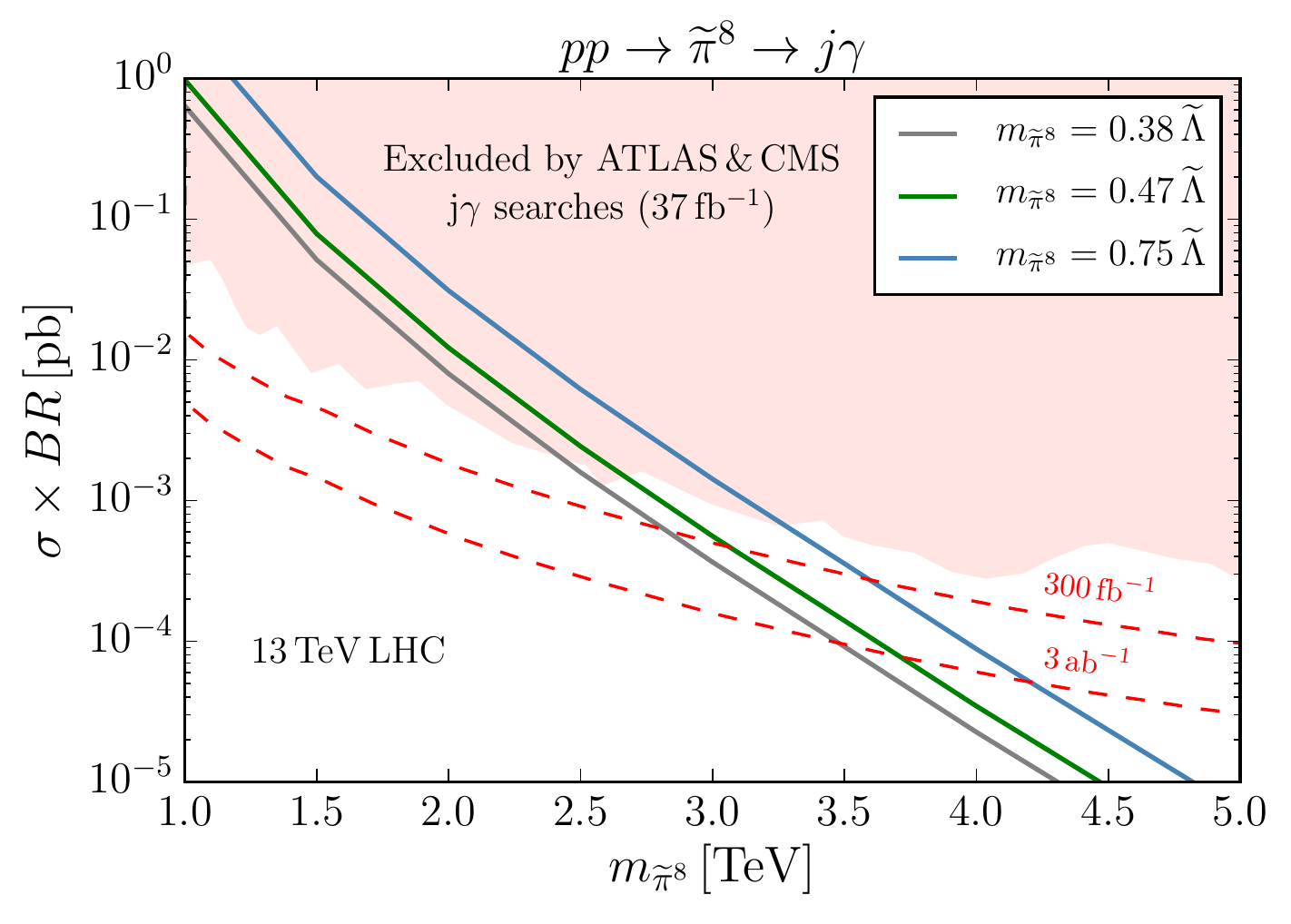}\,\includegraphics[width=0.5\linewidth]{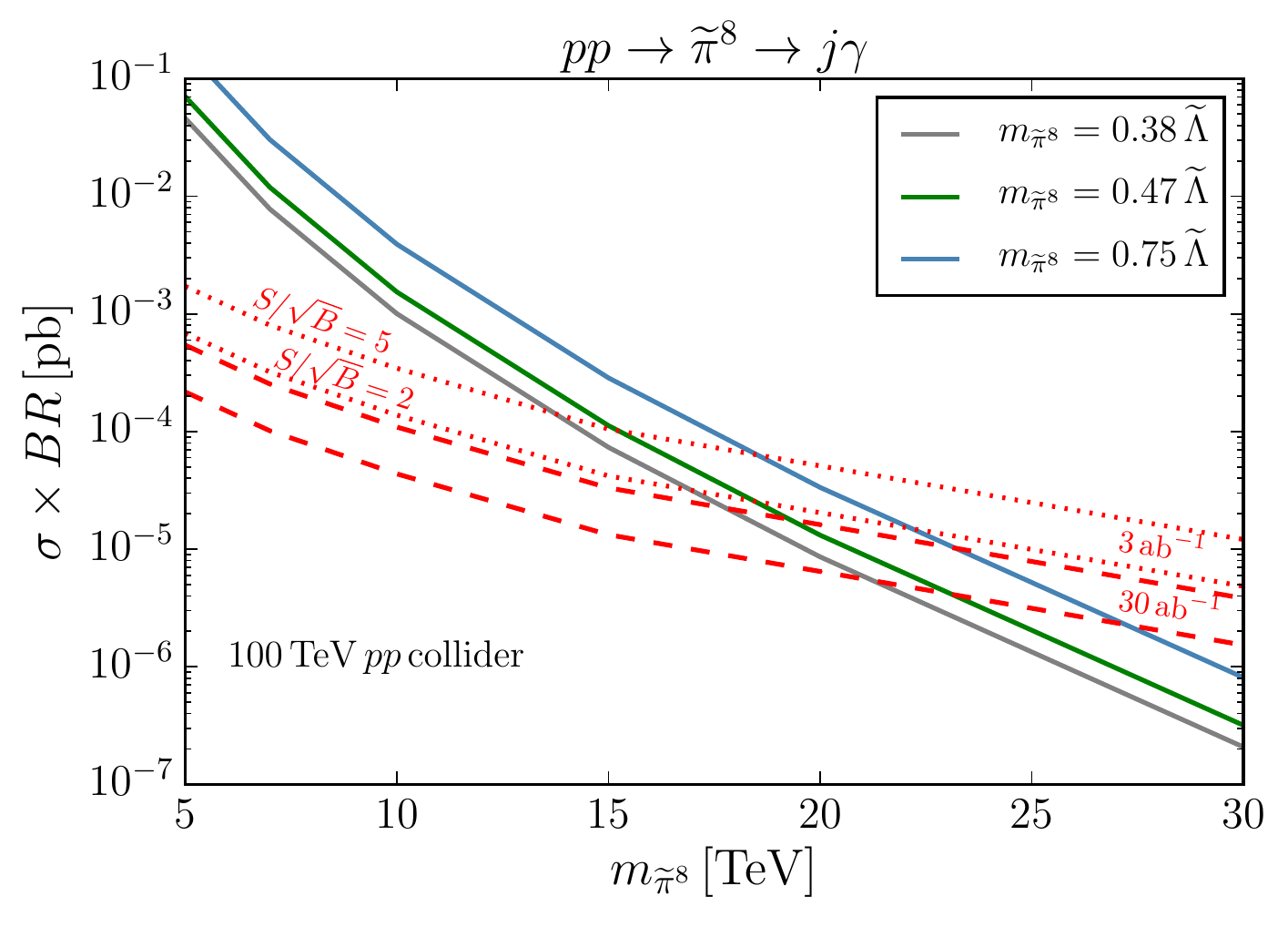}
\caption{Cross-sections, limits, and projected sensitivities for the process $pp \to \octet \to j \gamma$ at the 13 TeV LHC (left) and a 100 TeV $pp$ collider (right). Predictions are shown for several choices of $m_{\octet}/\hl$, corresponding to the various values assumed in our tetraboson analyses. The limits apply to narrow $q\gamma$ resonances~\cite{Sirunyan:2017fho, Aaboud:2017nak}, and are likely overly pessimistic for the wider $g\gamma$ final state of interest here. They should be interpreted as a conservative bound. The projections for higher luminosities at 13 TeV were obtained by rescaling the current expected limits in Ref.~\cite{Sirunyan:2017fho}.} 
\label{fig:jgamma}
\end{center}
\end{figure} 

Directly applying the bounds on $\sigma \times BR$ reported in Refs.~\cite{Sirunyan:2017fho, Aaboud:2017nak} to the process involving the $\octet$, we arrive at the results shown in Fig.~\ref{fig:jgamma}. Current results conservatively exclude octet masses below $\sim 2.5 - 3.5$ TeV. The projected sensitivities for higher luminosities were obtained in the same way as for dijets above (by rescaling the expected CMS limits). With 3 ab$^{-1}$, the exclusion reach of the LHC will be extended up to $\sim 4-4.5$ TeV masses. This channel is thus significantly more sensitive than dijets, but is also more model-dependent, as it depends quite sensitively on the value of the hyperquark hypercharge. Our choice of $Y=4/3$ illustrates the maximum reach in this channel such that there are dimension-6 operators allowing the triplets to decay. 

We also consider the 100 TeV reach for the octet in the $j \gamma$ channel.  We use the same simulation pipeline as in our dijet analysis, and the same selection criteria as Ref.~\cite{Sirunyan:2017fho}, except now requiring $p_T(j_1)>1.5$ TeV, $p_T(\gamma_1)>1.25$ TeV. For a given $m_{\octet}$, we compare the number of signal and background events in a window centered on $m_{\octet}$ and extending to where the expected number of signal events falls to half of its maximum value. The results are shown in the right hand side of Fig.~\ref{fig:jgamma}. A 100 TeV collider will have an impressive reach for the octet in this channel, with discovery potential extend up to $m_{\octet} \sim 15-25$ TeV (corresponding to $\hl \sim 30-45$ TeV), depending on the $m_{\octet}/\hl$ ratio. The signal-to-background ratio is $\sim 0.1$ or larger across the discovery range. Comparing to the dijet reach shown in Fig.~\ref{fig:FCC_dijet}, we see that the $j\gamma$ channel can be considerably more sensitive to the octet, although this depends on the electric charge assignments for the hyperquarks. 

\subsection{$\gamma \gamma$}

\begin{figure}[t!]
\begin{center}
\includegraphics[width=0.5\linewidth]{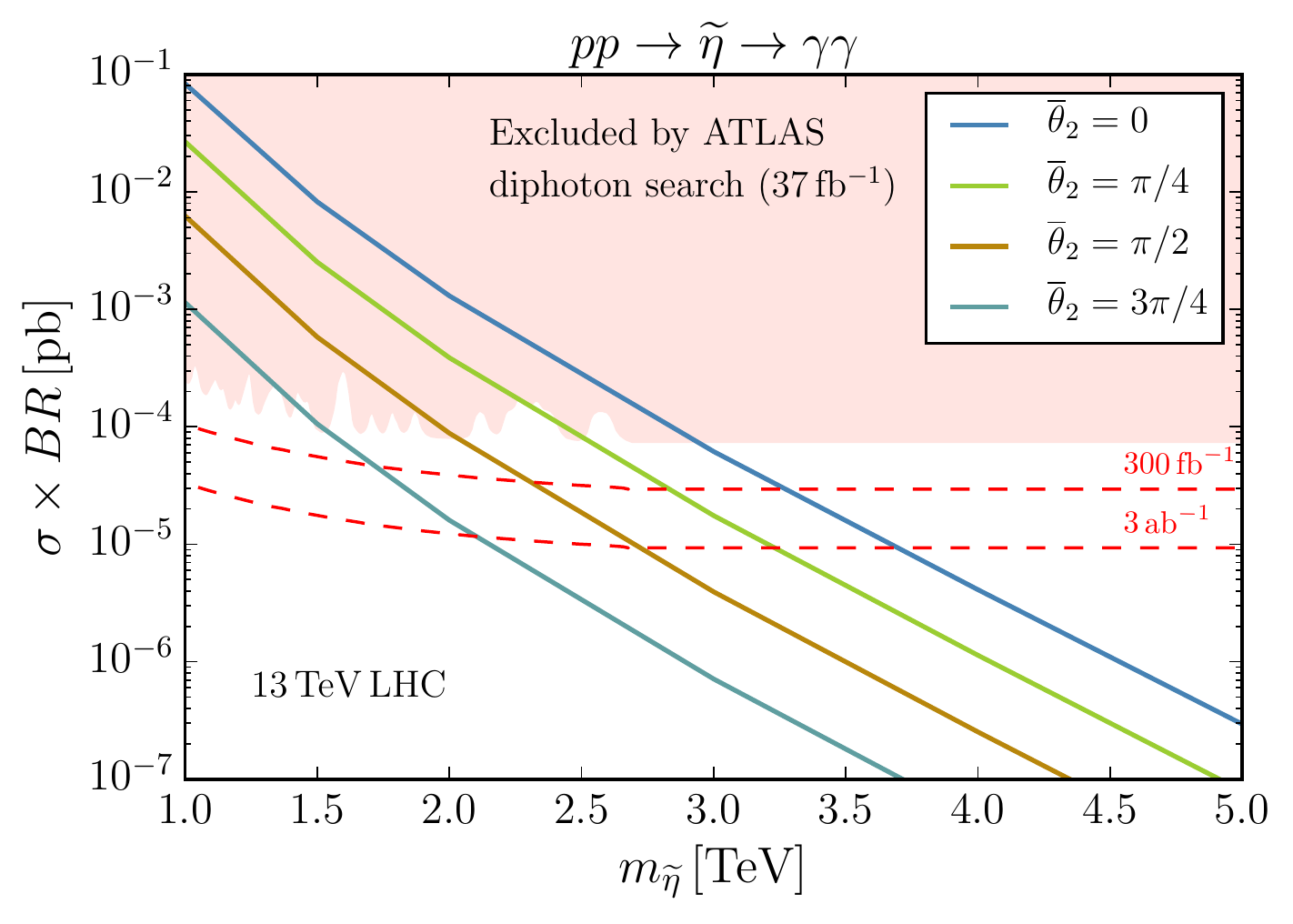}\,\includegraphics[width=0.5\linewidth]{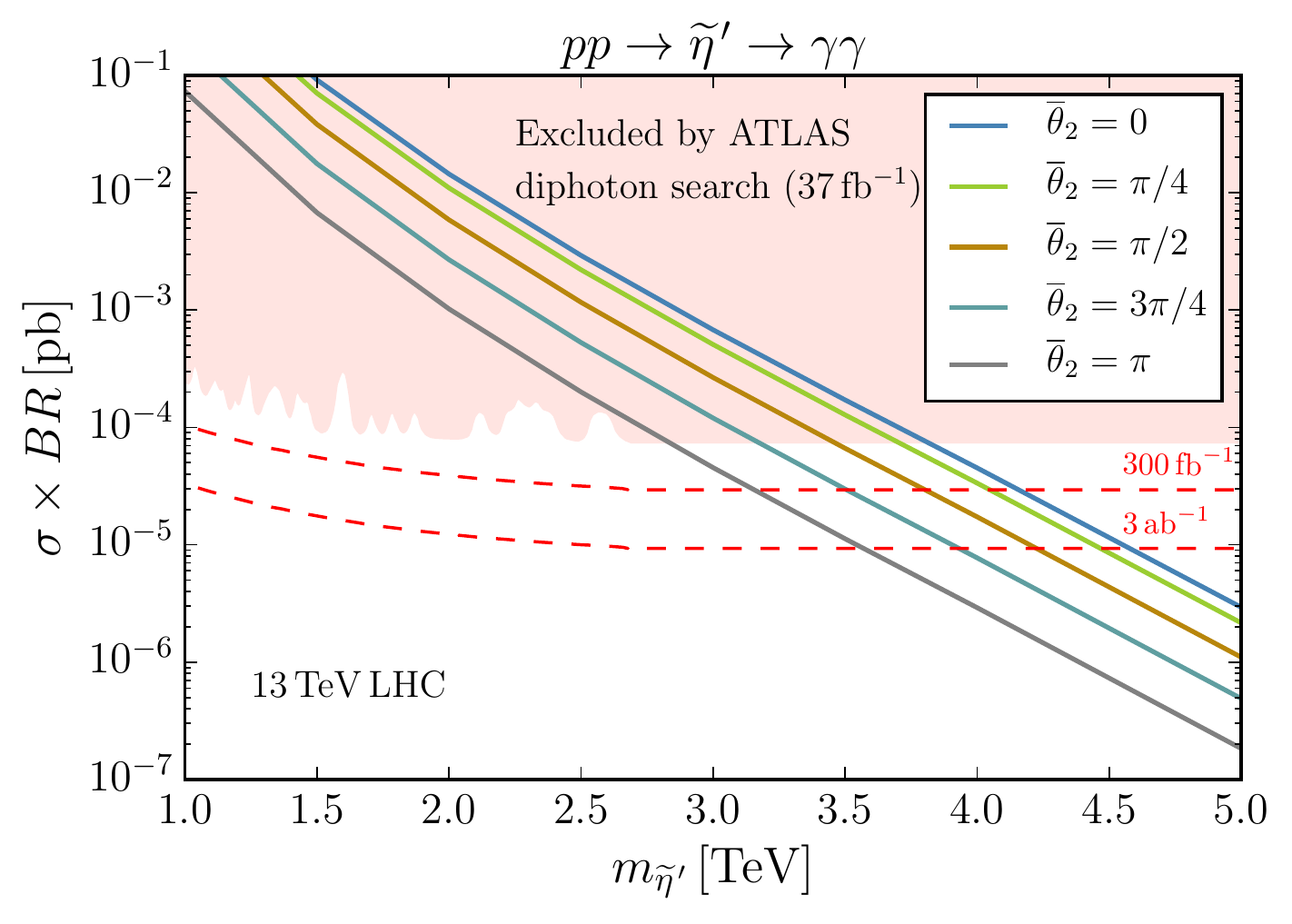}
\caption{Cross-sections, limits, and projected sensitivities for the process $pp \to \heta\, (\hetap) \to \gamma \gamma$ at the 13 TeV LHC. For the $\heta$ of the $U(5)$ model, predictions are shown for $m_{\heta}=\hl/2$, $m_{\pizero}=0.4m_{\heta}$. For the $\hetap$ of the $U(3)$ model, the predictions correspond to $m_{\hetap} =\hl$ and $m_{\octet}=0.47 m_{\hetap}$. In both cases we also show the dependence on $\overline{\theta}_2$. Note that for the higher luminosity projections we assume $\left(\int \mathcal{L} dt \right)^{1/2}$ scaling for both the low and high-mass regions, which will likely underestimate the sensitivity at high masses, where the search is essentially background-free. Scaling instead by the ratio of integrated luminosities, the reach would asymptote to $\sigma \times BR \sim 10^{-5}$ ($10^{-6}$) pb at 300 fb$^{-1}$ (3 ab$^{-1}$).} 
\label{fig:diphoton_LHC}
\end{center}
\end{figure}

The singlet pseudoscalars can also decay to diphotons. We show limits on the diphoton production cross-section through the $\heta$ in the $U(5)$ model and the $\etap$ in the $U(3)$ model in Fig.~\ref{fig:diphoton_LHC}, along with the predicted cross-sections for various values of $\thetatwo$. The $\thetatwo$-dependence arises from $CP$-violating decays to hyperpions, which is kinematically open in both cases (as before we take $m_{\heta}=\hl/2$, $m_{\pizero} = 0.4 m_{\heta}$ in the $U(5)$ case, and $m_{\hetap}=\hl$, $m_{\octet}=0.47 m_{\hetap}$ in the $U(3)$ model). The limits are taken from the ATLAS search in Ref.~\cite{Aaboud:2017yyg}, which only report limits out to masses around 2.5 TeV for the spin-0 case. Since the spin-2 results are reported out to 5 TeV and are roughly flat past 2.5 TeV, we simply extrapolate the spin-0 limits out to higher masses. This is also consistent with the CMS diphoton search in Ref.~\cite{Khachatryan:2016yec}, which features lower integrated luminosity and thus weaker limits. Note that for the higher luminosity projections we assume the same scaling for both the low and high-mass regions with luminosity, which will likely underestimate the sensitivity at high masses, where the search is essentially background-free. Scaling instead by the ratio of integrated luminosities, the reach would asymptote to $\sigma \times BR \sim 10^{-5}$ ($10^{-6}$) pb at 300 fb$^{-1}$ (3 ab$^{-1}$).

\begin{figure}[t!]
\begin{center}
\includegraphics[width=0.5\linewidth]{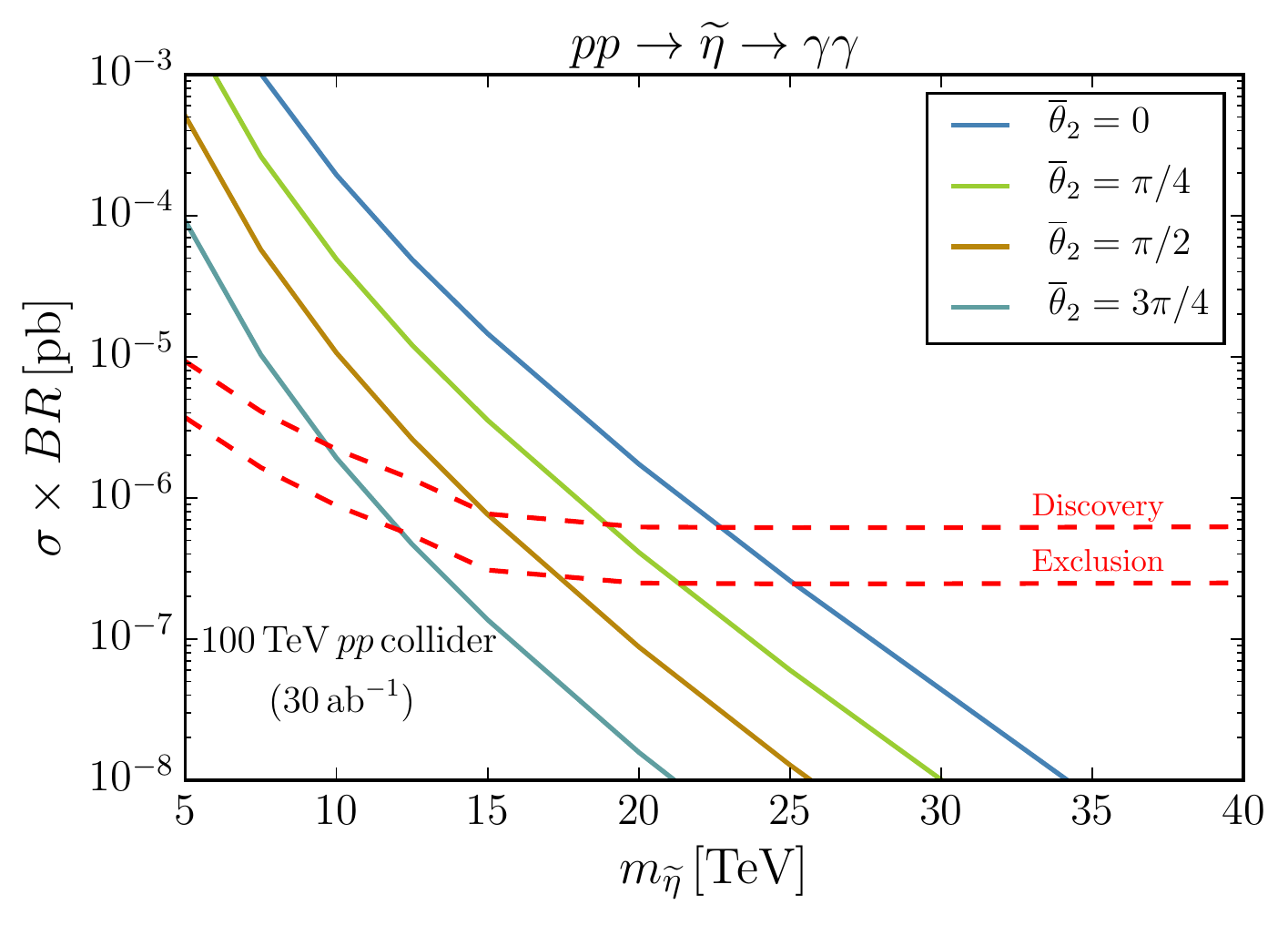}\,\includegraphics[width=0.5\linewidth]{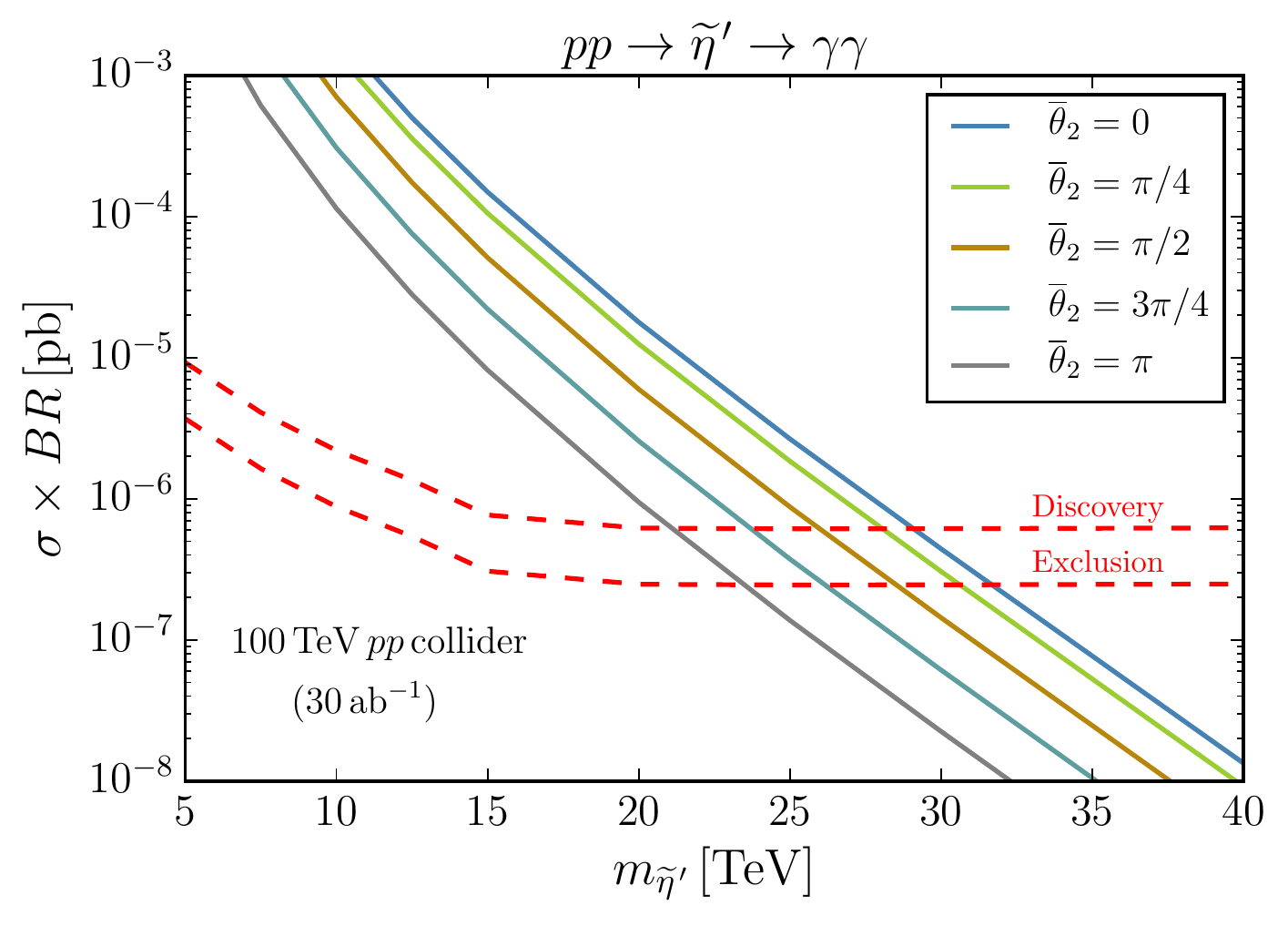}
\caption{Cross-sections and expected sensitivities for the process $pp \to \heta, \hetap \to \gamma \gamma$ at a future 100 TeV collider. The parameters are fixed as in Fig.~\ref{fig:diphoton_LHC}.} 
\label{fig:diphoton_FCC}
\end{center}
\end{figure}

We also estimate the reach of a future 100 TeV $pp$ collider in the diphoton channel. We generate signal and background events as above, considering only the irreducible Standard Model $\gamma \gamma$ background. We require two photons with $p_T(\gamma_1) > 0.4 m_{\gamma \gamma}$,   $p_T(\gamma_2) > 0.3 m_{\gamma \gamma}$, as in the 13 TeV ATLAS search~\cite{Aaboud:2017yyg}. For a given resonance mass, we cut in a window centered on the mass and extending out to the points where the number of expected signal events falls to 10\% of the maximum value. We define the discovery (exclusion) reach at low masses by requiring $S/\sqrt{B}=5$ (2) at 30 ab$^{-1}$. At large invariant masses, the Standard Model $\gamma \gamma$ background is very small, and so instead of a requirement on $S/\sqrt{B}$, we define the discovery (exclusion) reach by requiring more than 10 (4) signal events at 30 ab$^{-1}$, once all cuts are imposed. The results are shown in Fig.~\ref{fig:diphoton_FCC} for both the $\heta$ of the $U(5)$ model, and $\hetap$ of the $U(3)$ scenario, assuming the same relations between parameters as for the LHC analysis.

Fig.~\ref{fig:diphoton_FCC} shows that a 100 TeV collider can have excellent reach for the singlets in the diphoton channel. Comparing with Fig.~\ref{fig:jgamma}, for a fixed value of $\hl$ and small $\hth$, the sensitivity to new QCD-like sectors via diphotons can exceed that in $j\gamma$.

\subsection{Implications for Paired Diboson Resonance Searches} \label{sec:implications}

\begin{figure}[t!]
\begin{center}
\includegraphics[width=0.45\linewidth]{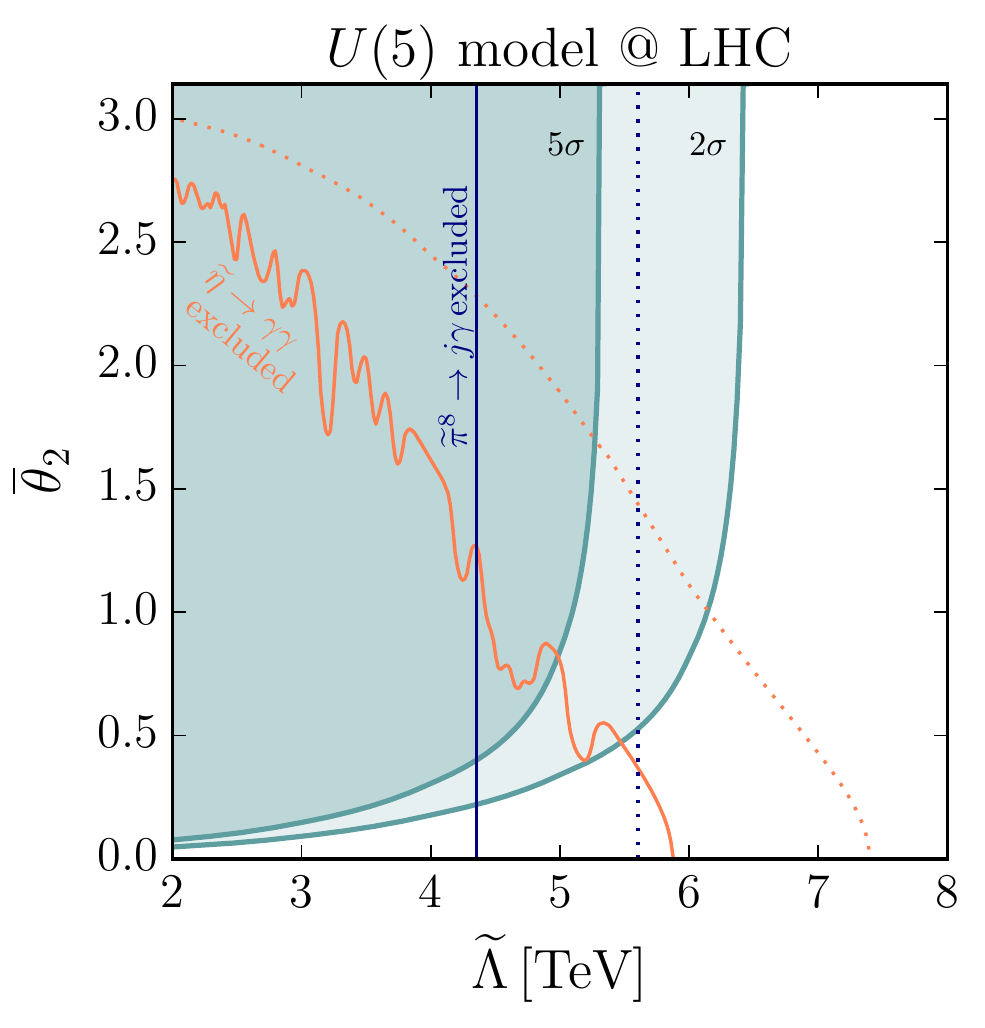} \, \includegraphics[width=0.45\linewidth]{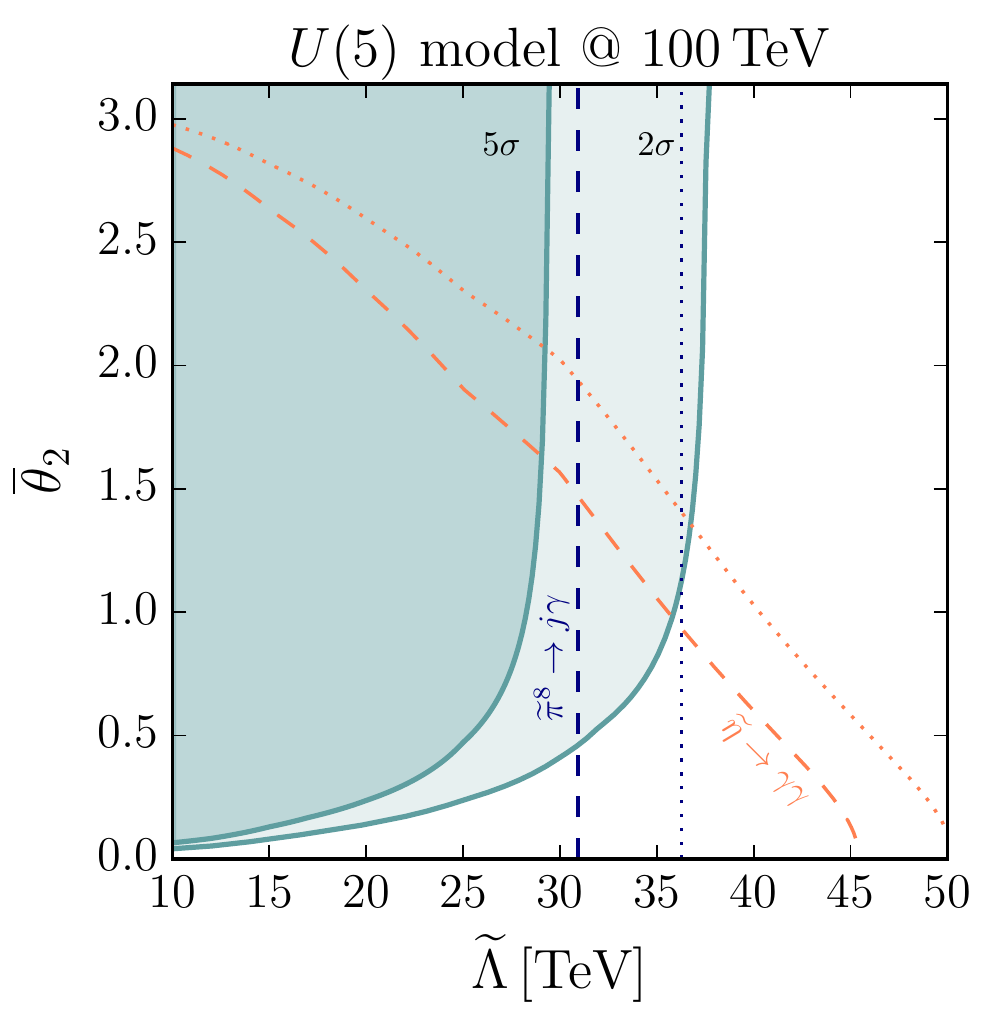}
\caption{Summary of the projected discovery and exclusion reach in  the $U(5)$ model at the 13 TeV LHC (left) and a 100 TeV collider (right).  We have assumed $m_{\heta} = \hl/2$, $m_{\pizero} = 0.4 m_{\heta}$, $m_{\octet} = 0.75 \hl$. The darker (lighter) shaded regions correspond to the approximate $5\sigma$ ($2\sigma$) sensitivity reach via the $ g g \to \heta \to \pizero \pizero \to gg +\gamma \gamma$ $CP$-violating process assuming 3 ab$^{-1}$ for the LHC and 30 ab$^{-1}$ for 100 TeV. We also show sensitivities to  $ g g \to \heta \to \gamma \gamma$ (orange) and $ g g \to \octet \to g \gamma$ (blue). For the LHC, the corresponding solid contours indicate current 95\% C.L.~exclusion limits, while dotted contours correspond to 95\% C.L.~projected sensitivities at 3 ab$^{-1}$. For the 100 TeV results, dashed (dotted) contours indicate $\sim 5\sigma$ discovery ($2\sigma$ exclusion) reach given 30 ab$^{-1}$. For all channels shown the experimentally accessible regions lie to the left of the corresponding contours.} 
\label{fig:model_2_money}
\end{center}
\end{figure} 

The limits and projections from diboson resonance searches have important ramifications for the discovery potential in paired diboson searches at the LHC and a future 100 TeV $pp$ collider. We show the combined results for our tetraboson and diboson sensitivity estimates for the $U(5)$ model in Fig.~\ref{fig:model_2_money} in the $\hl -\thetatwo$ plane, taking $m_{\heta} = \hl/2$, $m_{\pizero} = m_{\heta}/2.5$, $m_{\octet} = 0.75 \hl$, and $Y=4/3$ for the non-singlet hyperquarks. On the left-hand side, we show results for the 13 TeV LHC. The dark and lighter shaded regions correspond to the $S/\sqrt{B}=5$ (`$5\sigma$') and $S/\sqrt{B}=2$ (`$2\sigma$') reach in the $gg \to \heta \to \pizero \pizero \to jj+\gamma \gamma$ channel provided 3 ab$^{-1}$ integrated luminosity. Also shown are the current bounds and 3 ab$^{-1}$ $95\%$ C.L.~projections from the $gg \to \heta \to \gamma \gamma$ and $ gg \to \octet \to j \gamma$ channels. The other diboson sensitivities are subdominant. The right hand side shows the analogous results for a future 100 TeV $pp$ collider with 30 ab$^{-1}$ of integrated luminosity.

From Fig.~\ref{fig:model_2_money}, we conclude:
\begin{itemize}
\item Given the current constraints from $j\gamma$ and $\gamma \gamma$ searches, there is still discovery potential for $CP$-violating hypermeson decays in the $jj +\gamma\gamma$ final state at the LHC, of which the $U(5)$ model provides an example. The sensitivity in $jj +\gamma\gamma$ can exceed that provided by $\octet \to j\gamma$ at 3 ab$^{-1}$  and provide the most sensitive probe of new QCD-like sectors with $\mathcal{O}(1)$ vacuum angles. The $\heta \to \gamma \gamma$ channel provides complementary coverage to $CP$-violating $\heta$ decays at small vacuum angles. The $\heta \to \gamma \gamma$ and $\octet \to g \gamma$ channels are typically more sensitive than $\heta$, $\octet \to jj$ and processes involving resonant production of the $\hetap$ and $\hrho$.

\item A 100 TeV collider will be able to access an impressive range of $\hl$ and $\thetatwo$ values in a variety of channels. $CP$-violating $\heta \to \pizero \pizero$ decays can be probed for confinement scales up to $\sim 35-40$ TeV and $\thetatwo$ as small as $\sim 0.05$ for small $\hl$. Such a signal would likely be accompanied by a variety of corresponding diboson signals, most notably a $j \gamma$ resonance corresponding to the $\octet$ and possibly a diphoton signal from the $\heta$ \footnote{Note that the $j\gamma$ and $\gamma \gamma$ reach grows more rapidly than that for $jj \gamma \gamma$ with increasing center of mass energy, because the backgrounds for the former are dominated by quark-quark and quark-gluon--initiated processes, for which the parton luminosities increase more slowly with $\sqrt{s}$.}. Corresponding signals in the dijet channel and from the $\hetap$ and $\hrho$ will also be present for lower values of $\hl$. 

\end{itemize}

Although not included in Fig.~\ref{fig:model_2_money}, note also that $\heta$ decays to the $4j$ final state at a 100 TeV collider could provide sensitivity to $\hl$ up to $\sim 10$ TeV and $\hth$ as small as $\sim 0.3$ for low $\hl$. Such a signal would be accompanied by $\octet \to jj$ decays, and possibly $\heta \to jj$ for small $\hth$, along with signatures of the $\hrho$ and the various hypercharge-dependent processes outlined above.

\begin{figure}[t!]
\begin{center}
\includegraphics[width=0.45\linewidth]{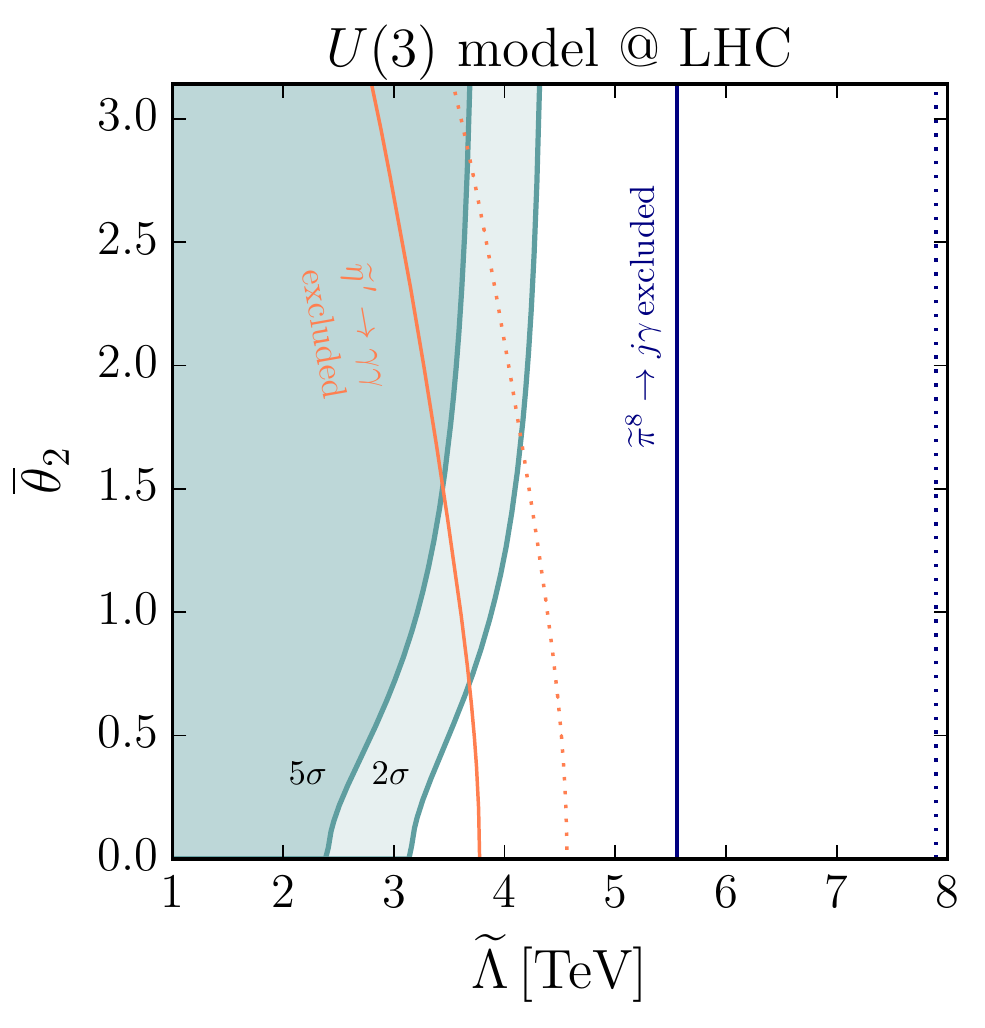} \, \includegraphics[width=0.45\linewidth]{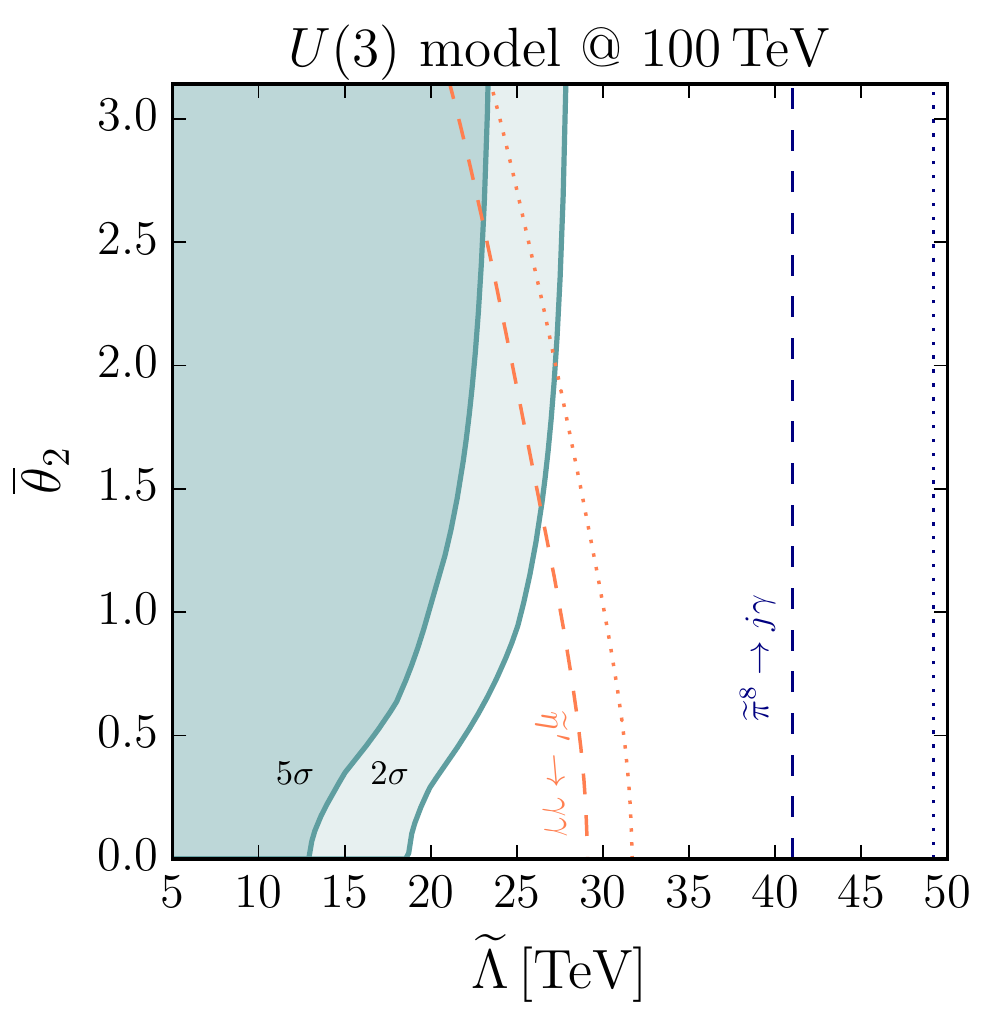}
\caption{As in Fig.~\ref{fig:model_2_money}, but for the $U(3)$ model at the 13 TeV LHC (left) and 100 TeV (right). We have assumed $m_{\hetap} = \hl$, $m_{\hrho}=0.8 m_{\hetap}$, $m_{\octet} = 0.47 \hl$, corresponding to the off-shell $\hrho$ benchmark in Eq.~(\ref{eq:benchmarks}). The darker (lighter) shaded regions correspond to the approximate $5\sigma$ ($2\sigma$) sensitivity reach via $ g g \to \hetap, \rho^* \to \octet \octet \to g\gamma +g \gamma$ processes assuming 3 ab$^{-1}$ for the LHC and 30 ab$^{-1}$ for 100 TeV. We also show sensitivities to  $ g g \to \hetap \to \gamma \gamma$ (orange) and $ g g \to \octet \to g \gamma$ (blue). For the LHC, the corresponding solid contours indicate current 95\% C.L.~exclusion limits, while dotted contours correspond to 95\% C.L.~projected sensitivities at 3 ab$^{-1}$. For the 100 TeV results, dashed (dotted) contours indicate $\sim 5\sigma$ discovery ($2\sigma$ exclusion) reach given 30 ab$^{-1}$. For all channels shown the experimentally accessible regions lie to the left of the corresponding contours.} 
\label{fig:model_1_money}
\end{center}
\end{figure} 

Analogous results for the processes involving $\octet$s in the $U(3)$ model are shown in Fig.~\ref{fig:model_1_money}. In this case, we have taken $m_{\hetap} = \hl$, $m_{\hrho}=0.8 m_{\hetap}$, $m_{\octet}=0.47 m_{\hetap}$. Because the $CP$-violating decays in this scenario involve the $\hetap$ with mass near the confinement scale, the reach for these processes at hadron colliders are not as strong as in the $U(5)$ case, but they can still provide a compelling signal for large vacuum angles. In particular, from Fig.~\ref{fig:model_1_money} we conclude:
\begin{itemize}
\item At the LHC, searches for $j \gamma$ resonances already constrain $\octet$ masses to be heavier than $\sim 2.5-3$ TeV. This would appear to exclude regions in which the LHC will have sensitivity to $CP$-violating $\hetap$ decays. One should bear in mind, however, that the interplay between these various searches depend on the charge assignments for the hyperquarks. Furthermore, the $j\gamma$ bound shown in Fig.~\ref{fig:model_1_money} is aggressive, and would be weakened in a more careful recast of the existing searches accounting for the smearing in the $j\gamma$ final state associated with gluons rather than quarks. Nonetheless, we expect that a discovery of the $\hetap$ in this mode at the LHC will become increasingly unlikely if the limits on $j \gamma$ resonances continue to improve with no evidence of the $\octet$. 

\item At 100 TeV, one can observe the effects of $CP$-violating $\hetap$ decays for $\hl$ up to $\sim 25-30$ TeV in the $U(3)$ model for large vacuum angles. In contrast to the $U(5)$ case, contributions from $\hrho^{(*)}  \octet \octet$ interactions comprise an irreducible contribution to this channel, even in the absence of $\hth$. We expect the $\octet$ to appear in $jj$ and $j\gamma$ resonance searches (and the $\hrho$ to appear in $q \bar{q}$ searches if $\hrho \to \octet \octet$ is kinematically forbidden) before seeing the tetraboson signals corresponding to the $\hrho^{(*)}$ and $\hetap$, for this benchmark. This is consistent with the our treatment of non-resonant $\octet$ pair production as a background to the resonant tetraboson signatures in Secs.~\ref{sec:jaja} and~\ref{sec:jjja}.

\item On-shell $\hrho \to \octet \octet$ production will also be difficult to observe via $2j2\gamma$ and $3j1\gamma$ at the LHC given current $j\gamma$ constraints on the $\octet$. In the analyses of Secs.~\ref{sec:jaja} and~\ref{sec:jjja}, we assumed $m_{\octet}=0.38 \hl$. From Fig.~\ref{fig:jgamma}, current bounds on $j\gamma$ resonances for this choice of parameters imply $m_{\octet} \gtrsim 2.5$ TeV, corresponding to $\hl \gtrsim 6.5$ TeV and thus $m_{\hrho} \gtrsim 5.2$ TeV, which is beyond the projected sensitivities shown in Figs.~\ref{fig:jaja_FCC} and~\ref{fig:3j1gam}. However, there is a significant uncertainty on the $\hrho$ production cross-section, in addition to the model-dependent interplay between the various bounds. We believe that the $2j2\gamma$ and $3j1\gamma$ channel are worthwhile to investigate for evidence of a $\hrho$ as the LHC program continues.

\item At 100 TeV, on-shell $\hrho \to \octet \octet$ production can be discovered for $m_{\hrho}$ up to $\sim 15-20$ TeV in the $3j1\gamma$ final state, and $25-30$ TeV in $2j2\gamma$. From Figs.~\ref{fig:FCC_dijet} and~\ref{fig:jgamma}, such a discovery would likely also be accompanied by $\octet$ and $\hetap$ signatures in the diboson channels.

\end{itemize}

We once again emphasize that these conclusions are influenced by our benchmark choices for the model parameters, in particular the hypercharge of the new vector-like quarks. Nevertheless, these benchmarks illustrate the interplay between the various channels and the concrete potential for discovering a new QCD-like sector, and possibly evidence for a new vacuum angle, at the LHC and a 100 TeV $pp$ collider. We investigate some of the other consequences of $\hth$ below.

\section{Cosmological Implications of $\hth$}
\label{sec:cosmo}
The $\hth$ term in new strongly coupled gauge sectors can lead to an interesting interplay with the strong $CP$ problem and play an important role in early universe cosmology. When the new sector contains states charged under QCD, as in the collider-accessible scenarios discussed in the previous sections, a nonzero $\hth$ implies that QCD must have an axion. Furthermore, in scenarios where a dark pion charged under a species symmetry is a component of dark matter, as in the $U(5)$ model above, a large $\hth$ can drive annihilation processes responsible for setting the relic abundance~\cite{strumia1}. If hypercolor has its own dark axion, $\hth$ is small, and dark matter can be a mixture of QCD and dark axions. We discuss each of these observations in the next sections.

\subsection{Strong $CP$ and the QCD Axion} \label{sec:strong_CP}
The solutions to the strong $CP$ problem can be classified as ``UV" and ``IR," referring to the scale at which the solution operates relative to $\Lambda_{\rm \tiny QCD}$. The most well-known UV solutions are Nelson-Barr (NB) models~\cite{nelsoncp,barrcp,barrcp2,bbp}, based on microscopic $CP$ symmetry, and left-right models~\cite{Beg:1978mt,mohapatrasenjanovic,Georgi:1978xz,Babu:1989rb,barrsenjanovic}, based on microscopic $P$ symmetry. Recently there have also been a number other UV solutions proposed~\cite{hook_cp_violation,prateek1,prateek2}. By contrast, with the lattice exclusion of the vanishing up quark mass, the only known viable IR solution is the Peccei-Quinn mechanism~\cite{Peccei:1977hh,Peccei:1977ur}.

Most UV solutions utilize the fact that radiative corrections to $\theta$ in the SM are extremely small~\cite{ellisgaillard,khriplovich}. In NB models, for example, a new sector spontaneously breaks $CP$ at high scales, which is then communicated to the SM by a mechanism that permits a CKM phase  but does not generate $\theta$. A relatively simple example was given in~\cite{bbp}.

By relying on the small SM renormalization of $\theta$, UV solutions are typically quite fragile. Other new dynamics between the SM and the scales of the UV solution can easily render the solution inoperative, either by introducing new phases that contribute to $\theta$ or new couplings that increase its radiative corrections. Supersymmetry is a well-studied example~\cite{Dine:1993qm,Dine:2015jga,Albaid:2015axa} (and highlights the more general and rather mysterious fact that it is difficult to solve both strong $CP$ and the electroweak hierarchy problem at the same time). Similarly, in some cases, new strong dynamics unaffiliated with the strong $CP$ problem can inadvertently eliminate UV solutions to it~\cite{drapermckeen}. 

This seemingly unfortunate property can be turned around. By searching for new physics coupled to QCD, colliders can potentially {\emph{rule out}} UV solutions to the strong $CP$ problem. In fact, since the scale of UV solutions need not be close to the weak scale, this is likely the easiest way to test these mechanisms. Since the PQ mechanism is the only known IR mechanism, such discoveries could be thought of as constituting an ``indirect detection" of the QCD axion.\footnote{This conclusion neglects the possibility that strong $CP$ has an anthropic origin; although little about macrophysics would change if $\theta$ were much larger (see, for example,~\cite{ubaldi1}), it is not inconceivable that $\theta$ is connected with other problems (for example, the cosmological constant~\cite{kaloperterning,ubaldi2}.}

New QCD-like sectors coupled to QCD are a particularly clean example, as we will now discuss. Coupling to QCD both permits strong production at hadron colliders and transmits a correction to $\tqcd$ of order $\hth$ that rules out UV solutions for any detectably large $\hth$. Let us exhibit this correction in the $U(3)$ and $U(5)$ models analyzed above. 

In the $U(3)$ model, as shown in the appendix, there are two field-redefinition-invariant vacuum angles. Parity violation in $\qcdp$ is controlled by $\thetatwo$. However, at low energies and for small $m$, parity violation in QCD is controlled by the combination 
\begin{align}
\theta_{eff}=\bto-\nhc\thetatwo/3\;.
\end{align}
This can be seen by performing a $U(1)_A^\prime$ transformation to move $\hth$ into the hyperquark mass and then integrating out $\qcdp$. Alternatively, it can be thought of as a threshold correction arising from a vev of order $\hth$ for the $\hetap$ meson of $\qcdp$, which couples to QCD through the anomaly. Using $\theta+{\rm arg\,det}\,m_{q\bar q}+\nhc\hth/3 = \bto-\nhc\thetatwo/3$, we recover $\theta_{eff}$. $\theta_{eff}$ is corrected by ${\cal O}(m)$ effects, and in the opposite limit of large $m$, $\theta_{eff}=\bto$. Thus, if parity-violating effects are seen in $\qcdp$, it is strong evidence that $\theta_{eff}\neq 0$ at the scale of $\qcdp$, and therefore the strong $CP$ problem must be solved further in the infrared.

A similar effect arises in the $U(5)$ model, and the threshold correction is calculable in ChPT. The $\heta$ state couples to the QCD topological charge through the  anomaly, and in the presence of $\hth$, $\heta$ obtains a vev of order $\hth f_{\hpi}$ times a ratio of hyperquark masses. Thus
\begin{align}
\Delta\tqcd\sim\langle\heta/f_{\hpi}\rangle\sim\hth\times (m/m_3)
\end{align}
in the QCD-like limit $m_1\sim m_2\ll m_3$.

There is therefore a close connection between the discoverability of these models at hadron colliders, the detectability of their vacuum angle-dependent processes, and large threshold corrections to $\tqcd$. Although new QCD-like sectors are far from the most general possibility for new physics, they are a sharp proof-of-concept that the discovery of BSM physics can provide useful insight regarding solutions the strong $CP$ problem. In these extensions of the SM, if strong $CP$ is solved dynamically, QCD must have an axion.

The QCD axion has long been known to be a compelling candidate for dark matter~\cite{dinefischler,preskillwilczekwise,abbottsikivie}, but new strong dynamics is also a setting for a variety of other DM candidates, which may then contribute alongside the axion to a sector of multi-component DM. Two candidates,  discussed in the subsequent sections, are particularly tied to the hypercolor $\hth$-term.

\subsection{Dark Hyperpions}
New confining gauge sectors can give rise to a host of possible dark matter candidates, including glueballs, hyperbaryons analogous to protons, hyperpions analogous to the $\pi^\pm$, and dark axions. In the case of dark hyperpions, $\hth$ can play an important role in freezeout processes~\cite{strumia1}.

The $U(5)$ model discussed above is the simplest model with couplings to QCD and a hyperpion dark matter candidate. The off-diagonal neutral $\pi$ states are stabilized by an accidental $U(1)$ species symmetry. For generic parameters, the relic abundance is set primarily by $2\rightarrow 2$ $\hpi\hpi$ annihilations into $\pizero \pizero$, which then decay into SM bosons. However, near the $\heta$ resonance, the abundance can also be determined by $s$-channel annihilations through the CPV $\hpi^*\hpi\heta$ vertex~\cite{strumia1},
\begin{align}
{\cal L}\supset \hpi^*\hpi\heta\;.
\end{align}
Although the $\heta$ is a narrow state, thermal broadening~\cite{Griest:1990kh} creates a larger band of $m_{\hpi} < m_{\heta}$ where the annihilations are resonantly enhanced. In the isospin-preserving limit, this is complementary to the regime with the best collider prospects for the 4-boson resonance processes studied above: maximizing the CPV coupling while still permitting on-shell $\heta\rightarrow\hpi_0\hpi_0$ decays favors near-threshold $\hpi$. 

\begin{figure}[t!]
\begin{center}
\includegraphics[width=0.6\linewidth]{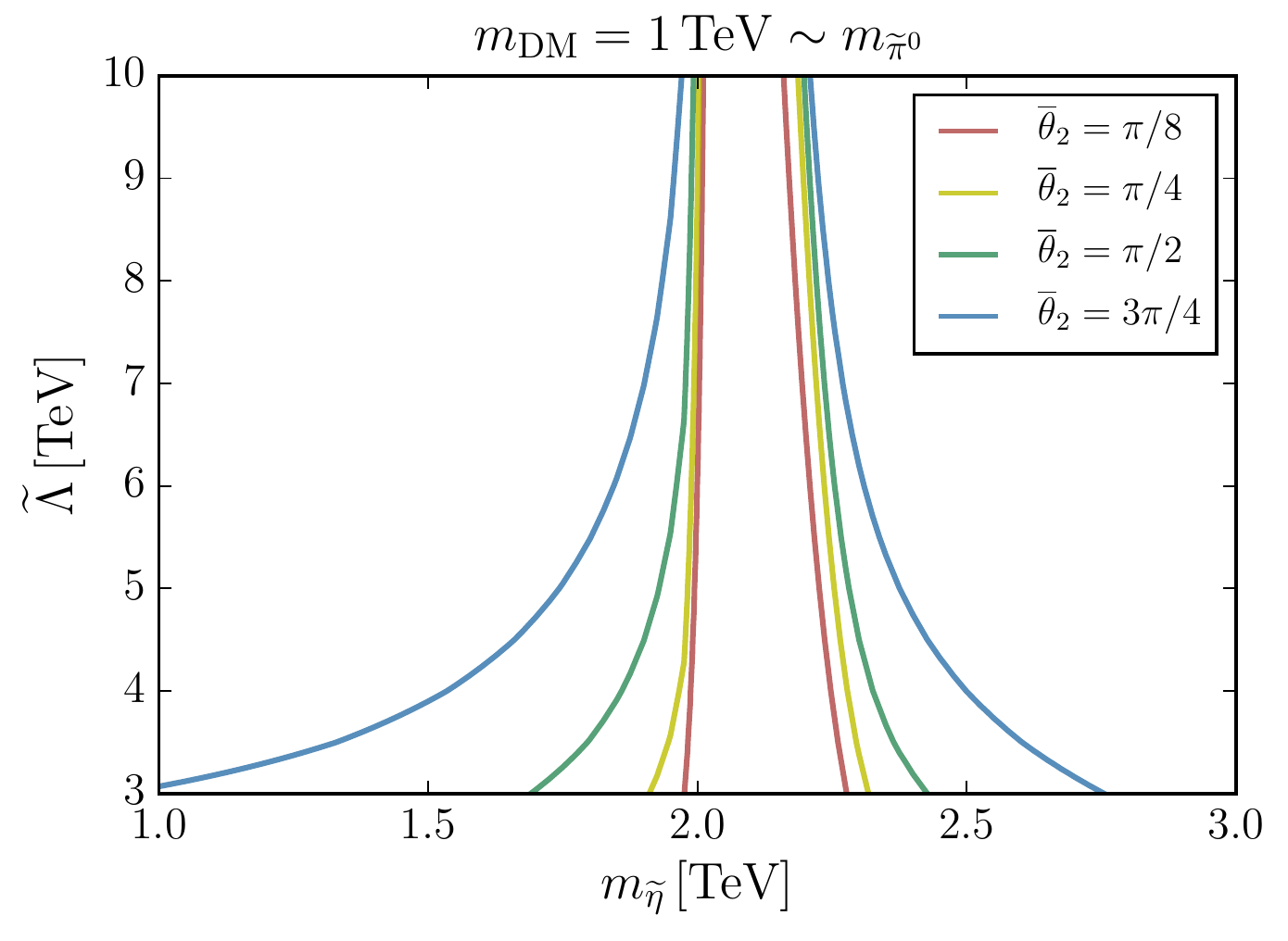}
\caption{Hypercolor sector scales for which the stable neutral hyperpion in the $U(5)$ model constitutes an ${\cal O}(1)$ fraction of dark matter for various values of $\thetatwo$. The freeze-out process is controlled by $CP$-violating $s$-channel annihilation through the $\heta$.} 
\label{fig:darkpionthermal}
\end{center}
\end{figure}

Fig.~\ref{fig:darkpionthermal} shows a portion of the parameter space where 1 TeV dark pions saturate the relic density, including thermal effects. Direct and indirect detection prospects were studied in~\cite{strumia1}, and rates are typically small. Colliders may well provide the most promising setting for probing a dark pion component of the relic density via searches for its visible counterparts\footnote{Note that decays to dark pions will also give rise to mono-$X$ topologies. However, we expect the sensitivity in these channels to be significantly reduced relative to those involving visible decays, due to large backgrounds and a lack of discriminating kinematic features.}. 

Nonthermal production of dark pions associated with early matter domination is also of interest, as we will see in the next section. 

\subsection{Dark Axions}
If $\tqcd$ is relaxed by an axion, the associated PQ symmetry must be of extremely good quality, broken only by the QCD anomaly to a part in $10^{10}$~\cite{kamionkowski}. Such an axion can arise in string theory compactifications~\cite{svrcekwitten}, and it has been suggested that a typical compactification producing a viable QCD axion will produce many other light axion-like particles as well~\cite{Arvanitaki:2009fg}. In that case, one (linear combination) of these fields might couple to the hypercolor sector.

If hypercolor has a ``dark axion", $\hth$ is partially or completely dynamically relaxed, and we do not expect large parity-violating signatures at colliders (although 4-boson resonances are associated with the $\hrho$ are still expected). In this case, it is interesting to consider dark matter with contributions from both a QCD axion and a dark axion. We will only consider coherent oscillations from misalignment and assume that all of the dark axion mass comes from hypercolor.

In a typical model, both axions couple to both QCDs with different anomaly strengths which we take to be ${\cal O}(1)$,
\begin{align} 
{\cal L}\sim \left(c_{1}\frac{a_1}{f_1}+c_{2}\frac{a_2}{f_2}\right)G\widetilde G+\left(\frac{a_1}{f_1}+\frac{a_2}{f_2}\right)H\widetilde H
\end{align}
In general there are also coefficients for $a_i/f_i$ in the couplings to hypercolor, but for simplicity we have absorbed them into the $f_i$.  

\begin{figure}[t!]
\begin{center}
\includegraphics[width=0.6\linewidth]{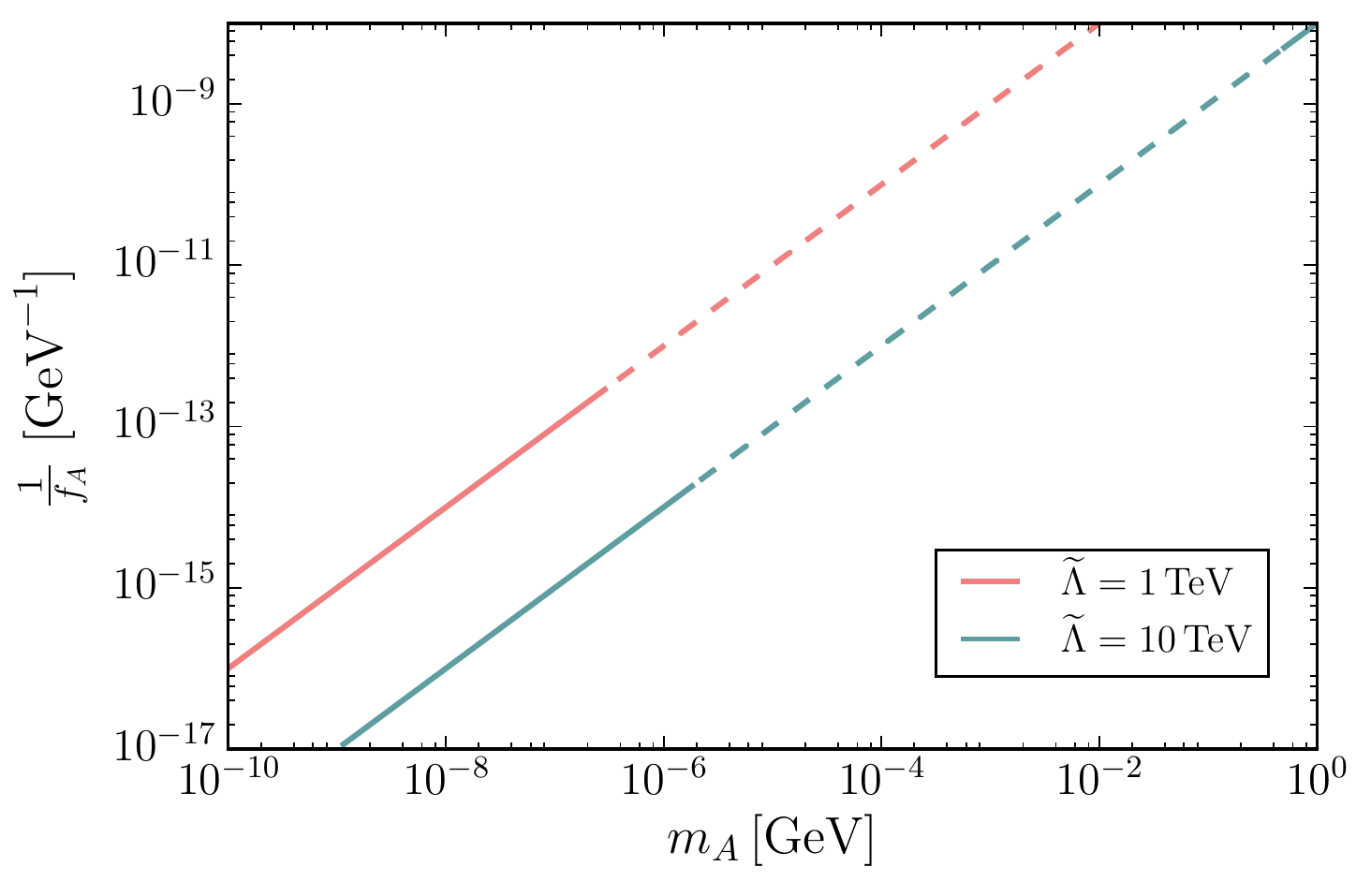}
\caption{The dark axion mass-decay constant relation for $\hl=1$ TeV (upper) and 10 TeV (lower). Dashed regions correspond to $a_A\rightarrow\gamma\gamma$ decays in conflict with either BBN or CMB observations.} 
\label{fig:darkaxionmass}
\end{center}
\end{figure}

Since we are interested in models with colored hyperquarks, $\hl$ is necessarily $\gg \Lambda_{\rm QCD}$. It is then convenient to change the basis, defining
\begin{align}
\frac{a_A}{f_A} &\equiv \frac{a_1}{f_1}+\frac{a_2}{f_2}\;\;\;\;\;\;\;\;\;\;\;\;\;f_A\equiv\frac{f_1f_2}{\sqrt{f_1^2+f_2^2}}\nonumber\\
\frac{a_B}{f_B} &\equiv \frac{-f_1 a_1+f_2a_2}{f_1^2+f_1^2}\;\;\;\;\;f_B\equiv\sqrt{f_1^2+f_2^2}\;.
\label{eq:fdefs}
\end{align}
The dark axion is mostly the $a_A$ state, while the QCD axion is mostly the $a_B$ state. Note that $f_A\sim \min(f_1,f_2)$ and $f_B\sim \max(f_1,f_2)$. The dark axion mass vs $f_A$ is shown in Fig.~\ref{fig:darkaxionmass}. The couplings to QCD are 
\begin{align}
{\cal L}\sim \left(c_A\frac{a_A}{f_A}+c_B\frac{a_B}{f_B}\right)G\widetilde G\;.
\label{eq:QCDcoupling}
\end{align}
where, with the definitions of Eq.~(\ref{eq:fdefs}), $c_{A,B}\sim 1$ for arbitrary confinement scales and decay constants.

As the universe cools the axion potentials start to turn on. Eventually the dark axion starts to oscillate\footnote{Note that for certain choices of the various scales and parameters, the interplay between the contributions to the potential and the resulting dynamics can be nontrivial; see e.g.~Ref.~\cite{Dienes:2015bka}.}, corresponding mostly to the lower decay constant and the larger $\Lambda$. The temperature dependence of the potential is relevant if oscillation begins during a conventional radiation-dominated phase.  For concreteness, we assume the same scaling as in QCD, namely 
\begin{align}
m_A(T)\sim m_A(0)\times\left(\frac{\hl}{T}\right)^4
\end{align}
above $\hl$, which approximately valid for both the $U(3)$ and $U(5)$ models for three hypercolors. Then the dark axion comes to dominate the energy density at
\begin{align}
T/\hl \sim \hth_0^2(f_A/M_p)^{\frac{7}{6}}\;,
\label{eq:Teq}
\end{align}
where $\hth_0$ is the initial misalignment angle. 

For $\hth_0\sim 1$,
preventing matter domination before 1 eV appears to require extremely low $f_A$ compared to  typical considerations for the QCD axion. For $\hl\sim 10$ TeV, Eq.~(\ref{eq:Teq}) suggests $f_A\sim 10^7$ GeV. On the other hand, if $a_A$ has $\sim\alpha/4\pi f_A$ coupling to QED and $f_A$ is so low, the dark axion typically decays before BBN and is not a part of the relic density. However, it is still subject to other astrophysical constraints through the couplings in Eq.~(\ref{eq:QCDcoupling}).

Larger decay constants are permissible with some fine-tuning of the initial misalignment angle. If we require the dark axion to contribute to the present-day dark matter desnity, then its lifetime must be much longer than the age of the universe to satisfy CMB constraints. These constraints are illustrated in Fig.~\ref{fig:darkaxionmass}. For $\hl\sim 10$ TeV, lifetimes consistent with the CMB require $f_A\gtrsim 10^{14}$ GeV. Then the initial misalignment angle must be fine-tuned to a part in $10^4$. 

The idea that both QCD and hypercolor have their own axion is most natural in the context of a string axiverse, in which case large decay constants are easier to understand. However, in this context, there is another route to achieve an acceptable cosmology without fine-tuning the initial misalignment angle: early matter domination driven by scalar moduli (saxions) can eliminate the relic density dependence on $\hl$.

Let us assume  saxion domination occurs, and eventually the saxion(s) decay at some $T_R\gtrsim 10$ MeV. Then both axions oscillate during matter domination. Because the Hubble scale is larger, the oscillation temperatures are lower than the corresponding  temperatures in radiation domination, and the temperature dependence of the axion masses is unimportant. Furthermore, each axion occupies a {\emph{fixed}} fraction $(\theta_{i0} f_i/M_p)^2$ of the energy budget until  $T_R$~\cite{Banks:2002sd}. This property applies both to the dark axion and the QCD axion. Correspondingly, the ratio of energy densities at late times is
\begin{align}
\frac{\Omega_A}{\Omega_B}=\left(\frac{\hth_0 f_A}{\theta_0 f_B}\right)^2
\end{align}
and axion domination around 1 eV is obtained if
\begin{align}
\max \left(\frac{\theta_{i0} f_i}{M_p}\right)^2\sim \frac{\rm eV}{T_R}\;.
\end{align}
In contrast to the conventional scenario without early matter domination, we see that the relic abundance of dark axions is insensitive to $\hl$. Furthermore, for typical couplings and comparable misalignment angles, the hierarchy $f_A\lesssim f_B$ implies that the QCD axion is a {\emph{larger}} component of the late-time energy density than the dark axion unless the decay constants are quite similar.

\subsection{Dark Hyperpions with Early Matter Domination}
Previously, we considered a species-stabilized hyperpion contribution to dark matter, with the relic abundance set by thermal freeze-out through annihilation processes involving $\hth$. As discussed in~\cite{strumia1}, the relic abundance can also be obtained through thermal freeze-out with annihilations $\hpi\hpi\rightarrow\hpi^0\hpi^0$, relevant for scenarios in which $\hth$ is small or vanishing. 

The dark axion scenario discussed in the previous section motivates reconsidering the hyperpion relic abundance with small $\hth$ and early matter domination. The total dark matter density in scenarios like our $U(5)$ model could then be a non-interacting mixture of QCD axions, dark axions, and hyperpions.

\begin{figure}[t!]
\begin{center}
\includegraphics[width=0.55\linewidth]{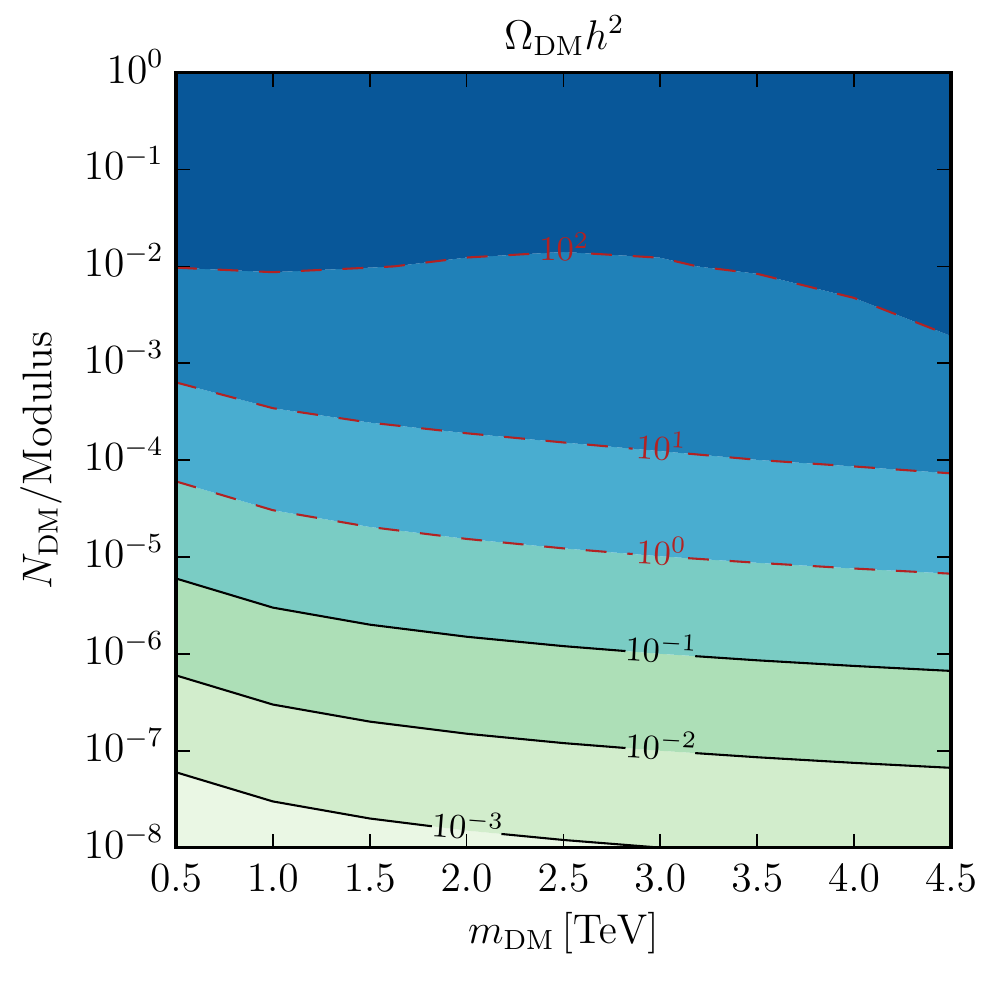}
\caption{Contours of the stable neutral hyperpion relic density, $h^2\Omega_{\rm DM}$, in the $U(5)$ model with early modulus domination. Red dashed contours correspond to relic abundances exceeding the observed cold dark matter density. Axes correspond to the hyperpion mass and the number produced per modulus decay. For this benchmark point, the modulus mass is 100 TeV (corresponding to a reheat temperature $\sim$10 MeV), the hypercolor scale is $\hl=10$ TeV, the $\heta$ mass is $5$ TeV, and the $\hpi^0$ is taken to be $2/3\,m_{DM}$ so that annihilations are not kinematically suppressed. Typically we require $N_{\rm DM}\sim 10^{-5}$, while annihilations are not significant until $N_{\rm DM}\gtrsim10^{-2}$.} 
\label{fig:earlymdpion}
\end{center}
\end{figure}

In the early matter domination scenario, the hyperpion relic abundance is set by the number $N_{\rm DM}$ produced in each modulus decay and the subsequent annihilation rate~\cite{moroirandall}. Na\"ively, $N_{\rm DM}$ should be of order $1$ per modulus, since the lightest modulus $S$ can couple to hypercolor at dimension 5, via $SH_{\mu\nu}H^{\mu\nu}/M_p$. However, as one might expect from experience with winos~\cite{moroirandall,dinedraperbose,Cohen:2013ama, Fan:2013faa, Blinov:2014nla}, the subsequent annihilation rate is much too low to prevent overclosure. Numerically integrating the Boltzmann equations, we find that we require $N_{\rm DM}\sim 10^{-5}$ per modulus, so that annihilations essentially play no role and every hyperpion produced in a modulus decay contributes to the relic abundance. A benchmark point is shown for illustration in Fig.~\ref{fig:earlymdpion}.

\subsection{Summary}
We have sketched some of the theoretical and cosmological implications of $\hth$ in a new hypercolor sector. When $\hth$ is large and hypercolor couples to QCD, UV solutions to strong $CP$ are not viable, and QCD must have an axion. The QCD axion can then contribute to dark matter in the usual ways. 

Apart from their interesting collider signatures, a motivation for studying hypercolor sectors in the first place is their realization of several dark matter candidates. Previous authors have shown that one scenario for obtaining a viable hyperpion relic density is through $\hth$-dependent resonant annihilations~\cite{strumia1}. With thermal effects included in the annihilation rates, parameters consistent with the relic density can be complementary to regions with good collider reach for the $\hth$-dependent processes studied above.

If we live in an axiverse, $\hth$ may be small if hypercolor is the primary source of another axion's mass. A dark axion coupled to a hypercolor sector with collider-relevant confinement scale is most easily realized if the universe undergoes a period of early matter domination with low reheating temperature. In this case, the relic axion abundance receives contributions from both dark and QCD axion oscillations with fractions set by the effective decay constants~(\ref{eq:fdefs}).

We have also seen that not all of these possible contributions to dark matter fit well together: with a dark axion and early matter domination, dark hyperpions are overproduced unless their occurrence in modulus decays is very rare. 

\section{Conclusions} \label{sec:concl}
Models of new QCD-like sectors are well-motivated, natural extensions of the Standard Model offering a rich spectrum of signatures at the high-luminosity LHC and future hadron colliders. In this context we have studied decays of heavier hypermeson states into lighter hypermesons, resulting in $4g$, $2g2\gamma$, and $3g1\gamma$ resonant final states. These processes are associated with $CP$-violating triple-pion interactions controlled by the vacuum angle $\hth$ as well as  $\hrho\hpi\hpi$ couplings. We provided sensitivity estimates for these signatures in benchmark models at the HL-LHC and a 100 TeV $pp$ collider, finding that the 4-boson resonances are an interesting and relevant probe at current and future colliders and can offer reach complementary to diboson final states.

We have also discussed axion, dark axion, and dark hyperpion candidates for dark matter in these models and the roles of $\hth$ in the cosmological history. In some cases collider searches offer complementary information: evidence for a large $\hth$ strongly supports the existence of a QCD axion, while small $\hth$ may indicate a dark axion, possibly with a period of early saxion domination to avoid overclosure constraints. 

There are numerous opportunities for generalization, including hypermeson cascade decays involving weak bosons (see also~\cite{baidobrescu}) and Higgs bosons, decays involving dark pions, and models with very light hyperpions. It would also be interesting to investigate the prospects for directly measuring the $CP$ properties of the various hypermesons at colliders. We hope to pursue some of these directions in future work.

\vskip 1cm

\section*{Acknowledgements}
We thank John Paul Chou,  Eva Halkiadakis, Scott Thomas, and Felix Yu for helpful discussions. The work of PD is supported by NSF grant PHY-1719642. JHY is supported by DOE Grant de-sc0011095.

\vskip 1cm

\appendix
\section{Model Details}
\label{app:models}

In this appendix we compute various properties of the $U(5)$ and $U(3)$ models utilized in the main text. In the first section we summarize our conventions that are common to both models. We then describe the invariant vacuum angles, spectra, couplings, and rates in each model.

\subsection{Conventions}

Chiral symmetry breaking in $\qcdp$ takes place at a scale $\hl$. The spectrum of each model includes a color octet $\octet$ and a singlet $\hetap$ analogous to the $\eta^\prime$ of QCD; the spectrum of the $U(5)$ model contains additional singlet and triplet states. The $\hetap$ is only a pseudo-Goldstone in the large-$\nhc$ limit, but the chiral description will be sufficient for modeling purposes.
We  introduce these hyperpions in the $\Sigma$ basis,
\begin{align}
\Sigma\equiv e^{2i\hpi^a T^a/\hfpi}\;,
\end{align}
where $T^a$ is a generator in the fundamental of SU(3) and we normalize $T(R)=1/2$. 

The leading-order chiral Lagrangian is
\begin{align}
{\cal L}=\frac{\hfpi^2}{4}\TR\left(D_\mu\Sigma^\dagger D^\mu\Sigma\right)+\frac{\mu\hfpi^2}{2}\TR\left(\Sigma^\dagger M+M^\dagger\Sigma\right)-V_0(\hetap)\;,
\end{align}
with $\mu\simeq \hl\simeq 4\pi \hfpi$.

Anomaly matching provides additional single-hyperpion couplings to SM field strengths.
Under an approximate global symmetry, the Lagrangian shifts by
\begin{align}
\Delta {\cal L}=\alpha^a \partial_\mu J^{\mu a}.
\end{align} 
Each chiral anomaly contributes 
\begin{align}
\partial_\mu J^{\mu a} \supset\frac{N_f}{8\pi^2}g_1g_2 D^{abc} G_{1\mu\nu}^b\widetilde G_2^{c\mu\nu}\;,
\end{align}
where
\begin{align}
D^{abc}\equiv\frac{1}{2}\TR(T^a\{T_1^b,T_2^c\})\;
\label{eq:dabc}
\end{align}
and $N_f$ is the number of Dirac flavors.

In addition to the ordinary kinetic term and gauge couplings to the SM, the UV hypercolor action contains a new source of parity violation,
\begin{align}
{\cal L}\supset \frac{\hth \hg^2}{32\pi^2}H_{\mu\nu}^a \widetilde H_{\mu\nu}^a+i {\rm Im}\,m_{ij}\,\psi_i\gamma^5\bar\psi_j
\end{align}
where $H$ is the $SU(\nhc)$ field strength. $\hth$ and the quark mass phases can be combined into vacuum angles invariant under chiral field redefinitions.

\subsection{$U(3)$ Model}

The $U(3)$ model contains three hyperquarks with SM quantum numbers $(3,1)_{4/3}$.
\subsubsection{Vacuum Angles}
In addition to the usual anomalous $U(1)_A$ of QCD, there is an additional $U(1)_A^\prime$  under which $\psi_L \rightarrow e^{-i\phi}\psi_L$, $\psi_R \rightarrow e^{i\phi}\psi_R$. This transformation is anomalous under both QCD and $\qcdp$. The two invariant vacuum angles are
\begin{align}
\bto&=\theta+{\rm arg\,det}\,m_{q\bar q}+\nhc{\rm arg}\,m\nonumber\\
\thetatwo&=\hth+3\,{\rm arg}\,m\;
\end{align}
where $m_{q\bar q}$ is the SM quark mass matrix and $m$ is the hyperquark mass, which must be universal since the $SU(3)_V$ symmetry is gauged. At low energies, $CP$-violation is described by a parameter-dependent combination of the two, discussed further in Sec~\ref{sec:strong_CP}.

\subsubsection{Spectra and Rates}
We  focus first on the $\octet$ and $\hetap$ states and two sets of their couplings: anomaly-induced couplings of the form $\hpi G_1 \widetilde G_2$, and parity-violating couplings of the form $\hetap\octet\octet$. The former provide production and decay modes for both states, while the latter generates decays sensitive to $\thetatwo$ that can provide the primary $\hetap$ decay mode. As stated above, including the $\hetap$ in ChPT is not expected to be particularly accurate, but it is sufficient for modeling purposes. 

The hyperpion matrix is
\begin{align}
\Sigma=e^{2i\left[(\octet)^a T^a+\hetap/\sqrt{6}\right]/\hfpi}
\end{align}
and the hyperquark mass matrix is
\begin{align}
M=me^{i\thetatwo/3}{\mathbb{I}}_{3\times 3}\;.
\label{eq:Mquark}
\end{align}
$V_0(\hetap)$ is modeled by its large-$\nhc$ form,
\begin{align}
V_0(\hetap) = a|\log\det(\Sigma)|^2 = a\left(\frac{\sqrt{6}\hetap}{\hfpi}+2\pi k\right)^2\;.
\end{align}
The branches $k$ are important for maintaining $2\pi$ periodicity of $\thetatwo$. Just as $\hth$ was moved into the quark mass matrix in Eq.~(\ref{eq:Mquark}), we can absorb the branch label into the definition of $\hetap$. The full $\hetap$ potential is then approximated by
\begin{align}
V(\hetap) = c\hl^2 {(\hetap)}^2 - 3\hl \hfpi^2m\cos\left(\frac{2\hetap}{\sqrt{6}\hfpi}-\frac{\thetatwo+2\pi k}{3}\right)\;.
\end{align}
We will assume that $\nhc$ is not large and that the first term in $V(\hetap)$ dominates. Then, in this basis, $\langle\hetap\rangle\sim {\cal O}(m)$ and can be neglected, while $m_{\hetap}\approx\hl$. 
The second term in $V$ controls the ordering of the branches: for $0\leq\thetatwo < \pi$, the branch with minimum energy is $k=0$, while for $\pi<\thetatwo < 2\pi$, it is $k=-1$. For $\thetatwo=\pi$ the two branches are degenerate and reflect spontaneous $CP$ violation. 

The octet mass is determined by a radiative contribution from gluon loops of order $\hl/3$ and a chiral contribution,
\begin{align}
m_8^2\approx (\hl/3)^2+2\mu m \cos\left(\frac{\thetatwo+2\pi k}{3}\right)\;.
\end{align}
The $\hetap\octet\octet$ coupling is
\begin{align}
\frac{2\mu m}{\sqrt{6}\hfpi}\sin\left(\frac{\thetatwo+2\pi k}{3}\right)\,\hetap\octet\cdot\octet\;.
\end{align}
The anomaly couplings,
\begin{align}
\beta^{abc}\hpi^aG_1^b\widetilde G_2^c\;,
\end{align}
 may be inferred from anomaly matching, 
\begin{align}
\beta^{abc}=-\frac{N_f g^2 D^{abc}}{8\pi^2 \hfpi}\;.
\label{eq:betaabc}
\end{align}

We can now compute the effective couplings  $\hetap G\widetilde G$, $\octet G\widetilde G$, $\hetap F\widetilde F$, $\octet G\widetilde F$ in the $U(3)$ model using Eqs.~(\ref{eq:betaabc}),~(\ref{eq:dabc}), and 
\begin{align}
\frac{1}{2}\TR\left[\left(\frac{1}{\sqrt{6}}\right)\left\{T^a,T^b\right\}\right]&=\frac{1}{2\sqrt{6}}\delta^{ab}\nonumber\\
\frac{1}{2}\TR\left[T^a\left\{T^b,T^c\right\}\right]&=\frac{1}{4}d^{abc}\nonumber\\
\frac{1}{2}\TR\left[\left(\frac{1}{\sqrt{6}}\right)\left\{Q,Q\right\}\right]&=\frac{16}{3\sqrt{6}}\nonumber\\
\frac{1}{2}\TR\left[T^a\left\{T^b,Q\right\}\right]&=\frac{2}{3}\delta^{ab}\;.
\end{align}
The number of Dirac flavors is $\nhc$, the number of hypercolors. Together these determine the anomaly couplings, which determine the rates $\hetap\rightarrow gg,\gamma\gamma$, $\octet\rightarrow gg,g\gamma$. 
The rates are given by:
\begin{align}
\Gamma(\hetap\rightarrow \gamma\gamma) &= \frac{1}{4\pi} m^3_{\hetap} A_{\hetap gg}^2\nonumber\\
\Gamma(\hetap\rightarrow gg) &= \frac{2}{\pi} m^3_{\hetap} A_{\hetap gg}^2\nonumber\\
\Gamma(\hetap\rightarrow\octet\octet) &= \frac{A_{\hetap\octet\octet}^2}{\pi m_{\hetap}}\sqrt{1-\frac{4m_{\octet}^2}{m_{\hetap}^2}}\nonumber\\
\Gamma(\octet\rightarrow g\gamma) &= \frac{1}{8\pi} m^3_{\octet} A_{\octet g\gamma}^2\nonumber\\\Gamma(\octet\rightarrow gg) &= \frac{1}{4\pi} m^3_{\octet} \frac{1}{8}\sum_{a,b,c=1}^8 |\beta^{abc}|^2
\end{align}
where the $A_{ijk}$ are coefficients of $\hpi G_1\widetilde G_2$ (and $\hetap\octet\octet$), and in the last case we have left the coupling in terms of $\beta^{abc}$ as given above.
Leading-order gluon-fusion cross sections are related to $\Gamma_{gg}$ rates by
\begin{align}
\sigma(gg\rightarrow \hetap)&=\frac{\pi^2}{8s m}{\cal L}\cdot\Gamma(\hetap\rightarrow gg)\nonumber\\
\sigma(gg\rightarrow \octet)&=\frac{\pi^2}{s m}{\cal L}\cdot\Gamma(\octet\rightarrow gg)
\end{align}
where 
\begin{align}
{\cal L}=\int^{\log(\sqrt{s}/m)}_{\log(m/\sqrt{s})}dy\,f(me^y/\sqrt{s})f(me^{-y}/\sqrt{s})\;.
\end{align}

In addition to the pseudoscalars, this setup also features a set of vector mesons with masses near the confinement scale. In particular, there is a vector octet state, $\hrho$, analogous to the $\rho$ of QCD, that kinetically mixes with the gluon and can thus be singly produced at colliders. The kinetic mixing induces a coupling of the $\hrho$ to Standard Model fermions, which we take to be parametrically $\sim \alpha_s/g_s$, as motivated in Refs.~\cite{kilic1,kilic2}. As in QCD, we also expect a large coupling of the $\hrho$ to $\octet \octet$, which we take to be $\sim 4\pi$. The $\hrho$ Lagrangian we consider for our collider studies is thus
\beq
\mathcal{L}_{\hrho} = -\frac{1}{4} \operatorname{Tr} F_{\mu \nu} F^{\mu \nu} + \frac{1}{2} m_{\hrho}^2 \hrho_{\mu}\, \hrho^{\mu} - 4\pi f^{abc} \, \hrho^{\mu}_a \, \octet_b \, \partial_{\mu} \octet_c + \alpha_s \hrho^{\mu}_a \, \overline{q} \gamma_{\mu} T^a q
\eeq
where $F_{\mu \nu}$ is the $\hrho$ field strength and $q$ denotes all Standard Model quarks. Note that, due to the large coupling, the $\hrho$ decays almost exclusively to $\octet \octet$ if kinematically allowed, and to $q \bar{q}$ otherwise. The Lagrangian above is of course schematic, and the precise values of the couplings, mass, and resulting cross-sections should be interpreted correspondingly.

\subsection{$U(5)$ Model}

The $U(5)$ model contains five hyperquarks with SM quantum numbers of $(3,1)_{4/3}+2\times (1,1)_0$.

\subsubsection{Vacuum Angles}
The two invariant angles in the $U(5)$ model are
\begin{align}
\bto&=\theta+{\rm arg\,det}\,m_{q\bar q}+\nhc{\rm arg}\,m_3\nonumber\\
\thetatwo&=\hth+3\,{\rm arg}\,m_3+{\rm arg\,det}\,m_{12}\;
\end{align}
where $m_{q\bar q}$ is the SM quark mass matrix, $m_3$ is the colored hyperquark mass, and $m_{12}$ is the neutral hyperquark mass matrix which we take to be proportional to ${\rm diag}(m_1,m_2)$.

In the limit $\det m_{12}\rightarrow 0$, $\thetatwo$ is unphysical and $CP$ violation in the QCD sector is controlled by $\bto$. In the opposite limit of large $m_{1,2}$, the model reduces to the $U(3)$ model of the previous section. We will be interested in the case $0<m_{1,2}<m_3<\hl$. In this case low-energy $CP$ violation in QCD is controlled by $\bto+c\times \langle\heta\rangle/\hfpi$, where $c$ is a constant and $\heta$ is a hypermeson analogous to the QCD $\eta$, which obtains a vev of order $\hth\times m_{1,2}$.

\subsubsection{Spectra and Rates}
We work in the approximations 
\begin{align}
m_1\approx m_2\equiv m\ll m_3
\label{eq:appx}
\end{align}
 which yields simple analytic formulae and is a reasonable first approximation in the parameter regimes relevant for our collider studies. By a combination of chiral rotations, $\thetatwo$ can be placed entirely in $\arg\det m_{12}$ so that the mass matrix takes the form
\begin{align}
M={\rm diag}(me^{i\thetatwo/2},me^{i\thetatwo/2},m_3,m_3,m_3)\;.
\end{align}
The ``isospin limit" $m_1\approx m_2$ suppresses neutral hyperpion mixing, so that the $\heta$ and $\hpi^0$ states correspond approximately to the $SU(5)$ generators 
\begin{align}
T_{\hpi^0}=\frac{1}{2}\left(\begin{array}{ccc}
1 &0&0_{1\times 3} \\
0 & -1&0_{1\times 3}\\
0_{3\times 1}&0_{3\times 1}&0_{3\times 3}
\end{array}\right)\;\;\;\;T_{\heta}=\frac{1}{\sqrt{15}}\left(\begin{array}{ccc}
-\frac{3}{2} &0&0_{1\times 3} \\
0 & -\frac{3}{2}&0_{1\times 3}\\
0_{3\times 1}&0_{3\times 1}&1_{3\times 3}
\end{array}\right)\;.
\end{align}
In the approximations~(\ref{eq:appx}), the relevant masses are
\begin{align} \label{eq:U5masses}
m_{\hpi}^2=&2 m\hl\cos\left(\frac{\thetatwo+2\pi k}{2}\right)\nonumber\\
m_{\heta}^2=&\frac{4}{5}m_3\hl+\frac{6}{5}m\hl\cos\left(\frac{\thetatwo+2\pi k}{2}\right)\nonumber\\
m_{\hpi^8}^2=&2 m_3\hl
\end{align}
where $m_{\rm DM}\approx m_{\hpi^0}$ near the isospin limit, and $k$ is again a branch label equal $0$ for $0\leq\thetatwo<\pi$ and $1$ for $\pi\leq\thetatwo<2\pi$.
The isospin-breaking $\heta-\hpi$ mixing angle is
\begin{align}
\theta_{\heta-\hpi}= \frac{\sqrt{15}(m_1-m_2)}{2m_3}\cos\left(\frac{\thetatwo-2\pi k}{2}\right)
\label{eq:etapimix}
\end{align}
and the $CP$-violating triple-pion coupling inducing $\heta\rightarrow\hpi\hpi$ decays is
\begin{align}
\lambda_{\heta\hpi\hpi}=-\sqrt{\frac{3}{5}}\frac{m\hl}{\hfpi}\sin\left(\frac{\thetatwo+2\pi k}{2}\right)\;.
\end{align}
In the presence of $\thetatwo$, the $\heta$ state receives a small vacuum expectation value,
\begin{align}
\langle\heta/\hfpi\rangle=-\frac{\sqrt{15}m \hfpi}{2 m_3}\sin\left(\frac{\thetatwo+2\pi k}{2}\right)\;.
\end{align}
The $\heta$ also couples to QCD and QED through anomalies with coefficients
\begin{align}
\frac{1}{2}\TR\left[T_{\heta}\left\{T_Q,T_Q\right\}\right]&=\frac{16}{3\sqrt{15}}\nonumber\\
\frac{1}{2}\TR\left[T_{\heta}\left\{T^a,T^b\right\}\right]&=\frac{1}{2\sqrt{15}}\delta^{ab}\;.
\end{align}
The $\hpi^0$ inherits small SM anomaly couplings through the mixing~(\ref{eq:etapimix}). Together the anomaly couplings and the $\heta$ vev lead to a threshold correction to the QCD $\theta$-angle when $\qcdp$ is integrated out,
\begin{align}
\Delta \theta_{QCD}=\nhc \frac{m}{m_3}\sin\left(\frac{\thetatwo+2\pi k}{2}\right)\;.
\end{align}

\bibliography{theta_and_new_QCD}{}
\bibliographystyle{JHEP}

\end{document}